\documentclass{aa}  
\usepackage{booktabs}
\usepackage{graphicx}
\usepackage{subcaption}
\usepackage{tabularx}
\usepackage{amsmath}
\usepackage{txfonts}
\usepackage{microtype}      
\fussy                    
\usepackage[]{hyperref}
\hypersetup{
    colorlinks,
    linkcolor={blue},
    citecolor={blue},
    urlcolor={blue}
}
\usepackage[dvipsnames]{xcolor}
\usepackage{float}
\usepackage{makecell}
\renewcommand{\arraystretch}{1.05}
\usepackage{placeins}
\usepackage{tabularx}
\usepackage{siunitx}
\usepackage{lastpage}
\usepackage{orcidlink} 

\DeclareCaptionFont{mycap}{\fontsize{8.8}{10}\selectfont}
\makeatletter
\renewcommand*\aa@tablefootfont{\fontsize{8.8}{10}\selectfont}
\renewcommand*\aa@authorfont{\large\leftskip=-1.cm plus 1fil \rightskip=-1.cm plus 1fil}
\AtBeginDocument{\renewcommand*\aa@pageof{, page \thepage{} of \pageref{LastPage}}}
\renewcommand\paragraph{
  \@startsection{paragraph}{4}{1.5em}{-16dd}
      {-\fontdimen2\font
      \@plus-\fontdimen3\font
      \@minus-\fontdimen4\font}{\normalsize\sffamily\itshape}}
\@ifpackagelater{siunitx}{2021/05/01}{
  \sisetup{separate-uncertainty = true,
           separate-uncertainty-units = single}
}{\sisetup{separate-uncertainty = true,
           multi-part-units = single}}
\makeatother

\DeclareSIUnit{\Myr}{Myr}
\DeclareSIUnit{\Msun}{\text{M}_\odot}
\DeclareSIUnit{\Mjup}{\text{M}_\text{Jup}}
\DeclareSIUnit{\Mearth}{\text{M}_\oplus}
\DeclareSIUnit{\Rjup}{\text{R}_\text{Jup}}
\DeclareSIUnit{\pc}{pc}
\DeclareSIUnit{\AU}{au}
\DeclareSIUnit{\degree}{deg}
\DeclareSIUnit{\arcmin}{arcmin}
\DeclareSIUnit{\arcsec}{arcsec}
\DeclareSIUnit{\mas}{mas}
\DeclareSIUnit{\mag}{mag}
\DeclareSIUnit{\micron}{\micro \meter}
\DeclareSIUnit{\sr}{sr}

\newcommand{\naco}{NaCo\xspace}
\newcommand{\sphere}{SPHERE\xspace}
\newcommand{\irdis}{IRDIS\xspace}
\newcommand{\ifs}{IFS\xspace}
\newcommand{\ispynaco}{ISPY--NaCo\xspace}
\newcommand{\ispysphere}{ISPY--SPHERE\xspace}
\newcommand{\snr}{\ensuremath{\text{S/N}}\xspace}
\newcommand{\fours}{\texttt{4S}\xspace}

\AtBeginDocument{
  \renewcommand{\figureautorefname}{Fig.}}

\newcommand{\Autoref}[1]{
  \begingroup
  \renewcommand{\figureautorefname}{Figure}
  \autoref{#1}
  \endgroup}

\begin{document} 

   \title{ISPY: NaCo Imaging Survey for Planets around Young stars}
   \subtitle{SPHERE follow-up observations reveal scattered-light disks, stellar binaries, and candidate exoplanets}

   \author{P.~V.~Steimle\inst{\ref{inst1},\ref{inst2}}\corrauth{paul.steimle@uni-goettingen.de}\orcidlink{0009-0001-4452-4944}
          \and E.~C.~Matthews\inst{\ref{inst1}}\email{matthews@mpia.de}\orcidlink{0000-0003-0593-1560}
          \and M.~J.~Bonse\inst{\ref{inst3},\ref{inst4}}\email{markus.bonse@eso.org}\orcidlink{0000-0003-2202-1745}
          \and R.~Launhardt\inst{\ref{inst1}}\email{rl@mpia.de}\orcidlink{0000-0002-8298-2663}
          \and E.~L.~Rickman\inst{\ref{inst5}}\email{erickman@stsci.edu}\orcidlink{0000-0003-4203-9715}
          \and G.~Cugno\inst{\ref{inst6}}\email{gabriele.cugno@uzh.ch}\orcidlink{0000-0001-7255-3251}
          \and Th.~Henning\inst{\ref{inst1}}\email{henning@mpia.de}\orcidlink{0000-0002-1493-300X}
          \and S.~Metchev\inst{\ref{inst7}}\email{smetchev@uwo.ca}\orcidlink{0000-0003-3050-8203}
          \and T.~D.~Pearce\inst{\ref{inst8}}\email{tim.pearce@warwick.ac.uk}\orcidlink{0000-0001-5653-5635}
          \and S.~Reffert\inst{\ref{inst9}}\email{sreffert@lsw.uni-heidelberg.de}\orcidlink{0000-0002-0460-8289}
          \and H.~Ruh\inst{\ref{inst9}}
          \and M.~Samland\inst{\ref{inst1}}\email{samland@mpia.de}\orcidlink{0000-0001-9992-4067}
          \and R.~van~Boekel\inst{\ref{inst1}}\email{boekel@mpia.de}\orcidlink{0000-0002-2190-3108}
          \and G.~Volakakis\inst{\ref{inst1}}\email{ivolakakis@physics.uoc.gr}\orcidlink{0009-0008-7164-5750}
          }

   \institute{Max-Planck-Institut für Astronomie, Königstuhl 17, 69117 Heidelberg, Germany\label{inst1}
   \and Institut für Astrophysik und Geophysik, Georg-August-Universität Göttingen, Friedrich-Hund-Platz 1, 37077 Göttingen, Germany\label{inst2}
   \and European Southern Observatory, Karl-Schwarzschild-str. 2, 85748 Garching, Germany\label{inst3}
   \and Max-Planck-Institut für Intelligente Systeme, Max-Planck-Ring 4, 72076 Tübingen, Germany\label{inst4}
   \and European Space Agency (ESA), ESA Office, Space Telescope Science Institute, 3700 San Martin Drive, Baltimore, MD 21218, USA\label{inst5}
   \and Department of Astrophysics, University of Zürich, Winterthurerstrasse 190, 8057 Zürich, Switzerland\label{inst6}
   \and Western University, 1151 Richmond Street London, Ontario N6A 3K7, Canada \label{inst7}
   \and Department of Physics, University of Warwick, Gibbet Hill Road, Coventry CV4 7AL, United Kingdom\label{inst8}
   \and Landessternwarte, Zentrum für Astronomie der Universität Heidelberg, Königstuhl 12, 69117 Heidelberg, Germany\label{inst9}
    }

   \date{Received 26 May 2026 / Accepted 29 October 2026}

 \abstract{
    From 2015 to 2019, the \ispynaco Survey scanned 204 disk-hosting systems to search for exoplanets and substellar companions at wide orbital separations ($s >$~\SI{50}{\AU}). To confirm and further characterize (candidate) companions, we carried out follow-up observations using the \sphere instrument at the ESO Very Large Telescope. We applied principal component analysis and the \fours algorithm to subtract the stellar PSF and reveal companions and circumstellar disks. For the first time, we characterized the contrast performance of \fours on VLT/SPHERE/IRDIS data. We find that contrary to the performance on VLT/NaCo data, no significant gain in contrast can be achieved, and we discuss the potential reasons for this. Further, we compared our detection limits to theoretical predictions for debris disk-shaping planets. We validated the companions HD\,101412 B and C as bound based on a common-proper-motion test and report their masses and spectral types. Here we report the first spatially resolved images of the close stellar binary HD\,131511. Our residual images resolve two known debris disks and one known protoplanetary disk in scattered light. In the case of HD\,121617, we recovered a tentative planetary candidate at $\sim$\SI{155}{\AU}. Using \fours, we detected close-in point-like features in the HD\,59659 and HD\,1461 systems. The latter was ruled out by further follow-up observations, and we attribute the initial detection to non-Gaussian residual noise. The validation of the candidates in the HD\,59659 and HD\,121617 systems is left open to future work. Apart from HD\,101412 B and C, we did not detect any of the initial candidate companions in \ispynaco, which led to the SPHERE follow-up in the first place. These findings validate the planet occurrence rates from the initial \ispynaco survey, demonstrate the importance of follow-up observations for tentative point sources, and reveal new images of stellar companions and disks.
    }


   \maketitle
   \nolinenumbers
 
\section{Introduction}
The occurrence of giant exoplanets and debris disks has long been thought to be correlated. Debris disks are tenuous belts of dust and gas that are actively produced by collisions of material in the circumstellar environment \citep[e.g.,][]{Wyatt_2008, Hughes_2018} and extrasolar analogs to the Solar System's Asteroid and Kuiper Belts. High-contrast direct imaging (HCI) in the optical and near and mid-infrared (IR) offers the possibility of studying these disks via scattered light \citep[e.g.,][]{Schneider_1999, Stapelfeldt_2004, Esposito_2020}. At the same time, HCI enables the detection and spectral characterization of giant planets via their thermal emission in the outer reaches of nearby stellar systems \citep[e.g.,][]{Bowler_2016, Follette_2023}, where indirect detection methods, such as radial velocity (RV) and transit observations, are insensitive \citep{Fischer_2014}. Therefore, HCI offers a unique view to study protoplanetary and debris disks alongside giant planets, which may sculpt any type of circumstellar disk through their gravitational influence \citep{Wyatt_2008, Krivov_2010, Lee_2016}.

It has been theorized that the occurrence rate of giant planets should be higher in debris disk systems since giant planets might stir up planetesimals and thus initiate a collisional cascade, resulting in observable dust \citep[e.g.,][]{Mustill_2009, Pearce_2022}. Many large-scale surveys have targeted debris disk systems (e.g., the Gemini NICI Planet-Finding Campaign, \citealt{Wahhaj_2013}; the Strategic Exploration of Exoplanets and Disks with Subaru, \citealt{Janson_2013}; the Large Binocular Telescope Interferometer Exozodi Exoplanet Common Hunt, \citealt{Skemer_2014}; and the $L'$-band Imaging Survey for Exoplanets in the North, \citealt{MussoBarcucci_2021}), and they have tentatively confirmed that giant-planet occurrence is higher for debris disk systems \citep[e.g.,][]{Meshkat_2017, Stone_2018}.

The interaction of planets with debris can leave observable signatures in the belts. These can include eccentric belts \citep{Faramaz_2014, Pearce_2014}, clumps \citep{Wyatt_2003, Booth_2023}, sharp edges \citep{ImazBlanco_2023, Pearce_2024}, gaps \citep{Pearce_2015, Shannon_2016, Sefilian_2021, Friebe_2022}, and warps \citep{Mouillet_1997}. The shapes and features of observed belts can therefore hint at the masses and orbits of unseen planets (e.g., \citealt{Pearce_2022}), which is of particular interest because it can probe otherwise-undetectable sub-Jupiter planets in the outer regions of planetary systems. Previous small-scale observing campaigns have targeted diverse architectures of debris disks, such as multi-belt disks, where the gap might be carved by giant planets \citep{Meshkat_2017, Matthews_2018, Lazzoni_2018, Lombart_2020}.

Some of the most well-known, directly imaged exoplanets reside in debris disk systems, such as $\beta$~Pic~b \citep{Lagrange_2010}; HR\,8799~b,c,d, and e \citep{Marois_2008, Marois_2010}; AF~Lep~b \citep{DeRosa_2023, Franson_2023, Mesa_2023}; HD\,95086~b \citep{Rameau_2013, Desgrange2022, Malin2024}; and TWA\,7~b \citep{Lagrange_2025, Crotts_2025}. However, the number of systems where both exoplanets and debris disks have been found remains small. Such simultaneous detections offer powerful insights into the dynamics and evolutionary histories of planetary systems, including how exoplanets migrated \citep{Poblete_2025}, the times when orbital instabilities occurred \citep{Smallwood_2023}, the processes driving disk evolution \citep{Franson_2023}, and the total disk mass in unseen planetesimals \citep{Pearce_2015, Sefilian_2021, Friebe_2022}, plus many more. However, while many planet-disk interactions are possible, very few have been directly observed. Hence, expanding this sample would be a crucial step toward understanding how planetary systems are shaped into the configurations we observe today.

Searching for planets in debris disk systems can also yield ancillary science outcomes, such as direct disk detections \citep[e.g.,][]{Janson_2013, Matthews_2017, Launhardt_2020}. The majority of debris disks are imaged via thermal emission from millimeter-sized dust grains in the submillimeter to centimeter wavelengths with radio telescopes such as ALMA or in the far IR with, for example, \textit{Herschel} \citep{Choquet_2018}. In contrast, HCI offers the possibility of imaging these disks with a higher spatial resolution in the optical and near to mid-IR, where light originating from the host star is scattered off the micron-sized dust particles. As this scattering leads to polarization and is enhanced at high incident angles, these detections are usually biased toward edge-on disks and favored for polarimetric observations. Nonetheless, disks detected with HCI provide important complementary insights into their structure, such as the presence of gaps \citep{Feldt_2017}, or the offset between the center of the disk and the location of the host star \citep{Ren_2019}. When combined with radio observations, HCI, with and without polarimetry, provides a complementary estimate of the dust grain properties, such as the grain size distribution \citep{Milli_2026}, giant impacts \citep{Engler2018, Jones_2023}, and the occurrence rate of ongoing collisions \citep{Kalas_2026}.

One of the largest surveys targeting stars harboring such debris disks is the Imaging Survey for Planets around Young stars (ISPY), which was carried out at VLT/NaCo (hereafter \naco, \citealt{Rousset_2003, Lenzen_2003}) using the $L'$-band imaging mode. In the mid-IR, the contrast between older, cooler companions and their host stars is more favorable compared to shorter wavelengths such as $H$ and $K$. Between 2015 and 2019, ISPY searched for planets around 204 targets, discovering several stellar and substellar companions (e.g., HD\,193571~B, \citealt{MussoBarcucci_2019}; PDS\,70~b, \citealt{Keppler_2018}; R~CrA~B, \citealt{Cugno_2019}; HD\,101412~B and C, \citealt{Cugno_2023}). In many cases, however, only a single epoch of data were not sufficient to distinguish between a background object and a co-moving gravitationally bound companion.

After the decommissioning of \naco at the end of ESO P103 (October 2019), we conducted follow-up observations with VLT/SPHERE (hereafter \sphere; \citealt{Beuzit_2019}) of several ISPY targets of interest, for example, those with companion candidates identified in the \naco images. While \sphere operates in a different wavelength regime ($Y$, $J$, $H$, and $K$ bands), it offers enhanced absolute sensitivity for our target sample compared to \naco \citep{Beuzit_2019}, allowing us to re-detect sources from \naco images regardless of their nature. This second epoch of observations allowed us to determine whether the signal in the \naco images was a data artifact or a true astrophysical source that would be re-detected. In the latter case, the source can either be a chance-aligned background star or a true bound companion. Due to the long time baseline between the \naco and \sphere epochs, we could classify sources not only via their colors but also distinguish between these two cases using a common-proper-motion (CPM) test \citep{Nielsen_2017, Herz2024}. Additionally, we could characterize detected objects via photometric and spectroscopic measurements from the $Y$ to $L$ band (0.95--\SI{3.79}{\micron}).\\ 
In this paper, we present (re-)observations of 13 disk-hosting targets from the ISPY survey with \sphere. The paper is structured as follows. We present our observations and data reductions in Sects.~\ref{sec: Observations} and \ref{sec: Data Reduction}. In Sect. \ref{sec: Results} we give a broad overview of our findings, describe each target individually, and conclude with a summary of the performance of our observations via the achieved contrast limits. A full suite of results in the form of point-spread function (PSF)-subtracted images and contrast curves is available as supplementary material online on Zenodo. We show examples in the appendix. In several systems, additional analyses are needed, which we summarize in the main text and describe in detail in the appendix. We compare our planetary-mass detection limits with theoretical predictions of debris disk-sculpting planets in Sect.~\ref{sec: Discussion} and assess the performance of \fours for \irdis. Lastly, we summarize our results and give a conclusion in Sect.~\ref{sec: Conclusions}.
\vspace{-5pt}
\section{Target selection and observations}\label{sec: Observations}
We observed a sample of protoplanetary and debris disk targets with the high-contrast imager \sphere \citep{Beuzit_2019}. The majority of our observations were collected in visitor mode as part of the ISPY guaranteed time observations (program ID 1101.C-0092(J), PI Launhardt -- carry-over compensation time for weather and technical losses, which could no longer be scheduled with \naco due to its decommissioning) between October 2022 and October 2023 in three separate observing runs. We collected an additional epoch of data for HD\,1461 in normal time and service mode (program 116.297U.001, PI Steimle) using an identical observing setup, to determine the nature of a promising candidate companion as described in Sect.~\ref{sec: HD 1461} below. We also collected several more epochs of observations for two promising candidates in the HD\,101412 system with \naco and the Gemini Planet Imager (GPI) at the Gemini South telescope \citep{Macintosh_2014} between 2017 and 2019, which we present in this paper.

Target selection was based on the results of the \ispynaco survey \citep{Launhardt_2020}: we selected stars with candidate companions (HD\,377, HD\,1461, HD\,101412, HD\,121617, HD\,131511, $\epsilon$~Sgr, HD\,179218, HD\,207129), and stars for which the previously determined disk properties suggest the presence of a giant planet within the detection limits of \sphere (HD\,181327, HD\,191089, see also \citealt{Pearce_2022}). The main observations were carried out in visitor mode, placing some restrictions on the available weather conditions and accessible right ascensions. As a consequence, we also observed two targets with candidates from the Radial Velocity Survey for Planets around Young stars (RVSPY; \citealt{Zakhozhay_2020}), which were chosen as backup targets (HD\,59659, HD\,105912). Lastly, we performed a deep search for the known RV companion $\epsilon$~Eri\,b. For the \naco candidate companions, a re-detection with \sphere allows CPM tests to be performed to confirm the bound nature of these objects as well as for spectral characterization based on the $L'$--$H$ band color. For the disk-shaping planet predictions, a point source consistent with the expected disk-sculpting location and mass would represent a promising candidate companion.

For all observations with \sphere, we used the IRDIFS mode, allowing dual-band imaging using VLT/SPHERE/IRDIS (hereafter \irdis; \citealt{Dohlen_2008, Vigan_2010}) and spectroscopy with VLT/SPHERE/IFS (\ifs hereafter; \citealt{Claudi_2008}) at the same time. The detailed observation setup, reference star if applicable, and filters used for \irdis are listed in \autoref{table: observations}. Removing the stellar PSF requires a set of reference images. For this purpose, we used both spectral differential imaging \citep[SDI;][]{Racine_1999, Sparks_2002} and angular differential imaging \citep[ADI;][]{Marois_2006}: the spectral diversity arises from using two filters (for \irdis) and an integral field unit (for IFS), while the angular diversity arises from observing in pupil-stabilized mode, in which the astronomical field rotates relative to the telescope pupil. Where possible, we also performed reference differential imaging \citep[RDI;][]{Lafreniere_2009}, which, combined with ADI, reaches better contrast performance closer to the host star and for disks that show extended emission around the host star. For this, we additionally observed a reference target without companions for around 10 minutes in between science observation blocks with the star-hopping mode \citep{Wahhaj_2021}.

For all stars, we collected unsaturated flux frames without the coronagraph to allow for relative photometric calibration. We used shorter exposure times compared to coronagraphic observations and a neutral density filter to prevent saturation of the detector. We also collected star-position calibration frames, in which a waffle pattern is introduced to the deformable mirror, producing four equidistant satellite spots. These were then used to calculate the exact position of the star behind the coronagraph, thus allowing us to perform ADI and highly precise astrometric measurements of any source relative to the target star.
\section{Data reduction}\label{sec: Data Reduction}
\subsection{IRDIS data processing}
\subsubsection{Pre-processing}
We downloaded all \irdis data from the ESO \sphere archive and performed the pre-processing data reduction using our custom reduction pipeline \texttt{REACHER}\footnote{\href{https://github.com/PaulSteimle/REACHER}{github.com/PaulSteimle/REACHER}} (Raw image Extraction And Calibration for High-contrast Exoplanet Research), which is optimized for star-hopping observations. We performed standard calibrations, including dark, flat, sky background, distortion, and true north correction according to the \sphere manual,\footnote{\href{https://www.eso.org/sci/facilities/paranal/instruments/sphere/doc/VLT-MAN-SPH-14690-0430_P114_july_2024_zwa.pdf}{\sphere Manual} (version P114.1, July 2024)} as well as the centering of the star behind the coronagraph. REACHER corrects the intensity scaling of the flux frames to match the science frames in order to allow straightforward relative flux calibration. Internally, REACHER relies on several routines from the \texttt{VIP} \citep{GomezGonzalez_2017, Christiaens_2023} Python package, for example, to perform fast Fourier transform shifts of images or the calculation of the star's position behind the coronagraph.
\vspace{-8pt}
\subsubsection{PSF subtraction}
When constructing the PSF reference, it is crucial to select only high-quality data, as outlier frames can bias the model and, consequently, the entire analysis. To remove the stellar PSF from the \irdis images, we used two complementary approaches. First, we applied principal component analysis \citep[PCA;][]{Soummer_2012, Amara_2012}, as implemented in the \texttt{VIP} \citep{GomezGonzalez_2017, Christiaens_2023} Python package, as it is easily tunable with a single hyperparameter: the number of PCA modes. Additionally, PCA is applicable to all three differential imaging methods (RDI, ADI, and SDI) used in the \irdis observations and is computationally fast, allowing for the full field of view (FoV) to be post-processed. Second, we also performed PSF subtraction for the central $1\times1$~\si{\arcsec} of \irdis images using the recently published algorithm \fours \citep{Bonse_2025}, which has been demonstrated to show improvements for \naco data of up to 1.5 mag in contrast for close-in regions ($< 3\lambda/D$) compared to PCA, but has not previously been applied to \sphere data. We compare the overall performance of the achievable contrasts between PCA and \fours on datasets from \irdis in Sect. \ref{sec: Limitations of 4S}.
\vspace{-8pt}
\paragraph{Preparation of the input data.}
We excluded bad frames within the science cube based on their similarity to the median frame, using the Pearson correlation coefficient. For the majority of our targets, we used only frames with coefficients above 0.7.
This threshold was chosen because it robustly excludes frames with bad seeing or when the telescope loses tracking, while still retaining the majority of the data. As $\epsilon$~Eri and $\epsilon$~Sgr are both bright, and short exposure times were used, the resulting cubes contain over 1000 frames and hence allowed the use of a more aggressive cut. Here we found that a threshold of 0.9 shows the best final results. 
\vspace{-8pt}
\paragraph{Principal component analysis.} \label{sec: IRDIS PCA}
We applied PCA to all our datasets. To reduce the influence of self-subtraction, a rotational threshold of 1.5 full width at half maximum (FWHM) is applied, excluding frames rotated by less than this when constructing the stellar PSF model for an individual frame. For stars where a reference star was observed, we perform a combined ADI and RDI PSF subtraction. The addition of a reference star allows for an improved detection of extended circular sources around the host star, such as disks, as well as an improved contrast limit close to the star when combining ADI and RDI.
\vspace{-8pt}
\paragraph{Signal-safe speckle subtraction.}
In addition to the PCA subtraction, we also performed post-processing with signal-safe speckle subtraction (\fours), a new PSF subtraction method, developed to reduce over- and self-subtraction of genuine point-like features and improve detection limits \citep{Bonse_2025}. The algorithm minimizes the temporal variance in the de-rotated residual sequence, making the loss function invariant to the signal of potential planets. This approach is combined with a mask based on explainable machine learning that is applied to the noise model parameters to prevent over-subtraction. Tests on \naco data have shown that \fours is able to improve detection limits by up to $\sim 1.5\,$ magnitudes compared to PCA, especially at close-in separations ($\leq 3\lambda/$D in the case of \naco). However, this improvement comes at a high computational cost, requiring GPU optimization to maintain a reasonable runtime. Due to GPU memory limitations, the FoV that can be processed with \fours is limited. In our analysis, we applied \fours to investigate the regions interior to the adaptive optics (AO) control radius ($\leq 0.5\,$\si{\arcsec}) where speckle noise dominates, and the expected gain of \fours is largest.
\vspace{-5pt}
\paragraph{Extraction of photometry and astrometry.}
We measured the positions of point-like sources by fitting a 2D symmetric Gaussian to the PSF-subtracted residual image. We calculated the signal-to-noise ratio (\snr) and detection significance as defined by \citet{Mawet_2014} and excluded negative lobes adjacent to the sources and quote detection significances based on a t-test \citep{student_1908}, thereby assuming the residual image noise to be Gaussian-distributed. We performed negative-injection-recovery tests, using a Nelder--Mead \citep{nelder1965simplex} minimization of the absolute residual around the source \citep{Wertz_2017} to measure the flux for sources that are not ruled out as background objects based on CPMs. If the final residuals deviate strongly from pure noise and suggest that the source position is inaccurate, we repeated the procedure and allowed all three parameters (separation, position angle, and flux) to vary. For HD\,101412, we used the Markov chain Monte Carlo (MCMC) sampler emcee \citep{Foreman-Mackey_2013} to obtain a full posterior sample for all three parameters. We used 100 walkers, limited to 1000 steps, and tested for convergence after 50 steps using an autocorrelation test. We quote positions of sources via their projected angular separation $\rho$ and position angle $\theta$. Using distances $d$ to the targets we quote projected physical separations as $s = \rho \cdot d$ in \si{\AU}. 
\subsection{IFS data processing}
The \ifs data were analyzed using the \texttt{spherical} framework \citep[][version 2.1.3]{Samland2025arXiv}. It enables end-to-end analysis by: (a) discovering, querying, and downloading data from the ESO science archive; (b) performing pre-processing and spectral cube building using the adapted \texttt{charis} pipeline \citep{Samland_2022}; (c) flux and image center calibration adapted from the \texttt{vlt-sphere} Python pipeline \citep{Vigan_2020}; and (d) high-contrast post-processing using the Temporal Reference Analysis for Planets \citep[\texttt{TRAP},][]{Samland_2021} algorithm. Companions are automatically detected using spectral template matching of substellar L-, T-type, and flat contrast spectra using the \texttt{species} package \citep{Stolker_2020}. The template matching works by fitting each spaxel of the detection cube in the wavelength direction with a linear combination of the spectral template, an offset, and in the case of T-dwarf templates, a slope (as the absorption features provide more spectral diversity to detrend the detection cube).  The contrast spectra of all detected point sources are automatically extracted by \texttt{TRAP} from the highest S/N position of the template-matched cube. The default parameters were used for the cube building step with \texttt{charis} using the optimal extraction algorithm, and a principal component fraction of 0.15 of the maximum number of components for \texttt{TRAP}.

\subsection{Archival data}
\subsubsection[NaCo]{\naco}
We used archival \naco data for HD\,1461, HD\,101412, HD\,121617, and $\epsilon$~Sgr. With the exception of HD\,121617, all datasets were obtained using a vortex coronagraph \citep{Mawet_2005, Absil_2013}. All datasets presented in this work are part of a larger effort to uniformly pre-process all \naco $L'$ observations which will be presented in a forthcoming publication (Bonse et al. in prep.). We here summarize the methods used for this: The database is built using \texttt{Pynpoint} \citep{Stolker_2019} and applies the following pre-processing steps: First, standard calibrations are performed, including dark subtraction and flat-field correction. Next, the thermal background is removed using averaged sky frames. Bad pixels are corrected via a $4\sigma$ clipping algorithm within a $5 \times 5$ pixel box. Afterward, all frames are aligned through sub-pixel cross-correlation. No temporal binning is applied prior to PSF subtraction to preserve the highest possible temporal resolution. A PCA-based frame selection criterion is employed to exclude low-quality frames, such as those affected by an open AO loop. The pre-processing steps for coronagraphic and non-coronagraphic data differ only in how they treat the thermal background. For the non-coronagraphic dataset of HD\,121617, a dithering pattern on the detector enables simultaneous capturing of sky frames alongside the science frames. In contrast, the vortex coronagraph data rely on dedicated sky frames in which the target star is positioned outside the FoV. We post-processed all cubes with PCA and \fours to subtract the stellar PSF and measure the positions of point-like sources analogous to the methods used for \irdis.

\subsubsection{GPI}
To further characterize the two companions around HD\,101412, we monitored the system in 2018 and 2019 with GPI installed at Gemini South. We processed all GPI raw data with the dedicated pipeline \citep{Maire_2010, Maire_2012, Perrin_2014}, including distortion and bad-pixel correction, to obtain the science-ready data cubes. Because the $H$-band was obtained in polarized light to search for disk signatures, it does not allow the extraction of absolute fluxes. Hence, we measured only the positions of the companions relative to the host star by fitting a 2D Gaussian to the signals in the residual PCA images.

For the spectrometric $J$- and $Y$-band data, we found that the satellite spots, used for centering and flux calibration, are only poorly visible, especially the first-order spots, which are strongly affected by the speckle halo, resulting in centering issues. Therefore, we adopted an alternative approach by fitting second-order spots with a Gaussian and determining the center from the positions of pairs of opposite satellite spots. Since the first-order satellite spots are usually also used as a PSF model for flux calibration and negative injection, we find that this is not sufficiently precise with the second-order spots. Hence, we did not perform PCA but classical ADI \citep{Marois_2006}, which is flux-conserving. This allowed us to measure the companion fluxes in the final images via aperture photometry and the positions by fitting a 2D Gaussian again. We find that the astrometry appears to agree with other datasets, while the photometry extracted from each spectroscopic bin is significantly higher than the measurements from our \ifs data. We attribute this excess in flux to the fact that the classical ADI PSF subtraction does not account for all stellar flux, and together with the performed aperture photometry, this leads to contamination in the companion fluxes. As our \ifs data cover the same wavelength range with higher precision, we neglected these flux measurements in our final spectral analysis of the companions.
\vspace{-7pt}
\subsubsection{SPHERE}
To validate the background origin of far-out sources in several datasets shown in Appendix \ref{sec: Common Proper Motion (CPM) Tests}, we used archival data obtained within the frameworks of the SpHere INfrared survey for Exoplanets \citep[SHINE;][]{Desidera_2021} and the Near-InfraRed SURvey for exoplanets (NIRSUR program ID 095.C-0298, PI: Beuzit) surveys. To maintain consistency across datasets, we reanalyzed them using the workflows described above for \irdis and \ifs.

\section{Results}\label{sec: Results}
\begin{figure*}
    \centering
    \makebox[\textwidth][c]{
    \includegraphics[width=\textwidth]{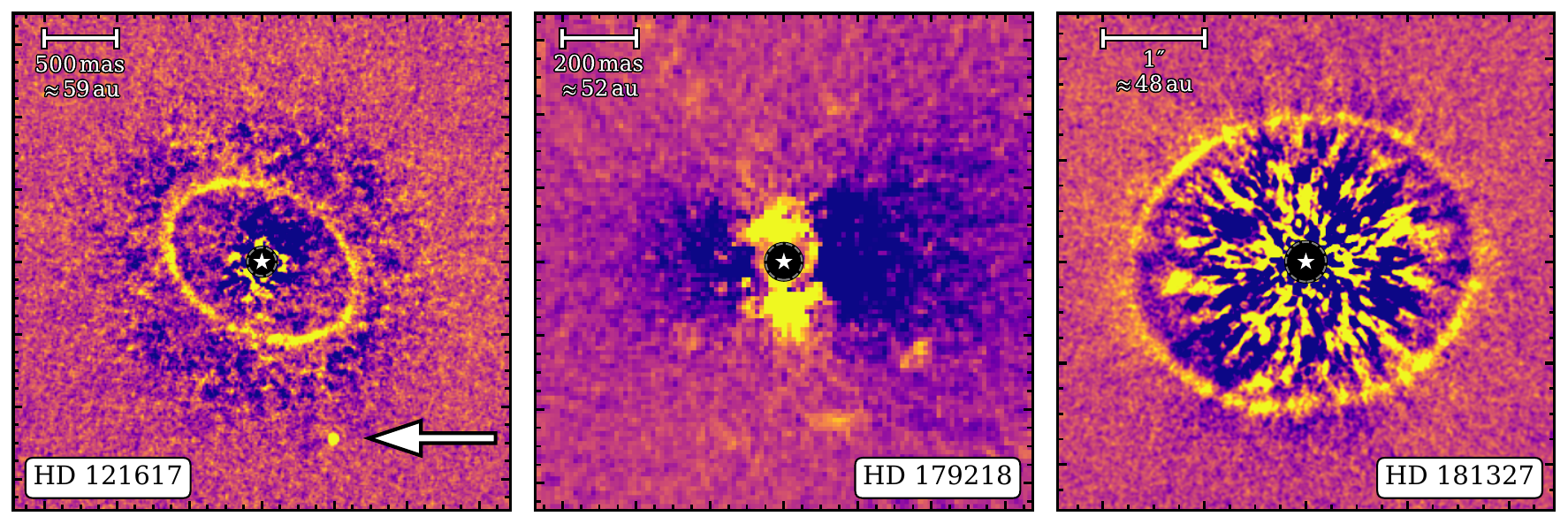}}
    
    \caption{Circumstellar disk detections for HD\,121617 (\textit{left}), HD\,179218 (\textit{center}), and HD\,181327 (\textit{right}). In the case of HD\,121617 and HD\,181327, the full ring of the debris disks is visible, while the nearly edge-on protoplanetary disk around HD\,179218 is visible north and south of the host star. Additionally, we mark the companion candidate HD\,121617~[b] with a white arrow. All images have been rotated so that true north is up and east is to the left. Host star locations are marked with the white star-shaped marker in the center; the black circle shows the inner working angle of the coronagraph.}
    \label{fig: all disks}
    \vspace{0.25cm}
\end{figure*}

\begin{figure*}
    \centering
    \makebox[\textwidth][c]{
    \includegraphics[width=\textwidth]{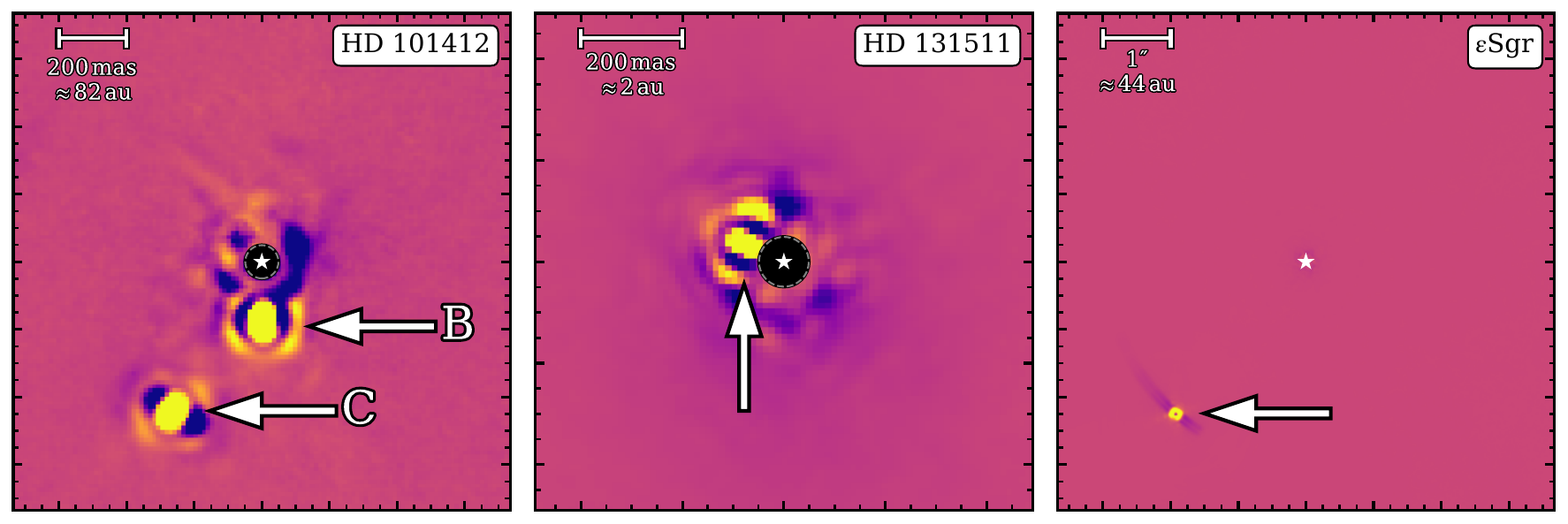}}
    \caption{PCA residual images of confirmed companions detected in the \ispysphere sample, all marked with arrows. \textit{Left}: HD\,101412 B and C, the discovered stellar-mass companions. \textit{Center}: First resolved images of the secondary HD\,131511~B in the spectroscopic binary system. \textit{Right}: Previously known and resolved $\epsilon$~Sgr~B (HD\,169022~B). Images are aligned to true north, with east to the left.}
    \label{fig: all companions}
    \vspace{0.25cm}
\end{figure*}

\begin{figure*}
  \begin{minipage}[c]{0.67\textwidth}
    \includegraphics[width=\linewidth]{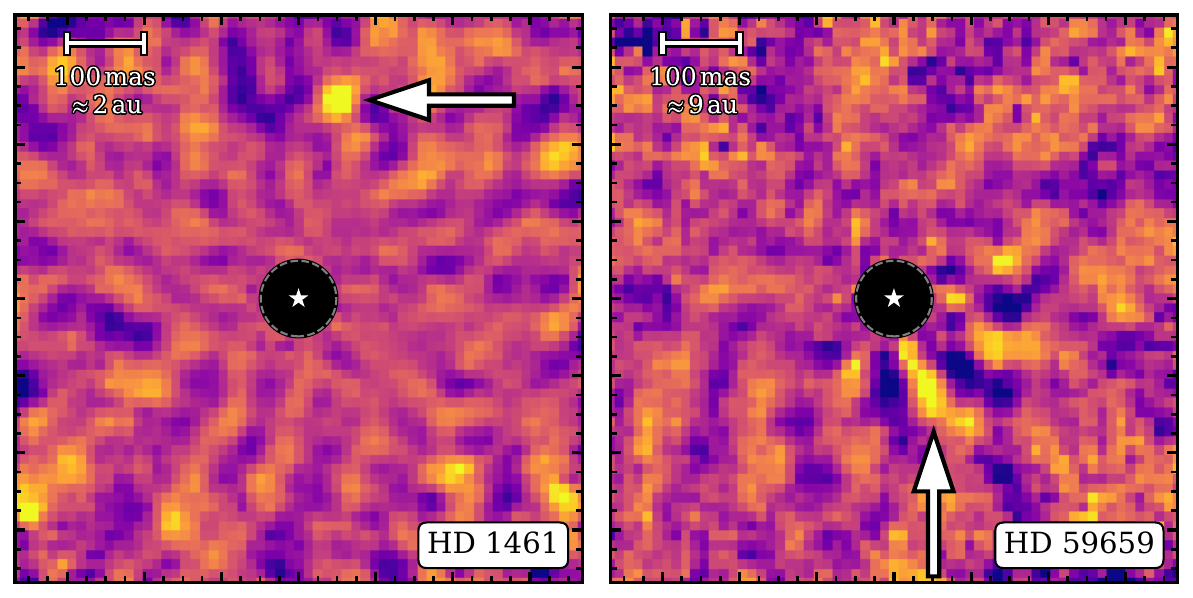}
  \end{minipage}\hfill
  \begin{minipage}[c]{0.3\textwidth}
    \caption{Companion candidates detected in the \fours residuals marked with arrows. In the HD\,1461 system (\textit{left}), the candidate is at a projected separation of around \SI{260}{\mas} while the candidate in the HD\,59659 (\textit{right}) system is somewhat closer at around \SI{120}{\mas} from the host star in the center. Images are also aligned to true north, with east to the left.}
    \label{fig: all candidates}
  \end{minipage}
\end{figure*}

Our cleaned and post-processed images (examples for three targets shown in \autoref{fig: all residual images}) probe the circumstellar environments of 13 targets. Several images contain point-like sources. We assess each source for astrophysical origin, either as a companion or a background object. Low-significance detections may still arise from speckle noise. We adopt standard nomenclature. Stellar-mass companions are quoted in alphabetical order using uppercase letters; substellar companions use lowercase letters. Unconfirmed sources which show signs of companion nature are labeled as candidates indicated with square brackets, for example, HD\,121617~[b].

We resolved extended disks in the HD\,121617, HD\,179218, and HD\,181327 systems, shown in \autoref{fig: all disks}. For HD\,121617 and HD\,181327, the reductions recover full circular disk structures. For HD\,179218, we detect emission from the nearly edge-on disk north and south of the star. In the HD\,121617 dataset, we also detect a point-like feature, shown in \autoref{fig: all disks}, left. Its position and separation tend to suggest a background origin. However, it satisfies the criteria for a planetary companion, and we therefore classify it as a candidate, HD\,121617~[b]. We image two known stellar companions shown in \autoref{fig: all companions} center and right. HD\,131511 B was previously unresolved but confirmed via radial velocities \citep{Nidever_2002} and astrometry \citep{Jancart_2005}. $\epsilon$~Sgr~B was previously resolved \citep{Golimowski_1993}. We also report the re-detection of two massive companions, HD\,101412 B and C \citep{Cugno_2023}. We further detect several unresolved sources at larger separations in the systems HD\,59659, HD\,101412, $\epsilon$~Sgr, HD\,179218, HD\,181327, HD\,191089, and HD\,207129. In most cases, the CPM tests and color analysis allowed us to conclude a background origin. In the final \fours residuals, we recovered HD\,101412~B and C as well as HD\,131511~B, which demonstrates that \fours performs reliably on \sphere datasets. In addition, \fours reveals two point-like features, not distinguishable from speckle noise in PCA reductions. The \fours images with detections of the signal close to HD\,1461 at \snr$=4.5$ (3.9$\sigma$) and the other in the HD\,59659 dataset at \snr$=6.0$ (4.2$\sigma$) are shown in \autoref{fig: all candidates}. Both features would be consistent with companion candidates. However, while the nature of the HD\,59659 signal remains inconclusive, follow-up observations show that the HD\,1461 \snr$=4.5$ signal is not real and likely arises from non-Gaussian noise. This raises doubts about the reliability of the \snr$=6.0$ signal in the HD\,59659 dataset.

In the following section, we describe all detected point-like and extended sources in detail and determine which of the point-like features are bona fide companions. We use values for distances and ages as reported in \autoref{table: stars astro physical prop}. We assess the performance of our observations by presenting contrast curves. For debris disk-hosting systems with predictions for disk-edge-shaping planets \citep{Pearce_2022}, we compare our detection limits and rule out nearly all scenarios in which a single giant planet is likely to be causing the observed disk gaps in the HD\,121617 and HD\,181327 systems.
\begin{table*}[]
\caption{Physical properties of all stars observed in \ispysphere.}
\label{table: stars astro physical prop}
\begin{tabular*}{\textwidth}{@{\extracolsep{\fill}}l c c c c c c}
\hline
\hline
Target & $d$  & Age  & Disk 1  & Disk 2 & M$_{\text{p, pred.}}$  & sma$_{\text{p, pred.}}$\\
       & [pc] & [Myr] & [au] & [au] & [\si{\Mjup}] &  [au] \\
\hline
HD\,377 & $38.48 \pm 0.09$ & $220 \pm 50$$^{\,e, q, r}$ & 2.40$^{ \, \alpha}$ & $60 \pm 10$$^{\,\epsilon}$ & $0.40 \pm 0.07$ & $46 \pm 8$ \\
HD\,1461 & $23.45 \pm 0.03$ & $6000 \pm 1000$$^{\,c, l, q}$ & 0.60$^{ \, \alpha}$ & $50 \pm 30$$\rightarrow$$150 \pm 40$$^{\, \epsilon}$ & $0.08 \pm 0.03$ & $50 \pm 20$ \\
$\epsilon$~Eri & $3.203 \pm 0.005$ & $1000 \pm 100$$^{\,p}$ &  ...  & $63.8 \pm 0.8$$\rightarrow$$75.0^{+0.9 }_{-0.7}$$^{\, \epsilon}$ & $0.19 \pm 0.04$ & $52.7^{+0.9 }_{-1.0}$ \\
HD\,59659 & $87.36 \pm 0.21$ & $45 \pm 5$$^{\,j, v}$ & $14.0 \pm 2.0$$^{\,\delta}$ &  ...  &  ...  &  ...  \\
HD\,101412 & $407 \pm 5$ & $2.0 \pm 1.0$$^{\,f, g}$ & $1.06 \pm 0.20$$^{\,\delta}$ & 92.50$^{ \, \gamma}$ &  ...  &  ...  \\
HD\,105912 & $48.18 \pm 0.20$ & $1000^{+600 }_{-700}$$^{\,b, c}$ & 0.80$^{ \, \alpha}$ & 22.30$^{ \, \alpha}$ &  ...  &  ...  \\
HD\,121617 & $116.5 \pm 1.0$ & $16 \pm 3$$^{\,n}$ &  ...  & $50 \pm 7$$\rightarrow$$106 \pm 7$$^{\, \epsilon}$ & $1.9 \pm 0.3$ & $37 \pm 5$ \\
HD\,131511 & $11.37 \pm 0.06$ & $1000 \pm 500$$^{\,k, r, s, t}$ & $100 \pm 70$$^{\,\epsilon}$ &  ...  & $0.20 \pm 0.10$ & $80 \pm 50$ \\
$\epsilon$~Sgr & $43.9 \pm 0.5$ & $130 \pm 30$$^{\,c}$ & 6.10$^{ \, \alpha}$ & $120 \pm 60$$^{\,\epsilon}$ & $2.2 \pm 0.8$ & $90 \pm 40$ \\
HD\,179218 & $264 \pm 3$ & $2.30 \pm 0.30$$^{\,m, u, d}$ & $131 \pm 22$$^{\,\beta}$ &  ...  &  ...  &  ...  \\
HD\,181327 & $48.15 \pm 0.13$ & $24 \pm 3$$^{\,a}$ & 29.40$^{ \, \alpha}$ & $73.3 \pm 0.4$$\rightarrow$$89.3 \pm 0.4$$^{\, \epsilon}$ & $1.60 \pm 0.10$ & $53.9 \pm 0.4$ \\
HD\,191089 & $50.06 \pm 0.11$ & $24 \pm 3$$^{\,a}$ & 25.40$^{ \, \alpha}$ & $37.0 \pm 2.0$$\rightarrow$$53.0 \pm 2.0$$^{\, \epsilon}$ & $0.89^{+0.07 }_{-0.06}$ & $28.0 \pm 1.0$ \\
HD\,207129 & $15.554 \pm 0.017$ & $3000^{+2000 }_{-1000}$$^{\,c, h, i, o, r, t}$ & 70.20$^{ \, \alpha}$ & $130 \pm 4$$\rightarrow$$170 \pm 4$$^{\, \epsilon}$ & $0.21^{+0.03}_{-0.07}$ & $109 \pm 3$ \\
\hline
\end{tabular*}
\tablefoot{Distances were calculated via the method proposed in \citet{Bailer-Jones_2018} and based on parallaxes measured by \textit{Gaia} DR3 \citep[][]{GaiaCollaboration_2023} or Hipparcos \citep[][]{Perryman_1997, vanLeeuwen_2007} in the case of $\epsilon$~Sgr. Disk detections are based on resolved images from ALMA or \textit{Herschel} or calculated from the spectral energy distribution. If only Disk 2 is listed, this means that it is resolved; if both disks are given, both can be resolved or unresolved. The terms M$_\text{p, pred}$ and sma$_\text{p, pred}$ correspond to the minimum mass and maximum semimajor axis for predicted single gap-carving companions in each system that would be responsible for the shaping of the disk, as determined by \citet{Pearce_2022}. We updated the disk semimajor axis from reference (f) \citet{Geers_2007} to \SI{92.5}{\AU} due to an incorrect pre-\textit{Gaia} distance for HD\,101412 used therein. Ages: (a) \citet{Bell_2015}; (b) \citet{Casagrande_2011}; (c) \citet{Chen_2014}; (d) \citet{Dent_2005}; (e) \citet{Esposito_2020}; (f) \citet{Geers_2007}; (g) \citet{Gray_2017}; (h) \citet{Gaspar_2013}; (i) \citet{Holland_2017}; (j) \citet{Houk_1978}; (k) \citet{Mamajek_2008}; (l) \citet{Nissen_2020}; (m) \citet{Pascual_2016}; (n) \citet{Pecaut_2016}; (o) \citet{Ramirez_2012}; (p) \citet{Sanghi_2026}; (q) \citet{Sierchio_2014}; (r) \citet{Stanford-Moore_2020}; (s) \citet{Stone_2018}; (t) \citet{Vican_2012}; (u) \citet{Vieira_2003}; (v) \citet{Zuckerman_2019} - Disks: ($\alpha$) \citet[][and citations therein]{Chen_2014}; ($\beta$) \citet{Dent_2005}; ($\gamma$) \citet{Geers_2007}; ($\delta$) Grant Kennedy's sdb library (\url{http://www.drgmk.com/sdb}); ($\epsilon$) \citet[][and citations therein]{Pearce_2022}}
\vspace{-5pt}
\end{table*}
\vspace{-7pt}
\subsection{Individual targets}
\subsubsection[HD 377 and HD 105912]{HD\,377 and HD\,105912}
For these targets, our residuals show no signals from any objects except the host stars. In the case of HD\,105912, only the first half of the observation was usable due to a telescope pointing problem, resulting in significantly shallower detection limits, which is especially reflected in the maximum of the contrast curve (see also \autoref{fig: joint contrast curves}).
\begin{figure}
    \centering
    \makebox[\linewidth][c]{
    \includegraphics[width=\linewidth]{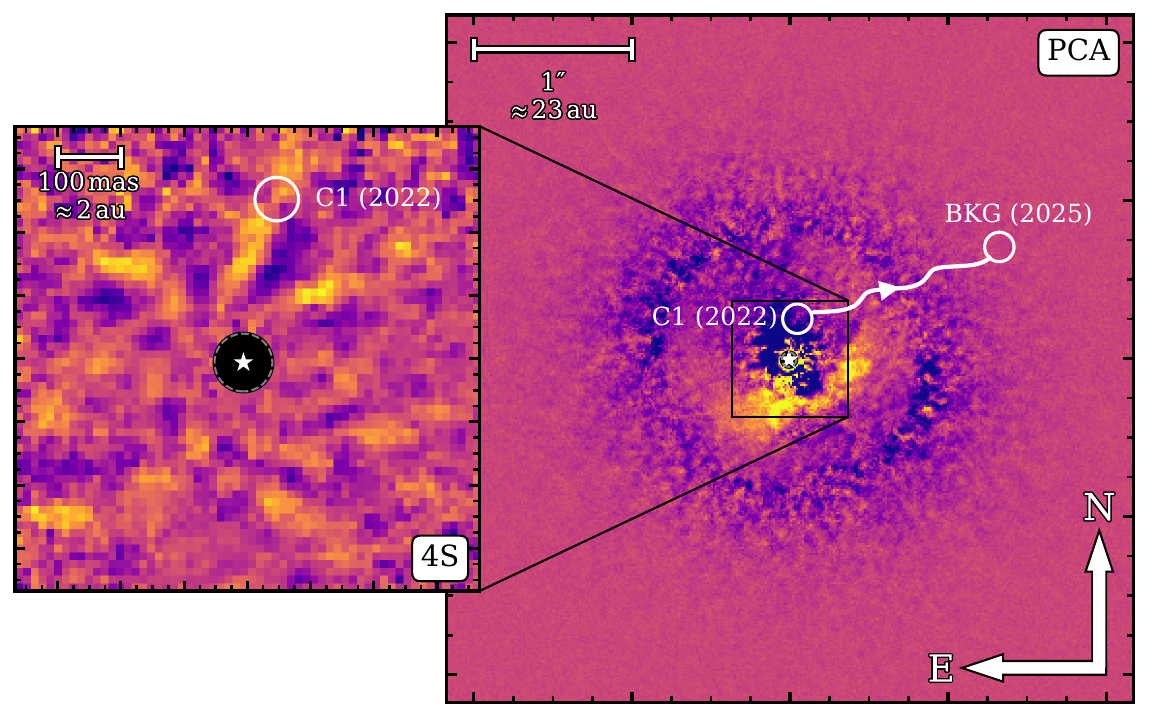}}
    \vspace{-0.5cm}
    \caption{Residual images of HD\,1461 with PCA and \fours for the new 2025 dataset taken with \sphere. The original location of the candidate detection is marked by C1 (2022). The alternative location, if the source is of background origin, is shown by the marker labeled BKG 2025. The motion between the two positions is shown by the connecting line. No signs of any source are visible in the new epoch.}
    \label{fig: HD1461 new epoch}
    \vspace{-0.65cm}
\end{figure}
\vspace{-7pt}
\subsubsection[HD 1461]{HD\,1461}\label{sec: HD 1461}
HD\,1461 is a solar-like (G3\,V) star with two known inner rocky planets, HD\,1461 b \citep{Rivera_2010} and c \citep{Diaz_2016}, and two belts of debris. For the inner belt, \citet{Chen_2014} derived a distance of \SI{0.6}{\AU} based on \textit{Spitzer} spectroscopy. The outer disk has been resolved with \textit{Herschel}, extending from \SI{50(30)}{\AU} out to \SI{150(40)}{\AU} \citep{Marshall_2021, Pearce_2022}.

In the first epoch of \sphere observations, we identified a  $3.9\sigma$ (\snr~$=4.5$) point-like feature at $\rho =$~\SI{262.57(45)}{\mas} north of the host star in the \fours residuals (\autoref{fig: all candidates} left). We measure a contrast of \SI{12.83(11)}{\mag}, placing the feature, if of bound nature, at around $M =$~\SI{42.1(37)}{\Mjup}. The PCA image shows a counterpart buried within speckles at a lower significance of $3.2 \sigma$ (\snr~$=3.5$), mimicking the behavior of the signal of AF~Lep b in the \naco data presented in \citet{Bonse_2025}. Given the close proximity to the host star, the likelihood of the signal being of background origin is quite low; however, the relatively low-significance detection prompts the question of whether this is a real companion or non-Gaussian random noise.

As our archival \ispynaco dataset has insufficient sensitivity to re-detect this signal, we collected an additional epoch of \sphere observations in November 2025 (see \autoref{table: observations}) to determine its true nature. While the observing conditions were excellent, with a seeing of $\approx$~\SI{0.5}{\arcsec}, and the achieved contrast limits are sufficient for re-detection, the data show no evidence of a point-like source (see \autoref{fig: HD1461 new epoch}). Given the high proper motion of HD\,1461, any stationary background object would now appear at a separation of $\rho \approx$~\SI{1.5}{\arcsec} from the host star. The absence of a detection at this location excludes a background source. We also determine the maximum possible on-sky motion for a bound companion to be $\approx$~\SI{470}{\mas}, based on the location and estimated mass of the candidate as well as stellar parameters. For this, we randomly drew from uniformly distributed orbital elements, fited the semimajor axis such that the orbit is consistent with our 2022 detection, and lastly calculated the position in this case for the 2025 epoch.

This consequently means that due to orbital motion, the companion could have moved closer to the host star, placing it below our detection limits. Such a scenario is consistent with systems such as $\mathrm{\beta}\,$Pictoris\,b \citep{Chauvin_2012} and the proposed companion around $\mathrm{\alpha}\,$Centauri\,A \citep{Beichman_2025}. To test this hypothesis, we performed orbital fits using \texttt{orvara} \citep{Brandt_2021_b}. We included our astrometric measurements, Hipparcos--Gaia Catalog of Accelerations (HGCA, \citealt{Brandt_2021_HGCA}), and 14 years of RV data from HARPS, CORALIE, APF, and HIRES.\footnote{Accessed via \href{https://dace.unige.ch/dashboard/}{dace.unige.ch}} We find no stable orbital solution that is consistent with the 2025 non-detection. This excludes a bound companion as the origin of the signal. Given the low detection significance of  $3.9\sigma$ and the small angular separation, we conclude that the signal is therefore likely a false positive caused by non-Gaussian noise. We examine this further in Sect. \ref{sec: Limitations of 4S}.
\subsubsection[ε Eri]{$\epsilon$~Eri}
At a distance of only \SI{3.2}{\pc} \citep{GaiaCollaboration_2023}, $\epsilon$~Eri (HD\,22049) is a K2\,V star and harbors a debris disk first discovered by IRAS IR excess \citep{Aumann_1985}. Starting in 1988, evidence of a Jupiter-like companion was found in RV data \citep{Campbell_1988}. 40 years later $\epsilon$~Eri\,b is a confirmed \SI{1.00(10)}{\Mjup} planet orbiting its host star on a circular, \SI{3.53(4)}{\AU} orbit nearly coplanar with the debris disk \citep{Thompson_2025}, which has been imaged multiple times, for example with ALMA \citep{Booth_2017, Booth_2023}. 
In our images we are not able to detect the planet. However, we report the deepest detection limits to date in the $H2$ band: during our observing epoch the planet has a projected separation of \SI{1.096(12)}{\arcsec}, and at this separation we reach a contrast of $\approx 4.4\times 10^{-6}$ ($\approx 13.4$~\si{\mag}) corresponding to a flux in the $H2$ band of $\approx$\,\SI{1.02e-15}{\W\per\m\squared\per\um}. This is consistent with other searches using ground-based observatories; for example, \citet{Mawet_2019} report a contrast of $\approx 13$~\si{\mag} using Keck/NIRC2 in the $M$ band. Recently, JWST revisited the system attempting to image the planet, which was unsuccessful, but reached contrasts of $\approx 3.0\times 10^{-7}$ ($\approx 16.3$~\si{\mag}) in the F444W filter ($\lambda_\text{mean} = 4.439\,$\si{\micron}) at the expected position of the planet, showcasing the challenges of imaging $\epsilon$~Eri\,b \citep{Sanghi_2026}.
\begin{table*}[t]
    \caption{Astrometric and photometric measurements of HD\,101412 B and C.}
    \label{tab: HD101412 Astrometry} 
    \centering
    \begin{tabular*}{\textwidth}{@{\extracolsep{\fill}}llcccccc}
        \hline
        \hline
         & & \multicolumn{3}{c}{HD\,101412 B} & \multicolumn{3}{c}{HD\,101412 C} \\
        \hline
        Night & Filter & $ \rho $ & $\theta$ & Contrast & $ \rho $ & $\theta$ & Contrast \\
         & & [mas] & [deg] & [mag] & [mas] & [deg] & [mag] \\
        \hline
        2017 March 17 & $L'$ & $173 \pm 19$ & $179.9 \pm 2.9$ & $7.08 \pm 0.28$ & $547 \pm 14$ & $148.93 \pm 0.68$ & $7.80 \pm 0.16$ \\
        2018 March 16 & $H$ & $171.5 \pm 7.1$ & $184.4 \pm 2.4$ &  ...  & $527.2 \pm 7.1$ & $149.38 \pm 0.77$ &  ...  \\
        2018 March 19 & $H$ & $172.1 \pm 7.1$ & $184.2 \pm 2.4$ &  ...  & $529.5 \pm 7.1$ & $149.43 \pm 0.77$ &  ...  \\
        2019 February 25 & $Y$ & $176.2 \pm 2.8$ & $183.49 \pm 0.93$ & ... & $529.4 \pm 2.9$ & $149.41 \pm 0.33$ & ... \\
        2019 February 26 & $J$ & $175.1 \pm 2.8$ & $182.58 \pm 0.94$ & ... & $530.6 \pm 2.9$ & $149.15 \pm 0.33$ & ... \\
        2023 March 30 & $K1$ & $174.5 \pm 1.3$ & $181.19 \pm 0.26$ & $5.203 \pm 0.035$ & $515.37 \pm 0.98$ & $149.26 \pm 0.14$ & $6.386 \pm 0.012$ \\
        2023 March 30 & $K2$ & $174.5 \pm 2.1$ & $181.46 \pm 0.36$ & $5.436 \pm 0.054$ & $515.0 \pm 1.0$ & $149.31 \pm 0.14$ & $6.611 \pm 0.014$ \\
        2023 March 30 & $YJH$ & $178.75 \pm 0.89$ & $181.79 \pm 0.28$ & $4.99 \pm 0.21$ & $510.0 \pm 1.6$ & $149.44 \pm 0.16$ & $6.16 \pm 0.23$ \\
        \hline
    \end{tabular*}
    \tablefoot{Observations in the $L'$ band are from \naco. $Y$, $J$, and $H$-band data were acquired using Gemini/GPI. $K1$ and $K2$-filter data are from \irdis, while the $YJH$ values represent the spectrally collapsed \ifs measurement. The quoted contrast for the IFS represents a mean over all used wavelength channels. Since the GPI $H$-band data are polarized, to search for emission from the protoplanetary disk, we do not compute the brightness of the companions.}  
\end{table*}
\begin{figure*}[!t]
    \vspace{-0.3cm}
  \begin{minipage}[c]{0.67\textwidth}
    \includegraphics[width=\linewidth]{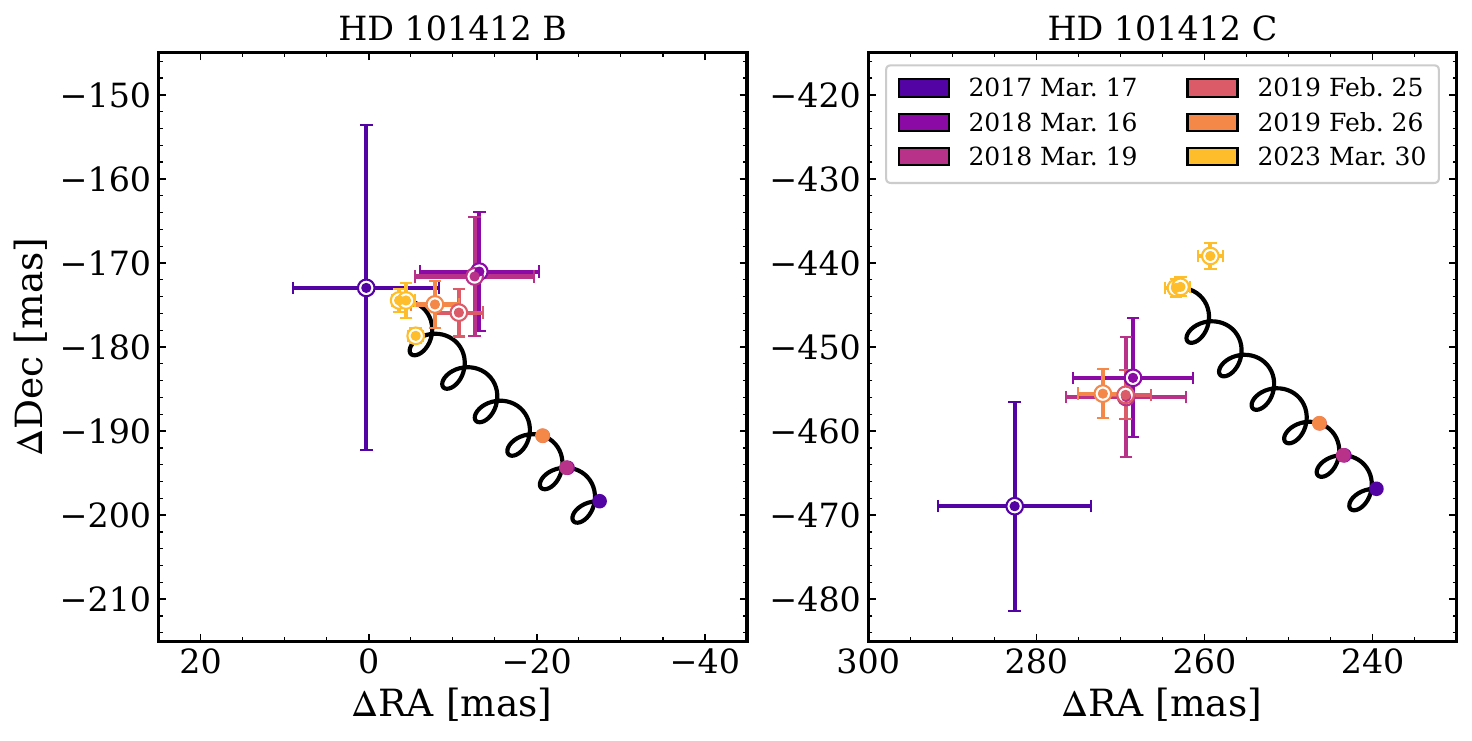}
  \end{minipage}\hfill
  \begin{minipage}[c]{0.3\textwidth}
    \caption{CPM test for HD\,101412 B and C. We anchored the black line, which illustrates how a static background object would move, at our most precise measurement from \sphere. Colored circles (without error bars) indicate the predicted background location of the target for each of our observation epochs. Dotted markers with error bars show measured astrometry for HD\,101412\,B and C. Both objects remain mostly stationary relative to the host star and do not move in the direction of the black line, indicating that they are in bound orbits around HD\,101412.} \label{fig: HD101412 astrometry}
  \end{minipage}
  \vspace{-0.35cm}
\end{figure*}
\subsubsection[HD 59659]{HD\,59659}
HD\,59659 is a young star of type F7\,V \citep{Houk_1978} in the Argus association. \citet{Torres_2008} suggest that HD\,59659 is a widely separated binary with CD$-$48~2972 as its companion, separated by $\rho \approx$\,\SI{81.7}{\arcsec} ($s\approx$\SI{7100}{\AU}). Our residual images show two point-like features.

Using PCA, we detect one far-out source visible toward the east from HD\,59659. Using archival SHINE data, we can determine that its motion over a six-year baseline is consistent with that of a background object (\autoref{fig: CPM}). The \fours reduction (shown in \autoref{fig: all candidates}, right) reveals a point-like feature very close ($\rho =$~\SI{120.96(20)}{\mas}) to the southwest ($\theta =$~\SI{202.11(14)}{\degree}) of HD\,59659. Given its proximity to the host star, it is unlikely to be a chance-aligned background object. It is detected at $4.2\sigma$ in the \fours reduction, and shows characteristic negative side lobes that are a typical result of self-subtraction in ADI analyses. The feature is also visible in the PCA reductions; however, as with HD\,1461, its peak is identifiable given its location but not distinguishable from the speckle noise surrounding it in the PCA analysis. We measure a contrast of \SI{10.22(11)}{\mag} between the feature and the host star, which corresponds to an absolute magnitude of \SI{13.20(13)}{\mag}, if bound to HD\,59659. In this case, ATMO planetary evolution models \citep{Phillips_2020} suggest a planetary mass of $M =$~\SI{10.03(41)}{\Mjup} at system age. 

We also performed a \fours reduction on the archival SHINE dataset; yet, we do not achieve as deep detection limits as with our 2023 dataset, and hence no further conclusions can be reached, as this is consistent with a non-detection of a background star as well as a true bound companion.

To test whether the detected signal corresponds to the candidate identified in the RVSPY data \citep{Zakhozhay_2022}, we performed orbital fits using \texttt{orvara} \citep{Brandt_2021_b}. A full description of this is provided in Appendix \ref{sec: HD 59659 orbital analyses}. Here we give a summary: We included 73 RV measurements, our astrometric detection, and HGCA. We fixed the primary mass to the literature value and the secondary mass to the value inferred from photometry.

The MCMC chains converge, but the inferred orbital parameters are not physically convincing: The solution requires a highly eccentric ($e > 0.8$) and inclined ($i \approx$~\SI{90}{\degree}) orbit with an orientation that places the detection near apastron. In addition, the high eccentricity is mainly driven by an offset in the first subset of RV data relative to the remaining measurements. Omitting this initial subset of 15 data points, the MCMC sampler fails to converge, casting doubt on the reliability of the orbital solution.

Taking all this evidence into account, we find that no definite conclusion about the true origin of the point-like feature is possible. We do find evidence of it being caused by a real point-like feature, such as the two negative side lobes next to the positive feature, and the fact that we are able to jointly fit an orbit using relative astrometry of the secondary, RVs, and HGCA. However, the high eccentricity and inclination, as well as the similarity between the point-like and false-positive cases in HD\,1461, cast doubt on the reliability of the detected signal. We therefore conclude that the only way to determine the true nature of this source is with additional follow-up observations reaching even deeper limits, close to the star. We propose that these should be carried out in the near future. If the orbital solution found in this work is true, it suggests the apastron passage will occur near the beginning of 2028.
\subsubsection[HD 101412]{HD\,101412}\label{HD 101412}
\begin{table}[]
   \caption{Derived parameters for companions B and C.}
    \centering
    \renewcommand{\arraystretch}{1.6}
    \begin{tabular*}{\columnwidth}{@{\extracolsep{\fill}}p{1.3cm}cccc}
    \hline
    \hline
    Companion & ${T}_{\mathrm{eff}}$ [K] & $\mathrm{log}\,g$ & $R$ [\si{\Rjup}] & $M$ [\si{\Mjup}] \\
    \hline
    \:B & $3960^{+270}_{-240}$ & $3.76^{+0.89}_{-0.82}$ & $6.39^{+0.52}_{-0.38}$ & $191 \pm 30$ \\
    \:C & $3980^{+170}_{-180}$ & $3.36^{+0.92}_{-0.58}$ & $3.75^{+0.16}_{-0.09}$ & $82 \pm 13$ \\  
    \hline
    \end{tabular*}
    \tablefoot{The effective temperature ($\mathrm{T}_{\mathrm{eff}}$), surface gravity ($\mathrm{log}\,g$), and radius ($R$) are drawn from the posterior sample of the \texttt{species} fit, while the mass ($M$) was calculated directly from the photometric values.}
    \label{tab: HD101412 properties}
\end{table}
HD\,101412 (V1052~Cen, PDS\,57) is an A3\,V Herbig Ae/Be \citep{The_1994} $\lambda$ Boötis star \citep{Gray_2017}. Based on its location and stellar type, HD\,101412 is associated with the Centaurus and Crux star-forming regions \citep{The_1994}. The age is estimated at \SI{2(1)}{\Myr} based on membership \citep{Wade_2005}. 

Our collaboration first discovered two candidate companions to HD\,101412 in \naco $L'$-band images in 2017, as already mentioned in \citet{Cugno_2023}. Since then, we have monitored the system using Gemini/GPI and, more recently, \sphere (\autoref{fig: all companions} left). \citet{Rich_2022} analyzed our Gemini/GPI datasets in the context of the protoplanetary disk, but also detected the companions and concluded that they are bound based on the statistical unlikelihood of such close background stars. In this work, we provide the first definitive evidence that both are companions to HD\,101412, rather than foreground or background stars (\autoref{fig: HD101412 astrometry}). We determine the spectral types of both objects to be late K through early M (\autoref{fig: Spectra HD101412}) and determine their physical properties (\autoref{tab: HD101412 properties}).
\begin{figure}[ht!]
    \centering
    \makebox[\linewidth][c]{
    \includegraphics[width=\linewidth]{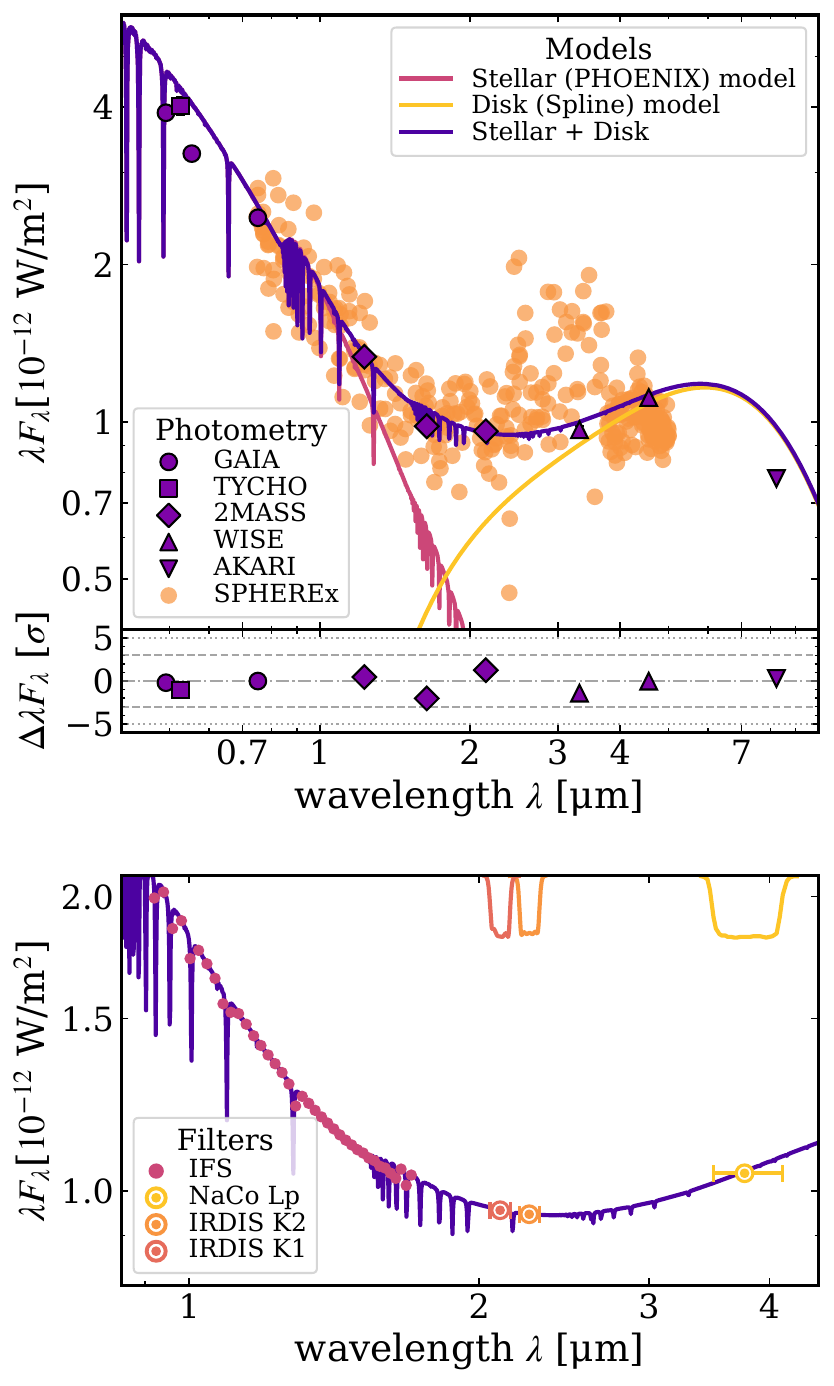}}
    \caption{\textit{Top}: Best-fit SED model for HD\,101412. Photometry from \textit{Gaia} DR3, Tycho, 2MASS, AKARI/IRC, and WISE appear as dark purple points with black borders. The orange data points show SPHEREx flux measurements of the star. We do not include them in our fit, but they agree with our best-fit spectrum. The stellar PHOENIX model is pink, the disk model is yellow, and the combined spectrum is purple. Residuals are shown below with 3$\sigma$ and 5$\sigma$ regions. \textit{Bottom}: Stellar flux estimates used to convert companion contrast to flux. The filter profiles for $K1$, $K2$, and $L'$ are shown at the top.}
    \label{fig: HD101412 SED} 
\end{figure}
We list all datasets collected by our collaboration, along with the measured position and brightness, in \autoref{tab: HD101412 Astrometry}. To validate the two companions as bound, we performed a CPM test shown in \autoref{fig: HD101412 astrometry}. Both companions clearly differ from the motion of a background object over six years. While HD\,101412 B remains stationary, the position of HD\,101412 C changes by $2.7\sigma$ between the \naco epoch and the \sphere observations, indicating possible orbital motion. We cannot, however, rule out that this change is due to calibration differences between instruments or statistical scatter. The final determination is left open to future work.

To study the properties of these objects, we also performed spectral characterization. We started by assembling all photometry and spectra in contrast units. As the stellar spectral energy distribution (SED; \autoref{fig: HD101412 SED}, upper purple markers) shows significant IR excess emission from the protoplanetary disk, which is not resolved and hence contributes to the reference PSFs used for relative photometry, we constructed our stellar+disk model as follows. First, we fitted optical photometry in the Geneva system from \cite{Paunzen_2022} and from \textit{Gaia} DR3 using the updated PHOENIX model grid by \cite{Hauschildt_2025}. We adopt a temperature of $T_\mathrm{eff}=8300$~K and a surface gravity of $\log g=3.8$ \citep{Cowley_2010}, as well as a photospheric metallicity of $[Fe/H]=-1.0$~dex \citep{Guimaraes_2006}. For the disk model, we tested a single-temperature and a two-temperature model, but neither reproduced the disk emission to the required level. Hence, we adopted an eight-parameter spline model to fit the photometry from the Tycho, \textit{Gaia} DR3, 2MASS, AKARI/IRC, and WISE surveys.\footnote{Accessed via \href{https://vizier.cds.unistra.fr/vizier/sed/?submitSimbad=Photometry&-c=11+39+44.4564911064-60+10+27.719780304&-c.r=5&-c.u=arcsec&show_settings=1}{vizier.cds.unistra.fr}} We then combined the disk and stellar models and use \texttt{pyphot} \citep{zenodopyphot} to extract the fluxes for the \ifs data points and for each filter. For photometric filters, this utilizes filter profiles from the SVO online library \citep{Rodrigo_2012, Rodrigo_2020, Rodrigo_2024}. We also show flux measurements of HD\,101412 from the SPHEREx satellite \citep{Bock_2026} for comparison with our final spectrum.
\begin{figure*}[!t]
  \begin{minipage}[c]{0.74\textwidth}
    \includegraphics[width=\linewidth]{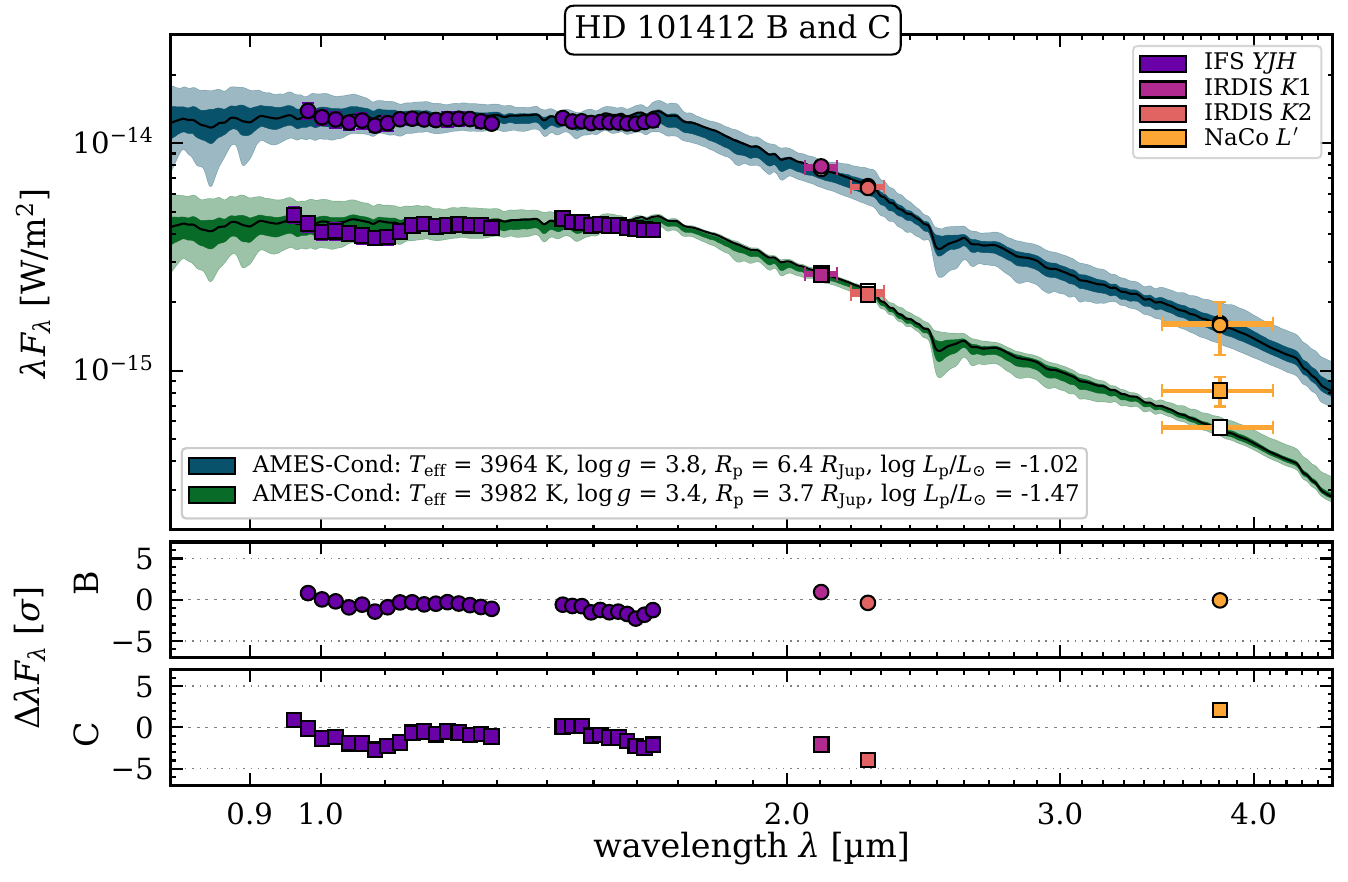}
  \end{minipage}\hfill
  \begin{minipage}[c]{0.24\textwidth}
    \caption{\textit{Top}: Best-fit model spectra for companions HD\,101412~B (blue) and HD\,101412~C (green), with the AMES-Cond model parameters listed in the legend \citep{Allard_2012}. The colored regions show the $1\sigma$ and $3\sigma$ uncertainties, while the black central line shows the best-fit model in each case. The colored markers show the respective instruments and filters. \textit{Bottom}: Individual residuals for companions B and C as marked on the y-axis.}
    \label{fig: Spectra HD101412}
  \end{minipage}

\end{figure*}
To characterize the atmospheric properties of HD\,101412 B and C, we fit their spectral and photometric data using the \texttt{species} package \citep{Stolker_2020}, which determines best-fit grid models via Bayesian inference with nested sampling. We selected the AMES-Cond model grid \citep{Allard_2012}, which covers the effective temperature expected for both companions. The resulting best-fit SEDs, together with the \ifs, \irdis\ $K1$ and $K2$, and \naco\ $L'$ data points, are shown in \autoref{fig: Spectra HD101412}. As the derived values for the surface gravity ($\log g$) have large uncertainties, the masses that are calculated from surface gravity and radius also show large uncertainties. Therefore, we adopted mass estimates based solely on evolutionary models, using the derived photometric fluxes. A detailed description of this, as well as of the atmospheric fitting procedure, including the choice of priors and the resulting corner plots, is provided in Appendix~\ref{sec: Atmospheric Characterization of HD 101412 B and C}. The best-fit values are given in \autoref{tab: HD101412 properties}.

Both companions appear inflated relative to main-sequence stars of comparable temperature. Notably, while they share similar effective temperatures, their masses differ significantly: companion~C lies close to the substellar--stellar boundary, whereas companion~B is clearly above it. Spectrally, both are consistent with late~K- to early~M-type objects. For companion~C, we find a $2\sigma$ discrepancy between the measured $L'$-band flux and the model prediction, which may indicate excess flux attributable to an accretion disk around the companion.

A robust orbital solution enabling a full assessment of the dynamical stability of both companions, with respect to each other and to the circumstellar disk at \SI{92.5}{\AU} \citep[][see also \autoref{table: stars astro physical prop}]{Geers_2007}, is beyond the scope of this work and warrants a dedicated future study. Nevertheless, assuming the projected physical separations of $s_\text{HD$\,$101412 B} \approx$~\SI{70}{\AU} and $s_\text{HD$\,$101412 C} \approx$~\SI{215}{\AU}, approximate the true semimajor axes and adopting coplanar, circular orbits, we estimate a mutual Hill radius of only $\approx\,$\SI{46}{\AU} for the two companions. This marginally exceeds the minimum separation required for mutual orbital stability of $\approx\,$\SI{42}{\AU} \citep{Chambers_1996}. This could suggest that the companions are on orbits with larger distances from the host star, eccentric or inclined orbits, or that the system is near the edge of dynamical stability. Given the young age of the system of only \SI{2(1)}{\Myr}, one or both companions could be about to be ejected from the system and become a free-floating star or brown dwarf. Under the same assumptions as above, the outer disk at \SI{92.5}{\AU} would be dynamically influenced by both companions, as each has an individual Hill radius exceeding one-third of its distance to the disk \citep{Gladman_1993, Kirsh_2009}. This could still be possible if the disk were in a mean-motion resonance, where it could reside safely within the unstable region. In this case, however, the disk would have a non-axisymmetric structure.

Future observations should revisit and monitor this system using high-spatial-resolution facilities, such as VLTI/GRAVITY+ \citep{GRAVITYCollaboration_2017}, VLT/ERIS-NIX \citep{Davies_2023}, and \sphere. Such observations would allow for a determination of whether companion~C exhibits measurable orbital motion, thereby constraining the companion's dynamical mass and enabling further study of the stability of this multi-star, disk-hosting system. Additional observations could also confirm whether the $L'$-band excess seen with \naco originates from a disk around HD\,101412~C. Furthermore, higher-resolution spectroscopy, particularly with VLTI/GRAVITY+, would provide sensitivity to the companions' surface gravity, enabling a robust determination of $\log g$ and, consequently, their individual masses. Lastly, observations with submillimeter and radio telescopes such as ALMA would be helpful if they are able to resolve the outer disk and constrain its properties. 

In addition to the host star and the two companions, we detect eleven point sources at projected separations $s>$\,\SI{500}{\AU}, within the \irdis FoV (\SI{11}{\arcsec}\,$\times$\,\SI{12.5}{\arcsec}). The brightest ones are also detected in the \naco $L'$ image, but their low \snr results in detections at $<3\sigma$, limiting astrometric precision. A CPM test between \naco and \sphere rules out both stationary background and bound-companion scenarios, favoring moving background stars. All eleven sources show no color, that is, $K1$--$K2 \approx 0$, which is consistent with stars \citep{Pecaut_2013} but also with a planetary nature, as at the young age of HD\,101412, planetary-mass companions are expected to be hot and show no methane feature causing the color difference between $K1$ and $K2$ \citep{Phillips_2020}. However, the background origin is further supported by the Galactic-center line of sight toward HD\,101412: Using \mbox{TRILEGAL} \citep{Girardi_2005}, we would expect $\approx 7$ background stars within the \irdis FoV at our sensitivity, consistent with the observed population. Background stars as close as the companions are not expected, consistent with \citet{Rich_2022} finding that the companions are likely to be bound within $3\sigma$, based on stellar population.
\begin{figure*}[t]
    \centering   
    \includegraphics[width=\textwidth]{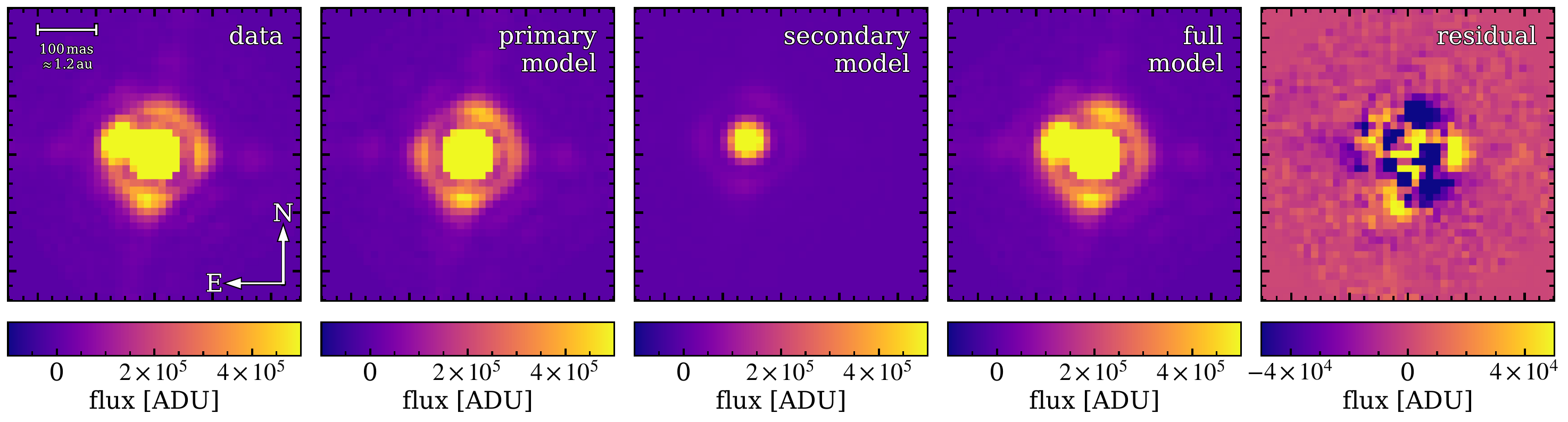}
    \vspace{-0.5cm}
    \caption{Best-fit result for the flux measurement of HD\,131511. \textit{Left to right}: Original reference PSF obtained for HD\,131511; the reference PSF of HD\,377, which we used as our single-star model and also acts as the model for the primary star; the shifted and scaled model for the secondary; the full model, adding the primary and secondary; and the residual when subtracting the joint model from the data.}
    \label{fig: HD131511 PSF model}
    \vspace{-0.3cm}
\end{figure*}
\subsubsection[HD 121617]{HD\,121617}
HD\,121617 is a young A1\,V \citep{Houk_1978} star and a member of the Upper Centaurus--Lupus association \citep{Hoogerwerf_2000, Gagne_2018}. The star hosts a debris disk, first identified via IR SED excess with IRAS \citep{Moshir_1989}. The disk has since been resolved with ALMA in continuum and CO gas \citep{Moor_2017, Marino_2026_ARKSOverview}, and with \irdis in dual-beam polarimetric imaging (DPI) mode in the $J$-band by \citet{Perrot_2023}. The ALMA continuum shows a southwest clump, interpreted as either dust--gas interaction or sculpting by an internal planet \citep{Weber_2026, Marino_2026_121617}, while \sphere scattered light shows no asymmetry \citep{Milli_2026}. The CO gas appears optically thick \citep{Brennan_2026}.

\Autoref{fig: all disks} (left) shows the PCA-ARDI residual image, clearly revealing the scattered-light ring. Its morphology matches polarized-light images from \citet{Perrot_2023}, with no overdensity at the ALMA clump location, but a bright scattered-light region in the southwest. Additionally, a point source at $\rho =$~\SI{1.315(19)}{\arcsec} south-southwest ($\theta =$~\SI{200.35(14)}{\degree}) of HD\,121617 is detected with a contrast of \SI{11.92(13)}{\mag} in $H2$ and \SI{11.74(12)}{\mag} in $H3$. If bound, this implies absolute magnitudes of \SI{13.76(13)}{\mag} ($H2$) and \SI{13.61(13)}{\mag} ($H3$) and a projected separation of $s=$~\SI{155.3(14)}{\AU}. Assuming coeval age, the brightness implies a young, hot ($T\approx$~\SI{1000}{\K}) planet of $M = 4.84_{-0.26}^{+0.22}$~\si{\Mjup} following ATMO models \citep{Phillips_2020}. If on a circular orbit, this planet would be at least three Hill radii exterior to the disk, consistent with orbital stability of the disk against perturbations by the planet (e.g., \citealt{Gladman_1993, Kirsh_2009}).

Archival \ispynaco data show no detection, reaching only a contrast of $\approx 9.5$~\si{\mag}, corresponding to a limit of apparent mag $\approx 16.4$ in $L'$. This is, however, consistent with a non-detection of the companion candidate as ATMO models predict an apparent magnitude of $\approx 17.0$ in $L'$ for optimistic cases, but also for a background star which can have an $L' - H$ color difference between $-$\SI{1.5}{\mag} and $+$\SI{0.1}{\mag} \citep{Pecaut_2013}, resulting in a magnitude limit of $\approx 17.5$ or fainter. The color ${H2} - {H3} = 0.15\pm0.18$~\si{\mag} is also consistent with either scenario, as a companion would show weak color at this temperature, especially with nonequilibrium chemistry. At $\rho =$~\SI{1.315(19)}{\arcsec} the source lies outside the \ifs FoV, and therefore, given this single epoch, its nature remains uncertain. A second-epoch observation in another filter with sufficient contrast is required for a conclusive determination.
\subsubsection[HD 131511]{HD\,131511}
HD\,131511 (DE Boo) is a single-lined spectroscopic binary with a K0\,V primary \citep{Gray_2017} in the Boötes constellation. Astrometric \citep{Jancart_2005} and RV \citep{Nidever_2002} data confirm the binary nature and determine a secondary mass of $\approx$\,\SI{0.45}{\Msun}.
\cite{Greaves_2014} first resolved a debris disk with \textit{Herschel}/PACS \citep{Poglitsch_2010}, previously inferred from the SED. This SED was partly incorrect due to saturated 2MASS data \citep{Koerner_2010, Gaspar_2013, Kennedy_2015}. Using a corrected SED, \cite{Pearce_2022} report a radius of \SI{100(70)}{\AU}. The \irdis FoV covers only $s\approx $~\SI{70}{\AU} and shows no disk signal.

The PCA residual image (\autoref{fig: all companions} center) shows a point-like source very close to HD\,131511. Its proximity causes a strong alteration of transmitted flux and the projected position due to the coronagraph \citep{Ceva_2025}. The source is visible in non-coronagraphic flux frames (\autoref{fig: HD131511 PSF model} left). We therefore forward-modeled host and companion using a two-PSF model based on HD\,377, selected for its clean reference (no additional source visible in the reduced coronagraphic images) and high counts (see \autoref{fig: HD131511 PSF model}, middle).

We fit six parameters, the x and y positions and intensities of both sources, by minimizing absolute residual flux (\autoref{fig: HD131511 PSF model} right). The residual shows minor systematics for the primary, due to a slight asymmetry in second-order PSF lobes, but none at the companion position. Using \texttt{scipy.optimize.minimize}, we derive a separation of $\rho =$~\SI{62.88(10)}{\mas} at $\theta =$~\SI{67.58(12)}{\degree}, corresponding to $s=$~\SI{0.7150(40)}{\AU}. The best fit is found for a contrast of \SI{2.2422(30)}{\mag}, implying an absolute magnitude of \SI{6.275(38)}{\mag} for the secondary. This corresponds to an M1.5\,V to M2\,V dwarf with $M = 0.47$--\SI{0.50}{\Msun} \citep{Pecaut_2013}, consistent with earlier estimates \citep{Nidever_2002, Jancart_2005}.
\subsubsection[ε Sgr]{$\epsilon$~Sgr} \label{HD 169022}
\begin{figure*}[t]
    \centering
    \makebox[\linewidth][c]{
    \includegraphics[width=\linewidth]{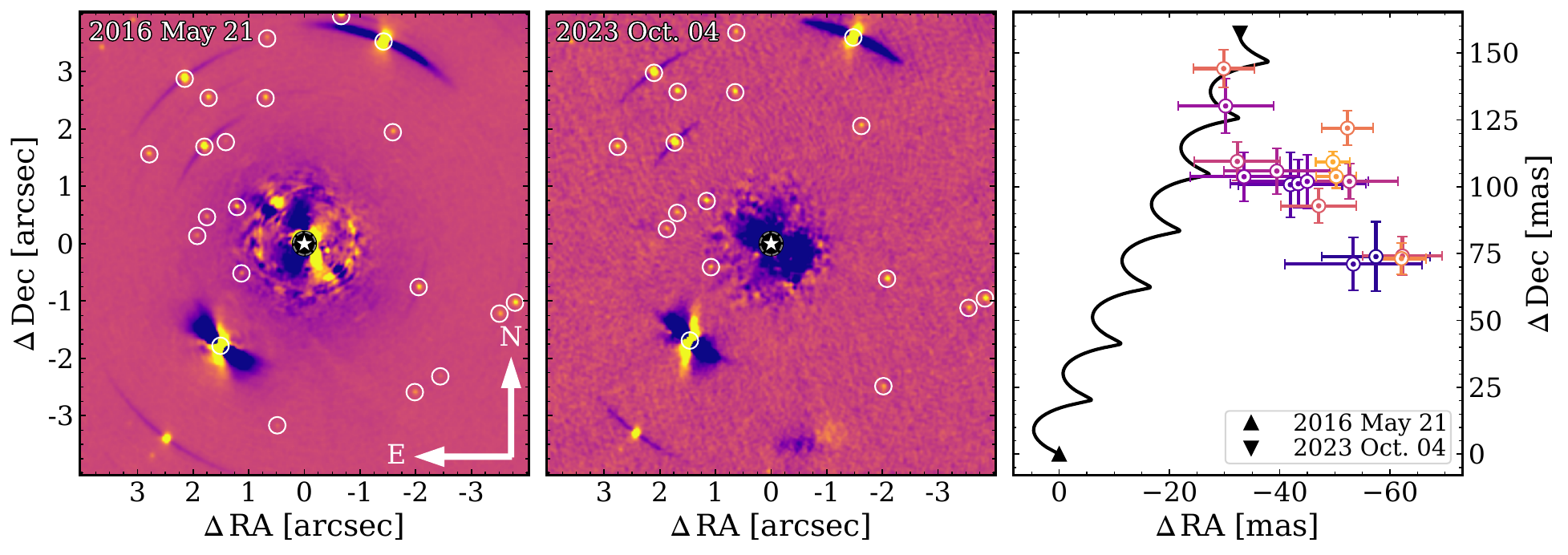}}
    \vspace{-0.5cm}
    \caption{CPM test for the background objects in the HD\,179218 images. All tested sources are again marked with white circles in the images. As multiple sources were tested for this target, the CPM test (\textit{right}) was performed in coordinates relative to the astrometric measurement in the first frame. The colored markers therefore correspond to the astrometric change relative to HD\,179218.}
    \label{fig: HD 179218 CPM}
\end{figure*}
$\epsilon$~Sgr (also known as HD\,169022 or Kaus Australis) is a binary system in Sagittarius with a B9\,IV primary \citep{Slettebak_1975}. Based on SED measurements, \citet{Chen_2014} infer the presence of a debris disk at \SI{120(60)}{\AU}. We do not detect scattered light from the disk, but resolve the secondary companion at $\rho = $~\SI{2.9555(59)}{\arcsec} and $\theta = $~\SI{139.54(14)}{\degree} east of north (\autoref{fig: all companions} right).\newpage

Comparison with our \ispynaco epoch reveals significant orbital motion. We measure a change in separation of $\Delta \rho =$~\SI{+83(14)}{\mas} and a change in position angle of $\Delta \theta =$~\SI{-0.95(32)}{\degree}, corresponding to a significance of 6.22$\sigma$. The motion is even more pronounced relative to the original detection: \citet{Golimowski_1993} report an upper limit of $\rho =$~\SI{2.22(5)}{\arcsec}, which differs from our \sphere measurement by around 15$\sigma$. This long temporal baseline provides partial constraints on the orbit, but we are not able to derive dynamical masses: The primary is too bright for \textit{Gaia}, which prevents the construction of an HGCA solution, and no public RV data are available.

If the disk edge were truncated by a giant planet, \citet{Pearce_2022} predict a minimum mass of $M=$~\SI{2.2(8)}{\Mjup} with a maximum semimajor axis of \SI{90(40)}{\AU} for such a companion. We detect two point-like features at $s \approx $ 95 and \SI{96}{\AU}, visible in the top-right panel of \autoref{fig: all residual images}. Their $H2$-band brightness implies masses of \SI{6.6}{\Mjup} and \SI{8.1}{\Mjup}. However, both sources show no color difference between $H2$ and $H3$, which is inconsistent with evolutionary models that predict $H2 - H3 \approx -3$~\si{\mag} at this age and mass. Consequently, both sources are likely to be of background origin.

At separations $\rho >$~\SI{3}{\arcsec}, we detect nine additional point-like sources. We exclude all of them as background objects based on their $H2 - H3$ color. This result agrees with predictions from the TRILEGAL Galactic model \citep{Girardi_2005}, similar to the case of HD\,101412. Given the position of $\epsilon$~Sgr, the \irdis FoV, and our detection limits, TRILEGAL predicts approximately 15 background sources in our images.
\subsubsection[HD 179218]{HD\,179218}
HD\,179218 (MWC\,614) is a young Herbig Ae/Be star of spectral class A0\,V \citep{Vieira_2003} and harbors a known protoplanetary disk. The disk was imaged in $L'$-band with \naco as part of the ISPY survey, and the respective images are published in \citet{Cugno_2023}. Further, \citet{Kluska_2018} report polarized-light detections with VLT/SPHERE/ZIMPOL, which agree with our detections. We report the detection of the disk in $H2$ and $H3$ as well as several point-like features.
Using archival data from SPHERE-NIRSUR allowed us to confirm all detected sources as background objects based on a CPM test (shown in \autoref{fig: HD 179218 CPM}): While most sources differ from a completely stationary background object, all strongly disagree with the expected relative non-motion of bound companions. As most of the objects exhibit a similar movement, we attribute this to an intrinsic CPM of these background objects.
\subsubsection[HD 181327]{HD\,181327}
HD\,181327 is a F6\,V \citep{Torres_2006} star and member of the $\beta$ Pictoris moving group. The debris disk, which is shown in \autoref{fig: all disks} (right), was first discovered in the SED IR excess measured by IRAS \citep{Mannings_1998}. Coronagraphic observations with HST imaged the system and were able to resolve the disk, discovering the nearly circular shape with a radial extent of \SI{25}{\AU} at a separation from the host star of around \SI{84}{\AU} \citep{Schneider_2006, Schneider_2014}. Further, ALMA continuum observations showed that the disk has a clean inner edge at \SI{73.3(4)}{\AU} and an inclination of \SI{30.0(5)}{\degree} \citep{Pawellek_2021}. Emission lines were also detected, indicating that the disk contains CO \citep{Marino_2016, Milli_2024}. Recent observations with JWST revealed that the disk contains H$_2$O in solid form, marking the first detection of water ice in any debris disk \citep{Xie_2025}.

While a full disk model is beyond the scope of this work, we determine the geometric properties by measuring the peak brightness at the shortest and longest projected radii, bisecting the disk through the star position. We find our detection is consistent with a disk of radius \SI{84}{\AU} and inclination \SI{30}{\degree}, and do not see any signs of asymmetries, consistent with previous measurements that the disk has a low eccentricity \citep[e.g.,][]{Milli_2024}. In future work, these images can help further characterize the disk by providing proper constraints on the offset of the disk center with respect to the star.

We detected a point-like feature toward the west of HD\,181327, at $\rho =$~\SI{4.908(84)}{\arcsec}. We validated this source as a background star using archival data and performing a CPM test, shown in \autoref{fig: CPM}.
\begin{figure*}[ht!]
    \centering
    \makebox[\linewidth][c]{
    \includegraphics[width=0.98\linewidth]{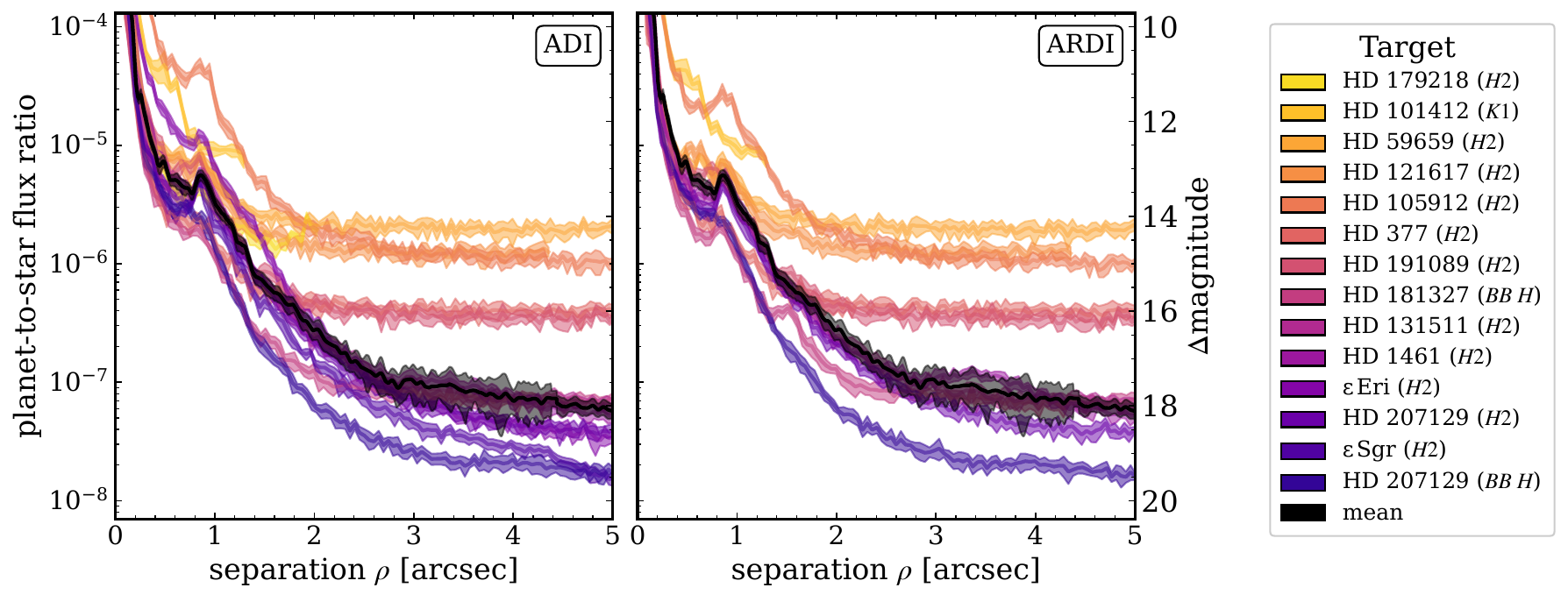}}
    \vspace{-0.73cm}
    \caption{Optimal contrast curves for all targets in our initial observed sample when performing PCA-based ADI (\textit{left}) or ARDI (\textit{right}) PSF subtraction. Targets are sorted and colored with respect to their maximum contrast at \SI{4}{\arcsec} separation from the star in increasing contrast ratio. Within the AO control radius, all datasets show consistency following the same curve. In the outer, background-limited regime, a higher contrast is achievable only for brighter stars. Curves for three individual targets are also shown as examples in \autoref{fig: individual contrast curves}. The corresponding data files for all targets are available online on Zenodo.}
    \label{fig: joint contrast curves}
\end{figure*}
\subsubsection[HD 191089]{HD\,191089}
HD\,191089 is a F5\,V star hosting a debris disk which consists of two distinct components, both imaged in scattered light with HST NICMOS \& STIS and Gemini/GPI \citep{Ren_2019}. We do not detect any of this emission from the disk in our $H2$/$H3$ images, but our reductions reveal a source far from HD\,191089 at $\rho =$~\SI{3.6353(62)}{\arcsec} at an angle of $\theta =$~\SI{138.22(14)}{\degree}. While at this distance, the probability of a chance-aligned background object is high, we confirm this hypothesis by reanalyzing archival data of the star from the SPHERE-SHINE survey: a CPM test (\autoref{fig: CPM}) validates the source as a background object. 
\subsubsection[HD 207129]{HD\,207129}
HD\,207129 is a nearby Sun-like (G2\,V; \citealt{Torres_2008}) star in the southern-hemisphere constellation Grus. A debris disk was first measured based on IRAS IR emission excess \citep{Walker_1988} and later resolved with HST \citep{Krist_2010} and \textit{Herschel}/PACS as well as SPIRE \citep{Griffin_2010, Poglitsch_2010, Marshall_2021}, revealing that the disk has a semimajor axis of around \SI{150(20)}{\AU} \citep{Marshall_2011, Pearce_2022}. This places the disk outside the FoV of \irdis, and no signal from the disk is visible in our images.

We detect a point source in the southwest corner in both epochs of our \ispysphere data. With a projected separation of $\rho =$~\SI{5.1642(88)}{\arcsec} ($s\approx$~\SI{80}{\AU}), it would be located inside the debris disk if bound to HD\,207129. However, a CPM test (see \autoref{fig: CPM}) rules out a companion, as the source shows motion identical to a stationary background object.

In addition, the \fours (\autoref{fig: all residual images}, bottom left and center) residuals show tentative detections of a very close-in point source in both epochs: In the 2023 dataset, a clear feature is visible to the northeast, which has moved slightly more inward and to the west, consistent with orbital motion for close-in companions, of the host star in the 2022 dataset, which also shows more noise in this region. We find that both features are consistent with speckles, as they appear at slightly wider separations in the longer-wavelength ($H3$) data than in the shorter-wavelength ($H2$) observations.
\subsection{Contrast curves}\label{sec: contrast curves}
\begin{figure*}[t]
    \centering
    \includegraphics[width=0.88\textwidth]{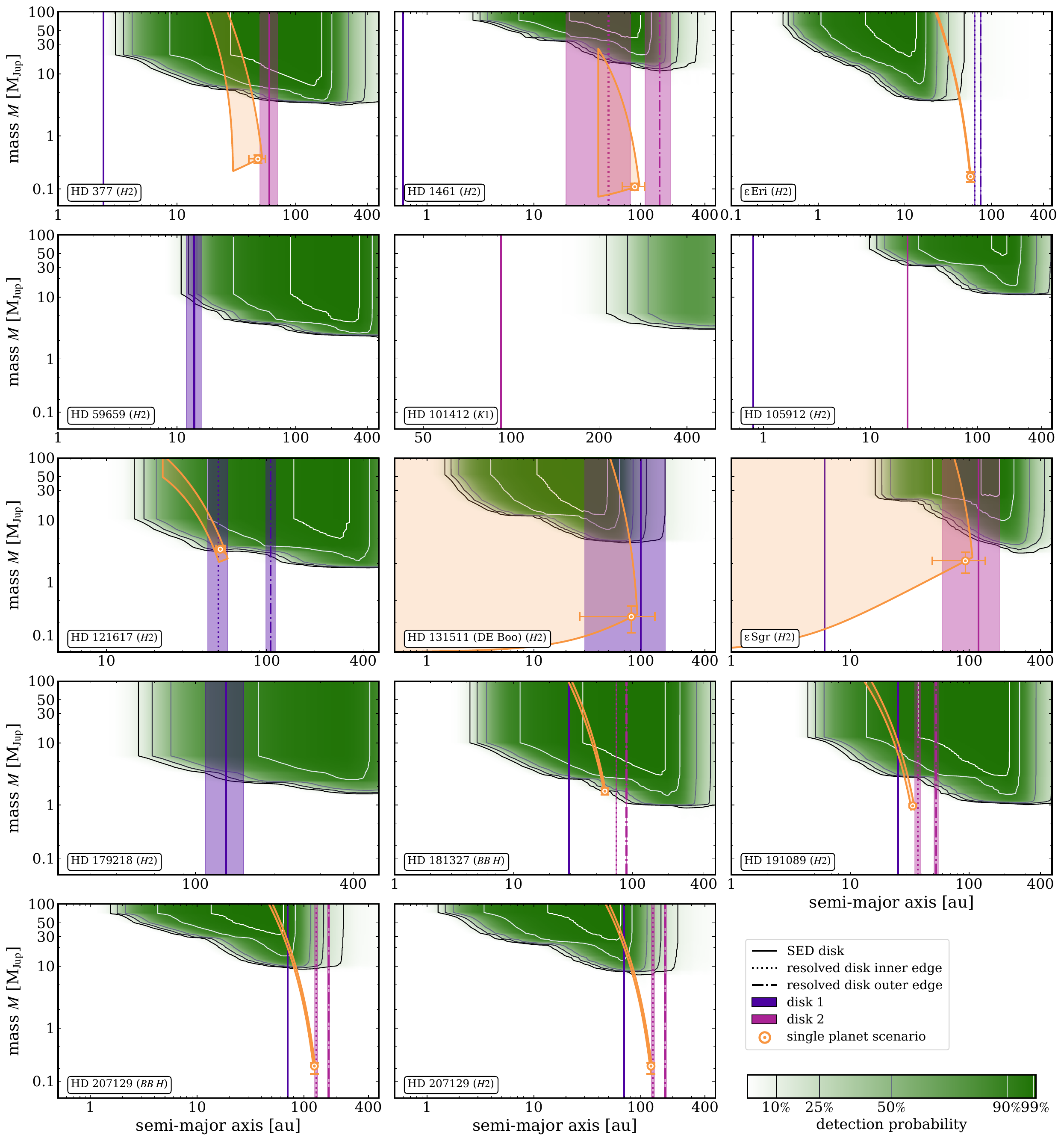}
    \caption{Detection completeness maps (green) compared to the parameter space for a single planet that could be responsible for shaping the edges of debris disks by \citet{Pearce_2022} (orange). The most extreme scenario for a single planet, with the minimum mass at the maximum semimajor axis, is marked with a dotted marker. Green shaded regions show the companion mass detection probability limits based on the contrast limits in the $H2$ band; see \protect\autoref{fig: individual contrast curves}. The conversion includes calculating masses from contrasts using ATMO planetary evolution models \citep{Phillips_2020}, and probabilities were calculated with the \texttt{Exo-DMC} \citep{Bonavita_2020}. The exact procedure is described in Sect. \ref{sec: Detection Limits and Planet Predictions}. Locations of known debris disks are shown by vertical extended lines in two shades of purple. If the extent of the disk is known, two lines of the same color indicate the inner edge and the outer edge. Note that the minimum orbital distances shown on the x-axis vary depending on the distance to each target.}
    \label{fig: all detection curves}
\end{figure*}
To assess the sensitivity of our images, we performed robust injection-recovery tests to derive the achievable contrast limits for our sample. We use the \texttt{applefy} package \citep{Bonse_2023}, which allows for individual injection-recovery tests, meaning that fake planets are injected one after another for a set of separations $n$ as well as parallactic angles $m$ rather than all in the same injection-recovery process. This then results in an $n \times m$ set of data cubes, which are individually post-processed using PCA with a variety of different numbers of PCA modes. In the last step, the S/N values are calculated from the residual images and converted to false-positive fractions expressed in terms of $\sigma$. For this last step, \texttt{applefy} features the possibility of using Student's t-test, assuming a Gaussian noise distribution, for small sample statistics \citep{student_1908, Mawet_2014}.

Before performing the injection tests, the cubes were cleaned of known point-like objects by injecting a PSF model with the inverted flux at the object's position. For all datasets, we calculated contrast curves for reductions with a variety of PCA modes in the range from 1 to 50 and for pure ADI and ARDI, which combines the RDI and ADI sets when building the PSF library. The contrast curves are always calculated up to a projected separation of $s =$~\SI{500}{\AU} around each star, or to the maximum \irdis FoV if the \SI{500}{\AU} were outside the FoV. As this \SI{500}{\AU} radius frequently contains large parts where the detector background is limiting, most of the residual noise is of a Gaussian nature. Hence, we used the t-test to calculate the contrast limits. \Autoref{fig: individual contrast curves} shows ADI contrast curve examples for three of the thirteen targets in the \ispysphere sample.

For a simple overview, \autoref{fig: joint contrast curves} shows the maximal achievable contrast at each separation for each host star for the pure ADI and ARDI cases. All curves follow roughly the same shape with increasing contrast at higher separations. Additionally, a bump at \SI{0.8}{\arcsec} is visible. This is caused by the outer edge of the AO control radius. Using ADI, we achieve a mean contrast at \SI{250}{\mas} and \SI{1}{\arcsec} of $(3.30\pm0.40)\times10^{-5}$ and $(3.71\pm0.49)\times10^{-6}$, respectively. Also taking into account the reference target data and performing ARDI, this slightly improves to $(3.17\pm0.42)\times10^{-5}$ at \SI{250}{\mas} and $(3.49\pm0.50)\times10^{-6}$ at \SI{1}{\arcsec}. In the background-limited regime, the maximum achievable contrast depends mainly on the brightness of the host star and thus varies widely. We note that, since our sample of observations is nonhomogeneous (e.g., filter setups or exposure times), these values should not be used to quantify the performance of \irdis in general.

\begin{figure*}
  \begin{minipage}[c]{0.68\textwidth}
    \includegraphics[width=\linewidth]{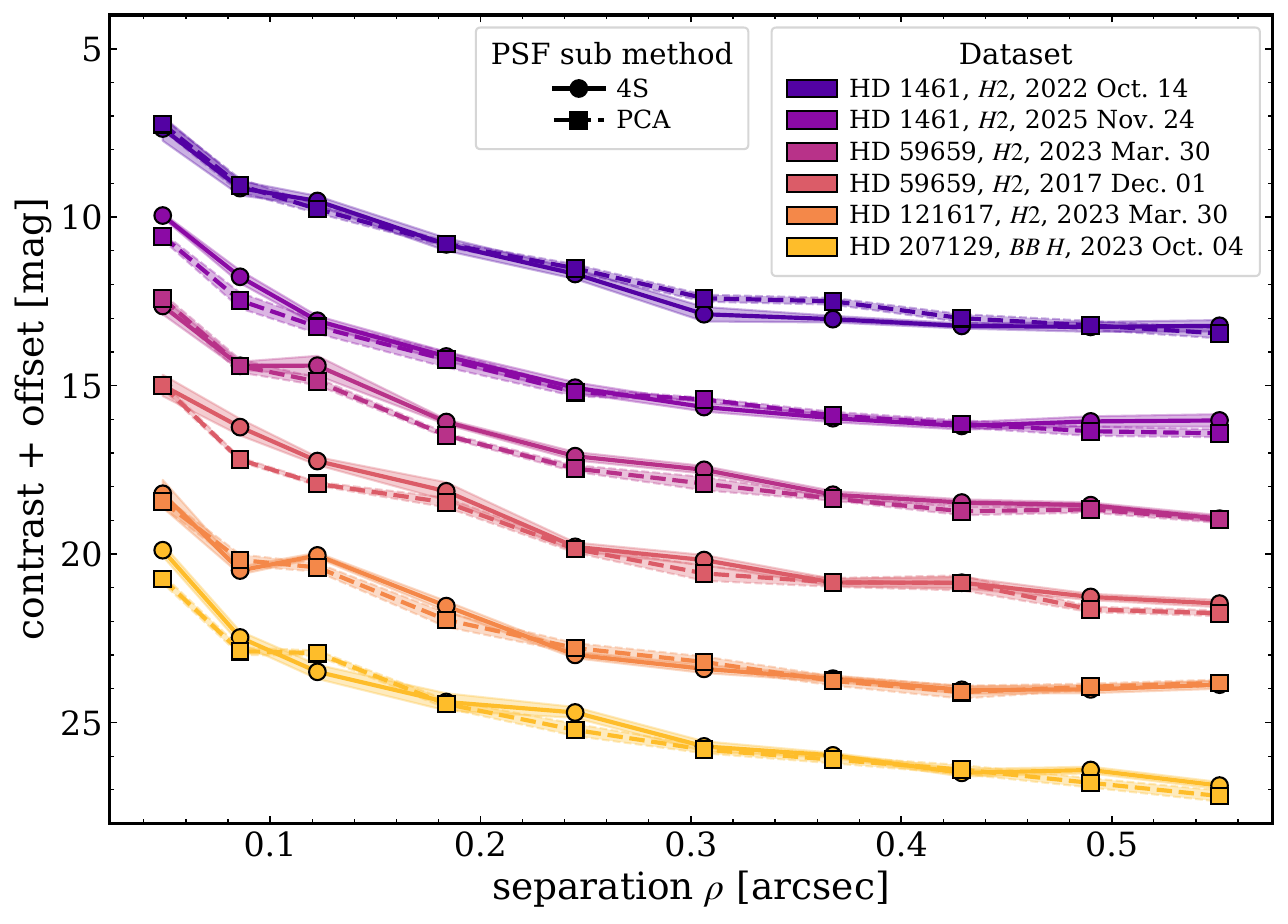}
  \end{minipage}\hfill
  \begin{minipage}[c]{0.3\textwidth}
    \caption{Comparison of achieved limits using \fours vs. PCA for all tested datasets. Colored markers show the best maximal achievable contrast at separations tested with PCA and \fours for each dataset, colored accordingly. The connecting lines are for visual purposes only. For each dataset, we added an offset to improve visibility, which was then applied to the \fours and PCA curves to allow for a direct comparison.}\label{fig: 4S performance}
  \end{minipage}
  \vspace{-0.4cm}
\end{figure*}

\subsection{Detection limits and planet predictions}\label{sec: Detection Limits and Planet Predictions}
Our analysis requires converting between measured magnitudes and planet masses to interpret the properties of measured point sources and determine detection limits for the survey. We performed this conversion in two steps. First, using ATMO strong nonequilibrium chemistry (NEQ strong) models \citep{Phillips_2020}, we converted the contrast curves into planet masses, accounting for system distances and host star magnitudes. In the second step, we used \texttt{Exo-DMC} \citep{Bonavita_2020} to calculate detection probabilities, which account for the possibility of a planet being at a current projected position where it is below the contrast limit, due to eccentric orbits, inclination, etc. \Autoref{fig: all detection curves} shows the resulting probabilities continuously as well as at discrete levels. We also show locations of known disks for each system. For debris disk-hosting systems, we also show limits for disk-edge-shaping companions by \citet{Pearce_2022}. In \autoref{table: stars astro physical prop}, we list the minimum mass at the maximum semimajor axis for these predictions, which are also marked by the dotted markers in \autoref{fig: all detection curves}. While in many cases we can rule out parts of the parameter space, there are only two stars (HD\,121617 and HD\,181327) where we reach the minimum predicted mass.
\section{Discussion}\label{sec: Discussion}
\subsection{Detections and further prospects}
In this work, we image several low-mass companions and disks, as highlighted in our discussion of each individual object in Sect. \ref{sec: Results} above. The discovery and confirmation of the two dwarf companions to HD\,101412 marks the start of an interesting system to investigate. Future monitoring of the companions with high-contrast facilities such as \sphere and VLTI/GRAVITY+ \citep{GRAVITYCollaboration_2017} would allow us to further constrain their orbits and, consequently, unravel the system architecture, in particular by providing constraints on whether the companion orbits are aligned with each other, with the outer disk, and their relative separation to the outer disk. Future work should also explore possible orbital configurations based on stability analysis. While a full determination of this is beyond the scope of this work, we find that, assuming the physical separations are comparable to the actual semimajor axes, the companions are close to the instability regime and for both, their individual Hill radii are most likely inconsistent with disk stability. This could indicate that the companions actually have larger semimajor axes, are not aligned with each other, are on eccentric orbits, or are in an unstable system configuration. The spectral analysis reveals that both companions appear inflated relative to main-sequence stars and have effective temperatures and spectral shapes similar to those of late-type K or early-type M dwarfs. A comparison of the photometric fluxes to evolutionary models suggests that companion C is near the substellar--stellar-mass border, while B is clearly in the stellar regime. The additional slightly enhanced $L'$-band flux hints at a disk around companion C. This could also reveal that the polycyclic aromatic hydrocarbon (PAH) emission found by \citet{Geers_2007} does not originate in a disk around the host star, but is instead emitted by dust and gas around companion C. Follow-up observations with high-resolution spectrographs such as VLTI/GRAVITY+ or VLT/CRIRES+ \citep{Kaeufl_2004, Dorn_2014, Dorn_2023}, or mid-IR instruments such as VLT/ERIS \citep{Davies_2023}, would yield additional information on these prospects and the chemical composition of the two companions. Additionally, high-resolution spectra will be sensitive to the surface gravity, allowing us to properly constrain their masses.
 
We also imaged two debris disks: HD\,121617 and HD\,181327. Both disks had previously been imaged in scattered light and thermal emission \citep{Lebreton_2012, Marino_2016, Perrot_2023, Milli_2024}, but our new data provide complementary information to the existing disk images. In particular, images of both targets have been published with SPHERE polarimetry \citep{Perrot_2023, Milli_2024}; our observations complement these previous efforts. A detailed, multi-wavelength disk modeling analysis is beyond the scope of this work, but could in the future provide constraints on the dust grain properties in each system \citep{Milli_2024, Olofsson_2026}. For HD\,121617, recent ALMA observations have revealed an arc of enhanced dust density \citep{Marino_2026_121617}, and our SPHERE observations hint at a brightening at a similar but not identical location. In the case of HD\,181327, meanwhile, HST/STIS imaging has previously revealed a significant asymmetry in scattered-light images \citep{Stark_2014} that is not obvious in our $H$-band observations. This asymmetry might indicate a grain pile-up, and a multi-wavelength analysis could confirm the grain sizes captured in this pile-up. For HD\,121617, a candidate companion is seen exterior to the disk; follow-up observations will determine whether the detected object outside the disk is a bound companion and, if so, whether it might influence the disk.
\subsection{Non-detection of predicted disk-shaping giant planets}
Comparing our detection probability maps to the parameter space for planets that can be responsible for shaping the edges of debris disks by \citet{Pearce_2022}, it becomes clear that, in most cases, the limits fall short and we are only able to rule out parts of the parameter space and miss out on the minimum mass, maximum semimajor-axis prediction. Relatively face-on systems provide the best constraints, since a planet can be detected at any orbital phase and cannot hide behind its host star. For both HD\,121617 and HD\,181327, which have rather face-on disks, our observations reach the minimum predicted mass internal to the disk at all position angles. Thus, to hide from detection, a giant planet would have to be misaligned with the disk: this is unlikely, and furthermore, such an inclined configuration would be expected to excite vertical structure in the disk, producing warps or an increased scale height \citep{Mouillet_1997, Pearce_2014}. In both systems, the predicted minimum-mass planet would have had time to perform this excitation through secular interactions, calculated via \texttt{GetSecularAndDiffusionTimes}\footnote{\href{https://github.com/TimDPearce/GetSecularAndDiffusionTimes}{github.com/TimDPearce/GetSecularAndDiffusionTimes}} \citep{Pearce_2014}. The absence of clear vertical perturbations may therefore argue against strong misalignment, although a sufficiently massive disk could in principle resist this dynamical heating \citep{Poblete_2023, Sefilian_2025}. At present, the lack of robust constraints on debris disk masses limits how strongly this argument can be applied \citep{Krivov_2021, Pearce_2025}. 

On the other hand, these non-detections, together with others such as the deep JWST imaging of Fomalhaut \citep{Gaspar_2023, Ygouf_2024}, HD\,53143 (Matthews et al. in prep.), or HD\,92945, HD\,107146, and HD\,206893 \citep{Bendahan-West_2026}, as well as JWST constraints based on modeling disk images for systems such as $\epsilon$~Eri \citep{Wolff_2025} and Vega \citep{Su_2024}, increasingly challenge the idea that single massive planets dominate the inner edges of all debris disks. In several of these systems, observations rule out giant companions over a wide range of separations. This growing body of evidence suggests alternative explanations might be at play. For example, debris disks could be sculpted by multiple, lower-mass planets, rather than a single large planet, if those lower-mass planets lie below detection limits \citep{Pearce_2022}. Alternatively, planet migration could be important; these disks could still have been sculpted by a single planet, but that planet has since migrated further inward \citep{Friebe_2022}. Disk self-gravity could also resist the influence of a planet \citep{Marshall_2026}, or it might be that many debris disks are simply unrelated to planets \citep{ImazBlanco_2023, Pearce_2024}. Further work and observations are required to explain this.
\subsection[Limitations of 4S for VLT/SPHERE]{Limitations of \fours for VLT/SPHERE}\label{sec: Limitations of 4S}
We performed additional injection-recovery tests for the inner $\approx$\,\SI{0.5}{\arcsec} with \fours as the PSF subtraction method to fully assess its performance and compare it to PCA. As \fours is computationally rather expensive, we limited the number of datasets to test to HD\,1461 (observation dates 2022 October 14 and 2025 November 24) and HD\,59659 (observation dates 2023 March 30 and 2017 December 01) due to the companion candidates in both datasets, as well as HD\,121617 (2023 March 30) and HD\,207129 (2023 October 04), as these represent the range of the \ispysphere datasets in terms of single and total exposure time, as well as field rotation and seeing. We show the results in \autoref{fig: 4S performance}. \fours achieves a contrast similar to PCA across all datasets we tested. The 2022 HD\,1461 dataset is the only one in our tested sample where we observe that \fours can reach deeper limits compared to PCA for separations of 200--\SI{500}{\mas}. Interestingly, this is exactly the range in which we detect the false-positive point-like feature in this dataset.

We therefore conclude that \irdis does not benefit from the higher-parametrized model offered by \fours. This can be attributed to the smaller number of temporal frames obtained by \sphere compared to \naco. Additionally, \sphere benefits from a much better AO system, which can achieve higher Strehl ratios and thus allows for simpler PSF models. Consequently, we find that in the case of HD\,1461 and, most likely, also HD\,59659, \fours does not reveal true companions; instead, non-Gaussian noise leads to low-\snr false-positive signals. 
\subsection{Non-detection of ISPY--NaCo candidates}
We robustly confirm the two companion candidates in the HD\,101412 system, while all other \ispynaco candidates (HD\,377, HD\,1461, HD\,121617, HD\,131511, $\epsilon$~Sgr, HD\,179218, and HD\,207129) are ruled out as genuine companions. This result is fully consistent with the giant-planet occurrence rates reported by \citet{Launhardt_2020}, reinforcing the broader statistical picture but raising questions about the origins of the signals:

While several \ispynaco candidates showed only marginal evidence and can therefore be readily dismissed as artifacts, others initially appeared robust and significant. The most compelling signal among these original \ispynaco candidates is in the HD\,179218 system. The close-in signal from the \ispynaco image was later confirmed but attributable to the emission of the circumstellar protoplanetary disk and not to a companion \citep{Cugno_2023}. As most of these original \ispynaco candidates are not re-detected in our survey, they are not only ruled out as genuine companions but also as background objects. This, together with non-detections of these candidates when reanalyzing the original \ispynaco data using \fours (Bonse et al. in prep.), provides a clear and self-consistent explanation: We attribute these candidates as false-positive detections to non-Gaussian residual noise in the \ispynaco PCA reductions. This consequently highlights a major difficulty in direct-imaging planet searches and also demonstrates that applying \fours to \naco observations yields more Gaussian-distributed noise. This then leads to a significant reduction in the false-positive rate and to improved robustness compared to PCA for detections with \naco.
\section{Conclusions}\label{sec: Conclusions}
In this paper, we have presented \sphere images of a sample of 13 disk-hosting stars observed as part of the ISPY survey. We detected two known and two new stellar-mass companions, three disks, and one planet candidate. Using \fours, we revealed two close-in point-like features mimicking planet candidates, which we dismiss as non-Gaussian residual noise. We also compared our achieved detection limits to theoretical predictions of disk-sculpting planets and assessed the performance of the new PSF subtraction method \fours for \irdis. Our main findings are summarized as follows:
\begin{enumerate}
    \item We identified two late-K to early-M-type stellar dwarf companions to HD\,101412, and we confirm that these are gravitationally bound and share CPM with the host star. Additionally, we found hints toward orbital motion in the astrometric measurements of companion C. We performed atmospheric characterization, showing that both companions are inflated compared to main-sequence stars. For the outer companion C, we measured a slightly ($2\sigma$) enhanced $L'$-band flux, hinting toward excess emission of an accretion disk. Future work should monitor the companions to determine the true orbital motion of this interesting hierarchy and to further characterize these young objects.
    \item We imaged two known stellar companions, HD\,131511~B (which has only previously been observed via RV and astrometric observations of the host star) and $\epsilon$~Sgr~B, at a separation of \SI{62.88(10)}{\mas} and \SI{2.9555(59)}{\arcsec}, respectively. We found a significant orbital motion of $\epsilon$~Sgr~B; however, as the host star is too bright to be included in \textit{Gaia} and no public RVs are available, we could only constrain the full orbit when including mass priors. Hence, no determination of the dynamical masses was possible. 
    \item We imaged the debris disk around HD\,121617. In our scattered-light images, this disk is brightest to the southwest; the bright location is offset from the bright ALMA clump imaged in \citet{Marino_2026_121617}. Additionally, we detected an astrophysical point source located to the southwest. Determining whether or not this is a giant bound $4.84_{-0.26}^{+0.22}$\si{\Mjup} planet is left open to future work.
    \item We imaged the debris disk in the HD\,181327 system and the protoplanetary disk in the HD\,179218 system. 
    \item In the \fours residuals, we recovered the companions in the HD\,101412 system as well as the secondary around HD\,131511, showing that \fours works on \sphere data. Additionally, we identified two close-in point-like features. In the HD\,1461 system, we identified the signal as non-Gaussian noise through follow-up observations. In the case of HD\,59659, we found that the candidate, together with the RVs, yields a stable orbital configuration that is, however, strongly eccentric. The current data are insufficient to fully determine the true nature of HD\,59659, and follow-up observations with improved detection sensitivity are required.
    \item When comparing the detection limits with the parameter space in which a single giant planet would shape the inner edge of a debris disk, we observed that in many systems we can rule out parts of this parameter space, but we cannot fully rule out any single-planet scenario. However, in the cases of HD\,121617 and HD\,181327, we excluded the single giant-planet sculpting scenario proposed in \citet{Pearce_2022}. Together with other recent planet searches that did not find sources consistent with the single giant-planet scenario (e.g., \citealt{Gaspar_2023, Su_2024, Ygouf_2024, Bendahan-West_2026, Wolff_2025}, Matthews et al. in prep.), a trend is emerging that suggests many debris disks are not sculpted by a single giant planet.
    \item In the case of \irdis, we found that \fours does not reach significantly improved contrast limits compared to PCA. Using \fours we found two new point-like features close to HD\,1461 and HD\,59659 that show signs of being tentative companion candidates. We can rule out the companion candidate of HD\,1461 as non-real; however, determining the true nature of the second one is left to future work. Until then, we find that close point-like features revealed by \fours are likely caused by non-Gaussian residual noise. 
\end{enumerate}
While we were not able to re-detect and thereby confirm any of the unclear original \ispynaco planet candidates, ruling these original candidates out as false-positive signals refines the overall \ispynaco statistics. By having analyzed all observed systems in detail, we maximize the value of our observations by almost completely ruling out a giant-planet scenario in two systems and contributing to the characterization of two known stellar- and substellar-mass companions. We find that the machine-learning based data reduction technique \fours achieves no significant improvement for VLT/SPHERE. Lastly, we identified new tentative planet candidates, which will hopefully be confirmed in future work.
\section*{Data availability}\label{data availability}
We make all \ispysphere pre-processed data cubes, reduced images, and computed contrast curves available online at \href{https://doi.org/10.5281/zenodo.22101966}{doi.org/10.5281/zenodo.22101966}.
\begin{acknowledgements}
We thank the anonymous referee for the constructive reports, which significantly improved the quality of the manuscript. Based on observations made with ESO Telescopes at the Paranal Observatory under program IDs 1101.C-0092(J) and 116.297U.001. TDP is supported by a UKRI Stephen Hawking Fellowship and a Warwick Prize Fellowship, the latter made possible by a generous philanthropic donation. Part of this work has been carried out within the framework of the NCCR PlanetS supported by the Swiss National Science Foundation under grant 51NF40\_205606. Computations were performed on HPC systems at the Max Planck Computing and Data Facility. We acknowledge the use of AI-assisted tools during the preparation of this manuscript. Specifically, Claude (Anthropic), ChatGPT (OpenAI), and GitHub Copilot (Microsoft) were utilized to support aspects of the writing and coding process. We take full responsibility for the content, accuracy, and integrity of this work. This work has made use of the following software: \texttt{astropy} \citep{AstropyCollaboration_2013,AstropyCollaboration_2022}, a community-developed core Python package for astronomy; \texttt{matplotlib} \citep{Hunter_2007}; \texttt{numpy} \citep{Harris_2020}; \texttt{pandas} \citep{ThepandasdevelopmentTeam_2025}; \texttt{scipy} \citep{2020SciPy-NMeth}; \texttt{Uncertainties}: a Python package for calculations with uncertainties (Eric O. Lebigot).
\end{acknowledgements}
\bibliography{bibliography}{}

@article{Absil_2013,
 adsurl = {https://ui.adsabs.harvard.edu/abs/2013A&A...559L..12A},
 archiveprefix = {arXiv},
 author = {{Absil}, O. and {Milli}, J. and {Mawet}, D. and {Lagrange}, A.-M. and {Girard}, J. and {Chauvin}, G. and {Boccaletti}, A. and {Delacroix}, C. and {Surdej}, J.},
 doi = {10.1051/0004-6361/201322748},
 eid = {L12},
 eprint = {1311.4298},
 journal = {\aap},
 month = {November},
 pages = {L12},
 primaryclass = {astro-ph.SR},
 title = {{Searching for companions down to 2 AU from {\ensuremath{\beta}} Pictoris using the L'-band AGPM coronagraph on VLT/NACO}},
 volume = {559},
 year = {2013}
}

@article{Allard_2012,
 adsurl = {https://ui.adsabs.harvard.edu/abs/2012RSPTA.370.2765A},
 archiveprefix = {arXiv},
 author = {{Allard}, F. and {Homeier}, D. and {Freytag}, B.},
 doi = {10.1098/rsta.2011.0269},
 eprint = {1112.3591},
 journal = {PTRSLSA},
 month = {June},
 number = {1968},
 pages = {2765-2777},
 primaryclass = {astro-ph.SR},
 title = {{Models of very-low-mass stars, brown dwarfs and exoplanets}},
 volume = {370},
 year = {2012}
}

@article{Amara_2012,
 adsurl = {https://ui.adsabs.harvard.edu/abs/2012MNRAS.427..948A},
 archiveprefix = {arXiv},
 author = {{Amara}, Adam and {Quanz}, Sascha P.},
 doi = {10.1111/j.1365-2966.2012.21918.x},
 eprint = {1207.6637},
 journal = {\mnras},
 month = {December},
 number = {2},
 pages = {948-955},
 primaryclass = {astro-ph.IM},
 title = {{PYNPOINT: an image processing package for finding exoplanets}},
 volume = {427},
 year = {2012}
}

@article{Aumann_1985,
 adsurl = {https://ui.adsabs.harvard.edu/abs/1985PASP...97..885A},
 author = {{Aumann}, H.~H.},
 doi = {10.1086/131620},
 journal = {\pasp},
 month = {October},
 pages = {885-891},
 title = {{IRAS observations of matter around nearby stars.}},
 volume = {97},
 year = {1985}
}

@article{Bailer-Jones_2018,
 adsurl = {https://ui.adsabs.harvard.edu/abs/2018AJ....156...58B},
 archiveprefix = {arXiv},
 author = {{Bailer-Jones}, C.~A.~L. and {Rybizki}, J. and {Fouesneau}, M. and {Mantelet}, G. and {Andrae}, R.},
 doi = {10.3847/1538-3881/aacb21},
 eid = {58},
 eprint = {1804.10121},
 journal = {\aj},
 month = {August},
 number = {2},
 pages = {58},
 primaryclass = {astro-ph.SR},
 title = {{Estimating Distance from Parallaxes. IV. Distances to 1.33 Billion Stars in Gaia Data Release 2}},
 volume = {156},
 year = {2018}
}

@article{Beichman_2025,
 adsurl = {https://ui.adsabs.harvard.edu/abs/2025AJ....169...17B},
 archiveprefix = {arXiv},
 author = {{Beichman}, Charles and {Bryden}, Geoffrey and {Llop-Sayson}, Jorge and {Ygouf}, Marie and {Greenbaum}, Alexandra and {Leisenring}, Jarron and {Gaspar}, Andras and {Krist}, John and {Rieke}, George and {Wolff}, Schuyler and {Su}, Kate and {Hodapp}, Klaus and {Meyer}, Michael and {Kelly}, Doug and {Boyer}, Martha and {Johnstone}, Doug and {Horner}, Scott and {Rieke}, Marcia},
 doi = {10.3847/1538-3881/ad890d},
 eid = {17},
 eprint = {2410.16551},
 journal = {\aj},
 month = {January},
 number = {1},
 pages = {17},
 primaryclass = {astro-ph.EP},
 title = {{Searching for Planets Orbiting Vega with the James Webb Space Telescope}},
 volume = {169},
 year = {2025}
}

@article{Bell_2015,
 adsurl = {https://ui.adsabs.harvard.edu/abs/2015MNRAS.454..593B},
 archiveprefix = {arXiv},
 author = {{Bell}, Cameron P.~M. and {Mamajek}, Eric E. and {Naylor}, Tim},
 doi = {10.1093/mnras/stv1981},
 eprint = {1508.05955},
 journal = {\mnras},
 month = {November},
 number = {1},
 pages = {593-614},
 primaryclass = {astro-ph.SR},
 title = {{A self-consistent, absolute isochronal age scale for young moving groups in the solar neighbourhood}},
 volume = {454},
 year = {2015}
}

@article{Bendahan-West_2026,
 adsurl = {https://ui.adsabs.harvard.edu/abs/2026MNRAS.546f2255B},
 archiveprefix = {arXiv},
 author = {{Bendahan-West}, R. and {Marino}, S. and {Carter}, A.~L. and {Squicciarini}, V. and {James}, A.~D. and {Sefilian}, A.~A. and {Pearce}, T.~D. and {Friebe}, M.~F. and {Lazzoni}, C. and {Lakeland}, B. and {Ray}, S. and {Wyatt}, M.~C. and {Matr{\`a}}, L. and {Milli}, J. and {Faramaz}, V.~C. and {Henning}, Th and {Hinkley}, S. and {Kennedy}, G.~M. and {Mesa}, D. and {Zurlo}, A.},
 doi = {10.1093/mnras/staf2255},
 eid = {staf2255},
 eprint = {2601.02156},
 journal = {\mnras},
 month = {March},
 number = {3},
 pages = {staf2255},
 primaryclass = {astro-ph.EP},
 title = {{JWST/MIRI coronagraphic search for planets in systems with gapped exoKuiper belts and proper-motion anomalies}},
 volume = {546},
 year = {2026}
}

@article{Beuzit_2019,
 adsurl = {https://ui.adsabs.harvard.edu/abs/2019A&A...631A.155B},
 archiveprefix = {arXiv},
 author = {{Beuzit}, J.-L. and {Vigan}, A. and {Mouillet}, D. and {Dohlen}, K. and {Gratton}, R. and {Boccaletti}, A. and {Sauvage}, J.-F. and {Schmid}, H.~M. and {Langlois}, M. and {Petit}, C. and {Baruffolo}, A. and {Feldt}, M. and {Milli}, J. and {Wahhaj}, Z. and {Abe}, L. and {Anselmi}, U. and {Antichi}, J. and {Barette}, R. and {Baudrand}, J. and {Baudoz}, P. and {Bazzon}, A. and {Bernardi}, P. and {Blanchard}, P. and {Brast}, R. and {Bruno}, P. and {Buey}, T. and {Carbillet}, M. and {Carle}, M. and {Cascone}, E. and {Chapron}, F. and {Charton}, J. and {Chauvin}, G. and {Claudi}, R. and {Costille}, A. and {De Caprio}, V. and {de Boer}, J. and {Delboulb{\'e}}, A. and {Desidera}, S. and {Dominik}, C. and {Downing}, M. and {Dupuis}, O. and {Fabron}, C. and {Fantinel}, D. and {Farisato}, G. and {Feautrier}, P. and {Fedrigo}, E. and {Fusco}, T. and {Gigan}, P. and {Ginski}, C. and {Girard}, J. and {Giro}, E. and {Gisler}, D. and {Gluck}, L. and {Gry}, C. and {Henning}, T. and {Hubin}, N. and {Hugot}, E. and {Incorvaia}, S. and {Jaquet}, M. and {Kasper}, M. and {Lagadec}, E. and {Lagrange}, A.-M. and {Le Coroller}, H. and {Le Mignant}, D. and {Le Ruyet}, B. and {Lessio}, G. and {Lizon}, J.-L. and {Llored}, M. and {Lundin}, L. and {Madec}, F. and {Magnard}, Y. and {Marteaud}, M. and {Martinez}, P. and {Maurel}, D. and {M{\'e}nard}, F. and {Mesa}, D. and {M{\"o}ller-Nilsson}, O. and {Moulin}, T. and {Moutou}, C. and {Orign{\'e}}, A. and {Parisot}, J. and {Pavlov}, A. and {Perret}, D. and {Pragt}, J. and {Puget}, P. and {Rabou}, P. and {Ramos}, J. and {Reess}, J.-M. and {Rigal}, F. and {Rochat}, S. and {Roelfsema}, R. and {Rousset}, G. and {Roux}, A. and {Saisse}, M. and {Salasnich}, B. and {Santambrogio}, E. and {Scuderi}, S. and {Segransan}, D. and {Sevin}, A. and {Siebenmorgen}, R. and {Soenke}, C. and {Stadler}, E. and {Suarez}, M. and {Tiph{\`e}ne}, D. and {Turatto}, M. and {Udry}, S. and {Vakili}, F. and {Waters}, L.~B.~F.~M. and {Weber}, L. and {Wildi}, F. and {Zins}, G. and {Zurlo}, A.},
 doi = {10.1051/0004-6361/201935251},
 eid = {A155},
 eprint = {1902.04080},
 journal = {\aap},
 month = {November},
 pages = {A155},
 primaryclass = {astro-ph.IM},
 title = {{SPHERE: the exoplanet imager for the Very Large Telescope}},
 volume = {631},
 year = {2019}
}

@article{Bock_2026,
 adsurl = {https://ui.adsabs.harvard.edu/abs/2026ApJ...999..139B},
 archiveprefix = {arXiv},
 author = {{Bock}, James J. and {Aboobaker}, Asad M. and {Adamo}, Joseph and {Akeson}, Rachel and {Alred}, John M. and {Alibay}, Farah and {Ashby}, Matthew L.~N. and {Bach}, Yoonsoo P. and {Bleem}, Lindsey E. and {Bolton}, Douglas and {Braun}, David F. and {Bruton}, Sean and {Bryan}, Sean A. and {Chang}, Tzu-Ching and {Chen}, Shuang-Shuang and {Cheng}, Yun-Ting and {Cheshire}, IV, James R. and {Chiang}, Yi-Kuan and {Choppin de Janvry}, Jean and {Condon}, Samuel and {Cook}, Walter R. and {Cooray}, Asantha and {Crill}, Brendan P. and {Cukierman}, Ari J. and {Dor{\'e}}, Olivier and {Dowell}, C. Darren and {Dubois-Felsmann}, Gregory P. and {Eifler}, Tim and {Everett}, Spencer and {Fabinsky}, Beth E. and {Faisst}, Andreas L. and {Fanson}, James L. and {Farrington}, Allen H. and {Fatahi}, Tamim and {Fazar}, Candice M. and {Feder}, Richard M. and {Frater}, Eric H. and {Grasshorn Gebhardt}, Henry S. and {Giri}, Utkarsh and {Goldina}, Tatiana and {Gorjian}, Varoujan and {Habib}, Salman and {Hart}, William G. and {Heinrich}, Chen and {Hora}, Joseph L. and {Huai}, Zhaoyu and {Hui}, Howard and {Jo}, Young-Soo and {Jeong}, Woong-Seob and {Kang}, Jae Hwan and {Kang}, Miju and {Kecman}, Branislav and {Kim}, Chul-Hwan and {Kim}, Jaeyeong and {Kim}, Minjin and {Kim}, Young-Jun and {Kim}, Yongjung and {Kirkpatrick}, J. Davy and {Kobayashi}, Yosuke and {Korngut}, Phil M. and {Krause}, Elisabeth and {Lee}, Bomee and {Lee}, Ho-Gyu and {Lee}, Jae-Joon and {Lee}, Jeong-Eun and {Lisse}, Carey M. and {Mariani}, Giacomo and {Masters}, Daniel C. and {Mauskopf}, Philip D. and {Melnick}, Gary J. and {Minasyan}, Mary H. and {Mirocha}, Jordan and {Miyasaka}, Hiromasa and {Moore}, Anne and {Moore}, Bradley D. and {Murgia}, Giulia and {Naylor}, Bret J. and {Nelson}, Christina and {Nguyen}, Chi H. and {Nguyen}, Hien T. and {Noh}, Jinyoung K. and {Padin}, Stephen and {Paladini}, Roberta and {Park}, Sung-Joon and {Penanen}, Konstantin I. and {Putnam}, Dustin S. and {Pyo}, Jeonghyun and {Ramachandra}, Nesar and {Ramanathan}, Keshav and {Rustamkulov}, Zafar and {Reiley}, Daniel J. and {Rice}, Eric B. and {Rocca}, Jennifer M. and {Seok}, Ji Yeon and {Smith}, Roger and {Stober}, Jeremy and {Susca}, Sara and {Teplitz}, Harry I. and {Thelen}, Michael P. and {Tolls}, Volker and {Torrini}, Gabriela and {Trangsrud}, Amy R. and {Unwin}, Stephen and {Velicheti}, Phani and {Wang}, Pao-Yu and {Wen}, Robin Y. and {Werner}, Michael W. and {Williams}, Abby E. and {Williamson}, Ross and {Wincentsen}, James and {Windhorst}, Rogier A. and {Yang}, Soung-Chul and {Yang}, Yujin and {Zemcov}, Michael},
 doi = {10.3847/1538-4357/ae2be2},
 eid = {139},
 eprint = {2511.02985},
 journal = {\apj},
 month = {March},
 number = {1},
 pages = {139},
 primaryclass = {astro-ph.IM},
 title = {{The SPHEREx Satellite Mission}},
 volume = {999},
 year = {2026}
}

@article{Bonavita_2020,
 adsurl = {https://ui.adsabs.harvard.edu/abs/2020ascl.soft10008B},
 archiveprefix = {ascl},
 author = {{Bonavita}, Mariangela},
 eid = {ascl:2010.008},
 eprint = {2010.008},
 howpublished = {Astrophysics Source Code Library, record ascl:2010.008},
 month = {October},
 title = {{Exo-DMC: Exoplanet Detection Map Calculator}},
 year = {2020}
}

@article{Bonse_2023,
author = {{Bonse}, Markus J. and
{Garvin}, Emily O. and
{Gebhard}, Timothy D. and
{Dannert}, Felix A. and
{Cantalloube}, Faustine and
{Cugno}, Gabriele and
{Absil}, Olivier and
{Hayoz}, Jean and
{Milli}, Julien and
{Kasper}, Markus and
{Quanz}, Sascha P.},
title = "{Comparing Apples with Apples:
Robust Detection Limits for Exoplanet High-contrast Imaging
in the Presence of Non-Gaussian Noise}",
journal = {\aj},
year = 2023,
month = aug,
volume = {166},
number = {2},
eid = {71},
pages = {71},
doi = {10.3847/1538-3881/acc93c},
archivePrefix = {arXiv},
eprint = {2303.12030},
primaryClass = {astro-ph.EP},
adsurl = {https://ui.adsabs.harvard.edu/abs/2023AJ....166...71B}}

@article{Bonse_2025,
 adsurl = {https://ui.adsabs.harvard.edu/abs/2025AJ....169..194B},
 archiveprefix = {arXiv},
 author = {{Bonse}, Markus J. and {Gebhard}, Timothy D. and {Dannert}, Felix A. and {Absil}, Olivier and {Cantalloube}, Faustine and {Christiaens}, Valentin and {Cugno}, Gabriele and {Garvin}, Emily O. and {Hayoz}, Jean and {Kasper}, Markus and {Matthews}, Elisabeth and {Sch{\"o}lkopf}, Bernhard and {Quanz}, Sascha P.},
 doi = {10.3847/1538-3881/adab79},
 eid = {194},
 eprint = {2406.01809},
 journal = {\aj},
 month = {April},
 number = {4},
 pages = {194},
 primaryclass = {astro-ph.EP},
 title = {{Use the 4S (Signal-Safe Speckle Subtraction): Explainable Machine Learning Reveals the Giant Exoplanet AF Lep b in High-contrast Imaging Data from 2011}},
 volume = {169},
 year = {2025}
}

@article{Booth_2017,
 adsurl = {https://ui.adsabs.harvard.edu/abs/2017MNRAS.469.3200B},
 archiveprefix = {arXiv},
 author = {{Booth}, Mark and {Dent}, William R.~F. and {Jord{\'a}n}, Andr{\'e}s and {Lestrade}, Jean-Fran{\c{c}}ois and {Hales}, Antonio S. and {Wyatt}, Mark C. and {Casassus}, Simon and {Ertel}, Steve and {Greaves}, Jane S. and {Kennedy}, Grant M. and et al.},
 doi = {10.1093/mnras/stx1072},
 eprint = {1705.01560},
 journal = {\mnras},
 month = {August},
 number = {3},
 pages = {3200-3212},
 primaryclass = {astro-ph.EP},
 title = {{The Northern arc of {\ensuremath{\in}} Eridani's Debris Ring as seen by ALMA}},
 volume = {469},
 year = {2017}
}

@article{Booth_2023,
 adsurl = {https://ui.adsabs.harvard.edu/abs/2023MNRAS.521.6180B},
 archiveprefix = {arXiv},
 author = {{Booth}, Mark and {Pearce}, Tim D. and {Krivov}, Alexander V. and {Wyatt}, Mark C. and {Dent}, William R.~F. and {Hales}, Antonio S. and {Lestrade}, Jean-Fran{\c{c}}ois and {Cruz-S{\'a}enz de Miera}, Fernando and {Faramaz}, Virginie C. and {L{\"o}hne}, Torsten and et al.},
 doi = {10.1093/mnras/stad938},
 eprint = {2303.13584},
 journal = {\mnras},
 month = {June},
 number = {4},
 pages = {6180-6194},
 primaryclass = {astro-ph.EP},
 title = {{The clumpy structure of ϵ Eridani's debris disc revisited by ALMA}},
 volume = {521},
 year = {2023}
}

@article{Bowler_2016,
 adsurl = {https://ui.adsabs.harvard.edu/abs/2016PASP..128j2001B},
 archiveprefix = {arXiv},
 author = {{Bowler}, Brendan P.},
 doi = {10.1088/1538-3873/128/968/102001},
 eprint = {1605.02731},
 journal = {\pasp},
 month = {October},
 number = {968},
 pages = {102001},
 primaryclass = {astro-ph.EP},
 title = {{Imaging Extrasolar Giant Planets}},
 volume = {128},
 year = {2016}
}

@article{Brandt_2021_b,
 adsurl = {https://ui.adsabs.harvard.edu/abs/2021AJ....162..186B},
 archiveprefix = {arXiv},
 author = {{Brandt}, Timothy D. and {Dupuy}, Trent J. and {Li}, Yiting and {Brandt}, G. Mirek and {Zeng}, Yunlin and {Michalik}, Daniel and {Bardalez Gagliuffi}, Daniella C. and {Raposo-Pulido}, Virginia},
 doi = {10.3847/1538-3881/ac042e},
 eid = {186},
 eprint = {2105.11671},
 journal = {\aj},
 month = {November},
 number = {5},
 pages = {186},
 primaryclass = {astro-ph.IM},
 title = {{orvara: An Efficient Code to Fit Orbits Using Radial Velocity, Absolute, and/or Relative Astrometry}},
 volume = {162},
 year = {2021}
}

@article{Brennan_2026,
 adsurl = {https://ui.adsabs.harvard.edu/abs/2026A&A...705A.201B},
 archiveprefix = {arXiv},
 author = {{Brennan}, A. and {Matr{\`a}}, L. and {Mac Manamon}, S. and {Marino}, S. and {Cataldi}, G. and {Hughes}, A.~M. and {Weber}, P. and {Han}, Y. and {Marshall}, J.~P. and {Zawadzki}, B. and {Luppe}, P. and {Sefilian}, A.~A. and {Mo{\'o}r}, A. and {MacGregor}, M.~A. and {Lovell}, J.~B. and {K{\'o}sp{\'a}l}, A. and {Bonduelle}, M. and {Mansell}, E. and {Wyatt}, M.~C. and {Pearce}, T.~D. and {Carpenter}, J.~M. and {Wilner}, D.~J. and {del Burgo}, C. and {P{\'e}rez}, S. and {Henning}, T. and {Milli}, J. and {Chiang}, E.},
 doi = {10.1051/0004-6361/202556566},
 eid = {A201},
 eprint = {2601.11824},
 journal = {\aap},
 month = {January},
 pages = {A201},
 primaryclass = {astro-ph.EP},
 title = {{The ALMA survey to Resolve exoKuiper belt Substructures (ARKS): VII. Optically thick gas with broad CO gaussian local line profiles in the HD 121617 disc}},
 volume = {705},
 year = {2026}
}

@article{Campbell_1988,
 adsurl = {https://ui.adsabs.harvard.edu/abs/1988ApJ...331..902C},
 author = {{Campbell}, Bruce and {Walker}, G.~A.~H. and {Yang}, S.},
 doi = {10.1086/166608},
 journal = {\apj},
 month = {August},
 pages = {902},
 title = {{A Search for Substellar Companions to Solar-type Stars}},
 volume = {331},
 year = {1988}
}

@article{Casagrande_2011,
 adsurl = {https://ui.adsabs.harvard.edu/abs/2011A&A...530A.138C},
 archiveprefix = {arXiv},
 author = {{Casagrande}, L. and {Sch{\"o}nrich}, R. and {Asplund}, M. and {Cassisi}, S. and {Ram{\'\i}rez}, I. and {Mel{\'e}ndez}, J. and {Bensby}, T. and {Feltzing}, S.},
 doi = {10.1051/0004-6361/201016276},
 eid = {A138},
 eprint = {1103.4651},
 journal = {\aap},
 month = {June},
 pages = {A138},
 primaryclass = {astro-ph.GA},
 title = {{New constraints on the chemical evolution of the solar neighbourhood and Galactic disc(s). Improved astrophysical parameters for the Geneva-Copenhagen Survey}},
 volume = {530},
 year = {2011}
}

@article{Ceva_2025,
 adsurl = {https://ui.adsabs.harvard.edu/abs/2025A&A...701A..78C},
 archiveprefix = {arXiv},
 author = {{Ceva}, W. and {Matthews}, E.~C. and {Rickman}, E.~L. and {S{\'e}gransan}, D. and {Vigan}, A. and {Bowler}, B.~P. and {Forveille}, T. and {Franson}, K. and {Hagelberg}, J. and {Udry}, S.},
 doi = {10.1051/0004-6361/202555310},
 eid = {A78},
 eprint = {2507.08654},
 journal = {\aap},
 month = {September},
 pages = {A78},
 primaryclass = {astro-ph.EP},
 title = {{Spectral analysis of two directly imaged benchmark L dwarf companions at the stellar-substellar boundary}},
 volume = {701},
 year = {2025}
}

@article{Chauvin_2012,
 adsurl = {https://ui.adsabs.harvard.edu/abs/2012A&A...542A..41C},
 archiveprefix = {arXiv},
 author = {{Chauvin}, G. and {Lagrange}, A.-M. and {Beust}, H. and {Bonnefoy}, M. and {Boccaletti}, A. and {Apai}, D. and {Allard}, F. and {Ehrenreich}, D. and {Girard}, J.~H.~V. and {Mouillet}, D. and {Rouan}, D.},
 doi = {10.1051/0004-6361/201118346},
 eid = {A41},
 eprint = {1202.2655},
 journal = {\aap},
 month = {June},
 pages = {A41},
 primaryclass = {astro-ph.EP},
 title = {{Orbital characterization of the {\ensuremath{\beta}} Pictoris b giant planet}},
 volume = {542},
 year = {2012}
}

@article{Chen_2014,
 adsurl = {https://ui.adsabs.harvard.edu/abs/2014ApJS..211...25C},
 author = {{Chen}, Christine H. and {Mittal}, Tushar and {Kuchner}, Marc and {Forrest}, William J. and {Lisse}, Carey M. and {Manoj}, P. and {Sargent}, Benjamin A. and {Watson}, Dan M.},
 doi = {10.1088/0067-0049/211/2/25},
 eid = {25},
 journal = {\apjs},
 month = {April},
 number = {2},
 pages = {25},
 title = {{The Spitzer Infrared Spectrograph Debris Disk Catalog. I. Continuum Analysis of Unresolved Targets}},
 volume = {211},
 year = {2014}
}

@article{Choquet_2018,
 adsurl = {https://ui.adsabs.harvard.edu/abs/2018ApJ...854...53C},
 archiveprefix = {arXiv},
 author = {{Choquet}, {\'E}. and {Bryden}, G. and {Perrin}, M.~D. and {Soummer}, R. and {Augereau}, J.-C. and {Chen}, C.~H. and {Debes}, J.~H. and {Gofas-Salas}, E. and {Hagan}, J.~B. and {Hines}, D.~C. and {Mawet}, D. and {Morales}, F. and {Pueyo}, L. and {Rajan}, A. and {Ren}, B. and {Schneider}, G. and {Stark}, C.~C. and {Wolff}, S.},
 doi = {10.3847/1538-4357/aaa892},
 eid = {53},
 eprint = {1801.05424},
 journal = {\apj},
 month = {February},
 number = {1},
 pages = {53},
 primaryclass = {astro-ph.EP},
 title = {{HD 104860 and HD 192758: Two Debris Disks Newly Imaged in Scattered Light with the Hubble Space Telescope}},
 volume = {854},
 year = {2018}
}

@inproceedings{Claudi_2008,
 adsurl = {https://ui.adsabs.harvard.edu/abs/2008SPIE.7014E..3EC},
 author = {{Claudi}, R.~U. and {Turatto}, M. and {Gratton}, R.~G. and {Antichi}, J. and {Bonavita}, M. and {Bruno}, P. and {Cascone}, E. and {De Caprio}, V. and {Desidera}, S. and {Giro}, E. and {Mesa}, D. and {Scuderi}, S. and {Dohlen}, K. and {Beuzit}, J.~L. and {Puget}, P.},
 booktitle = {Ground-based and Airborne Instrumentation for Astronomy II},
 doi = {10.1117/12.788366},
 editor = {{McLean}, Ian S. and {Casali}, Mark M.},
 eid = {70143E},
 month = {July},
 pages = {70143E},
 series = {Society of Photo-Optical Instrumentation Engineers (SPIE) Conference Series},
 title = {{SPHERE IFS: the spectro differential imager of the VLT for exoplanets search}},
 volume = {7014},
 year = {2008}
}

@article{Cowley_2010,
 adsurl = {https://ui.adsabs.harvard.edu/abs/2010A&A...523A..65C},
 archiveprefix = {arXiv},
 author = {{Cowley}, C.~R. and {Hubrig}, S. and {Gonz{\'a}lez}, J.~F. and {Savanov}, I.},
 doi = {10.1051/0004-6361/201015361},
 eid = {A65},
 eprint = {1008.1601},
 journal = {\aap},
 month = {November},
 pages = {A65},
 primaryclass = {astro-ph.SR},
 title = {{Abundances in the Herbig Ae star HD 101412. Abundance anomalies; {\ensuremath{\lambda}} Boo-Vega characteristics?}},
 volume = {523},
 year = {2010}
}

@article{Crotts_2025,
 adsurl = {https://ui.adsabs.harvard.edu/abs/2025ApJ...987L..41C},
 archiveprefix = {arXiv},
 author = {{Crotts}, Katie A. and {Carter}, Aarynn L. and {Lawson}, Kellen and {Mang}, James and {Biller}, Beth and {Booth}, Mark and {Ferrer-Chavez}, Rodrigo and {Girard}, Julien H. and {Lagrange}, Anne-Marie and {Liu}, Michael C. and {Marino}, Sebastian and {Millar-Blanchaer}, Maxwell A. and {Skemer}, Andy and {Strampelli}, Giovanni M. and {Wang}, Jason and {Absil}, Olivier and {Balmer}, William O. and {Bendahan-West}, Rapha{\"e}l and {Bogat}, Ellis and {Bowens-Rubin}, Rachel and {Chauvin}, Ga{\"e}l and {Fontanive}, Cl{\'e}mence and {Franson}, Kyle and {Kammerer}, Jens and {Leisenring}, Jarron and {Morley}, Caroline V. and {Rebollido}, Isabel and {Skaf}, Nour and {Sutlieff}, Ben J. and {Bruinsma}, Evelyn L. and {Hinkley}, Sasha and {Hoch}, Kielan and {James}, Andrew D. and {Kane}, Rohan and {Mawet}, Dimitri and {Meyer}, Michael R. and {Palatnick}, Skyler and {Perrin}, Marshall D. and {Ray}, Shrishmoy and {Rickman}, Emily and {Sanghi}, Aniket and {Stephenson}, Klaus Subbotina},
 doi = {10.3847/2041-8213/ade798},
 eid = {L41},
 eprint = {2506.19932},
 journal = {\apjl},
 month = {July},
 number = {2},
 pages = {L41},
 primaryclass = {astro-ph.EP},
 title = {{Follow-up Exploration of the TWA 7 Planet{\textendash}Disk System with JWST NIRCam}},
 volume = {987},
 year = {2025}
}

@article{Cugno_2019,
 adsurl = {https://ui.adsabs.harvard.edu/abs/2019A&A...624A..29C},
 archiveprefix = {arXiv},
 author = {{Cugno}, G. and {Quanz}, S.~P. and {Launhardt}, R. and {Musso Barcucci}, A. and {Brems}, S.~S. and {Cheetham}, A. and {Godoy}, N. and {Kennedy}, G.~M. and {Henning}, T. and {M{\"u}ller}, A. and {Olofsson}, J. and {Pepe}, F. and {Quirrenbach}, A. and {Reffert}, S. and {Rickman}, E.~L. and {S{\'e}gransan}, D.},
 doi = {10.1051/0004-6361/201935142},
 eid = {A29},
 eprint = {1902.04092},
 journal = {\aap},
 month = {April},
 pages = {A29},
 primaryclass = {astro-ph.SR},
 title = {{ISPY - NaCo Imaging Survey for Planets around Young stars. A young companion candidate embedded in the R CrA cloud}},
 volume = {624},
 year = {2019}
}

@article{Cugno_2023,
 adsurl = {https://ui.adsabs.harvard.edu/abs/2023A&A...669A.145C},
 archiveprefix = {arXiv},
 author = {{Cugno}, G. and {Pearce}, T.~D. and {Launhardt}, R. and {Bonse}, M.~J. and {Ma}, J. and {Henning}, T. and {Quirrenbach}, A. and {S{\'e}gransan}, D. and {Matthews}, E.~C. and {Quanz}, S.~P. and {Kennedy}, G.~M. and {M{\"u}ller}, A. and {Reffert}, S. and {Rickman}, E.~L.},
 doi = {10.1051/0004-6361/202244891},
 eid = {A145},
 eprint = {2211.15434},
 journal = {\aap},
 month = {January},
 pages = {A145},
 primaryclass = {astro-ph.EP},
 title = {{ISPY: NACO Imaging Survey for Planets around Young stars. The demographics of forming planets embedded in protoplanetary disks}},
 volume = {669},
 year = {2023}
}

@article{Dent_2005,
 adsurl = {https://ui.adsabs.harvard.edu/abs/2005MNRAS.359..663D},
 archiveprefix = {arXiv},
 author = {{Dent}, W.~R.~F. and {Greaves}, J.~S. and {Coulson}, I.~M.},
 doi = {10.1111/j.1365-2966.2005.08938.x},
 eprint = {astro-ph/0502544},
 journal = {\mnras},
 month = {May},
 number = {2},
 pages = {663-676},
 primaryclass = {astro-ph},
 title = {{CO emission from discs around isolated HAeBe and Vega-excess stars}},
 volume = {359},
 year = {2005}
}

@article{DeRosa_2023,
 adsurl = {https://ui.adsabs.harvard.edu/abs/2023A&A...672A..94D},
 archiveprefix = {arXiv},
 author = {{De Rosa}, Robert J. and {Nielsen}, Eric L. and {Wahhaj}, Zahed and {Ruffio}, Jean-Baptiste and {Kalas}, Paul G. and {Peck}, Anne E. and {Hirsch}, Lea A. and {Roberson}, William},
 doi = {10.1051/0004-6361/202345877},
 eid = {A94},
 eprint = {2302.06332},
 journal = {\aap},
 month = {April},
 pages = {A94},
 primaryclass = {astro-ph.EP},
 title = {{Direct imaging discovery of a super-Jovian around the young Sun-like star AF Leporis}},
 volume = {672},
 year = {2023}
}

@article{Desgrange2022,
 adsurl = {https://ui.adsabs.harvard.edu/abs/2022A&A...664A.139D},
 archiveprefix = {arXiv},
 author = {{Desgrange}, C. and {Chauvin}, G. and {Christiaens}, V. and {Cantalloube}, F. and {Lefranc}, L.-X. and {Le Coroller}, H. and {Rubini}, P. and {Otten}, G.~P.~P.~L. and {Beust}, H. and {Bonavita}, M. and {Delorme}, P. and {Devinat}, M. and {Gratton}, R. and {Lagrange}, A.-M. and {Langlois}, M. and {Mesa}, D. and {Milli}, J. and {Szul{\'a}gyi}, J. and {Nowak}, M. and {Rodet}, L. and {Rojo}, P. and {Petrus}, S. and {Janson}, M. and {Henning}, T. and {Kral}, Q. and {van Holstein}, R.~G. and {M{\'e}nard}, F. and {Beuzit}, J.-L. and {Biller}, B. and {Boccaletti}, A. and {Bonnefoy}, M. and {Brown}, S. and {Costille}, A. and {Delboulbe}, A. and {Desidera}, S. and {D'Orazi}, V. and {Feldt}, M. and {Fusco}, T. and {Galicher}, R. and {Hagelberg}, J. and {Lazzoni}, C. and {Ligi}, R. and {Maire}, A.-L. and {Messina}, S. and {Meyer}, M. and {Potier}, A. and {Ramos}, J. and {Rouan}, D. and {Schmidt}, T. and {Vigan}, A. and {Zurlo}, A.},
 doi = {10.1051/0004-6361/202243097},
 eid = {A139},
 eprint = {2206.00425},
 journal = {\aap},
 month = {August},
 pages = {A139},
 primaryclass = {astro-ph.EP},
 title = {{In-depth direct imaging and spectroscopic characterization of the young Solar System analog HD 95086}},
 volume = {664},
 year = {2022}
}

@article{Desidera_2021,
 adsurl = {https://ui.adsabs.harvard.edu/abs/2021A&A...651A..70D},
 archiveprefix = {arXiv},
 author = {{Desidera}, S. and {Chauvin}, G. and {Bonavita}, M. and {Messina}, S. and {LeCoroller}, H. and {Schmidt}, T. and {Gratton}, R. and {Lazzoni}, C. and {Meyer}, M. and {Schlieder}, J. and {Cheetham}, A. and {Hagelberg}, J. and {Bonnefoy}, M. and {Feldt}, M. and {Lagrange}, A.-M. and {Langlois}, M. and {Vigan}, A. and {Tan}, T.~G. and {Hambsch}, F.-J. and {Millward}, M. and {Alcal{\'a}}, J. and {Benatti}, S. and {Brandner}, W. and {Carson}, J. and {Covino}, E. and {Delorme}, P. and {D'Orazi}, V. and {Janson}, M. and {Rigliaco}, E. and {Beuzit}, J.-L. and {Biller}, B. and {Boccaletti}, A. and {Dominik}, C. and {Cantalloube}, F. and {Fontanive}, C. and {Galicher}, R. and {Henning}, Th. and {Lagadec}, E. and {Ligi}, R. and {Maire}, A.-L. and {Menard}, F. and {Mesa}, D. and {M{\"u}ller}, A. and {Samland}, M. and {Schmid}, H.~M. and {Sissa}, E. and {Turatto}, M. and {Udry}, S. and {Zurlo}, A. and {Asensio-Torres}, R. and {Kopytova}, T. and {Rickman}, E. and {Abe}, L. and {Antichi}, J. and {Baruffolo}, A. and {Baudoz}, P. and {Baudrand}, J. and {Blanchard}, P. and {Bazzon}, A. and {Buey}, T. and {Carbillet}, M. and {Carle}, M. and {Charton}, J. and {Cascone}, E. and {Claudi}, R. and {Costille}, A. and {Deboulb{\'e}}, A. and {De Caprio}, V. and {Dohlen}, K. and {Fantinel}, D. and {Feautrier}, P. and {Fusco}, T. and {Gigan}, P. and {Giro}, E. and {Gisler}, D. and {Gluck}, L. and {Hubin}, N. and {Hugot}, E. and {Jaquet}, M. and {Kasper}, M. and {Madec}, F. and {Magnard}, Y. and {Martinez}, P. and {Maurel}, D. and {Le Mignant}, D. and {M{\"o}ller-Nilsson}, O. and {Llored}, M. and {Moulin}, T. and {Orign{\'e}}, A. and {Pavlov}, A. and {Perret}, D. and {Petit}, C. and {Pragt}, J. and {Puget}, P. and {Rabou}, P. and {Ramos}, J. and {Rigal}, F. and {Rochat}, S. and {Roelfsema}, R. and {Rousset}, G. and {Roux}, A. and {Salasnich}, B. and {Sauvage}, J.-F. and {Sevin}, A. and {Soenke}, C. and {Stadler}, E. and {Suarez}, M. and {Weber}, L. and {Wildi}, F.},
 doi = {10.1051/0004-6361/202038806},
 eid = {A70},
 eprint = {2103.04366},
 journal = {\aap},
 month = {July},
 pages = {A70},
 primaryclass = {astro-ph.EP},
 title = {{The SPHERE infrared survey for exoplanets (SHINE). I. Sample definition and target characterization}},
 volume = {651},
 year = {2021}
}

@article{Diaz_2016,
 adsurl = {https://ui.adsabs.harvard.edu/abs/2016A&A...585A.134D},
 archiveprefix = {arXiv},
 author = {{D{\'\i}az}, R.~F. and {S{\'e}gransan}, D. and {Udry}, S. and {Lovis}, C. and {Pepe}, F. and {Dumusque}, X. and {Marmier}, M. and {Alonso}, R. and {Benz}, W. and {Bouchy}, F. and {Coffinet}, A. and {Collier Cameron}, A. and {Deleuil}, M. and {Figueira}, P. and {Gillon}, M. and {Lo Curto}, G. and {Mayor}, M. and {Mordasini}, C. and {Motalebi}, F. and {Moutou}, C. and {Pollacco}, D. and {Pompei}, E. and {Queloz}, D. and {Santos}, N. and {Wyttenbach}, A.},
 doi = {10.1051/0004-6361/201526729},
 eid = {A134},
 eprint = {1510.06446},
 journal = {\aap},
 month = {January},
 pages = {A134},
 primaryclass = {astro-ph.EP},
 title = {{The HARPS search for southern extra-solar planets. XXXVIII. Bayesian re-analysis of three systems. New super-Earths, unconfirmed signals, and magnetic cycles}},
 volume = {585},
 year = {2016}
}

@inproceedings{Dohlen_2008,
 adsurl = {https://ui.adsabs.harvard.edu/abs/2008SPIE.7014E..3LD},
 author = {{Dohlen}, Kjetil and {Langlois}, Maud and {Saisse}, Michel and {Hill}, Lucien and {Origne}, Alain and {Jacquet}, Marc and {Fabron}, Christophe and {Blanc}, Jean-Claude and {Llored}, Marc and {Carle}, Michael and {Moutou}, Claire and {Vigan}, Arthur and {Boccaletti}, Anthony and {Carbillet}, Marcel and {Mouillet}, David and {Beuzit}, Jean-Luc},
 booktitle = {Ground-based and Airborne Instrumentation for Astronomy II},
 doi = {10.1117/12.789786},
 editor = {{McLean}, Ian S. and {Casali}, Mark M.},
 eid = {70143L},
 month = {July},
 pages = {70143L},
 series = {Society of Photo-Optical Instrumentation Engineers (SPIE) Conference Series},
 title = {{The infra-red dual imaging and spectrograph for SPHERE: design and performance}},
 volume = {7014},
 year = {2008}
}

@article{Engler2018,
 adsurl = {https://ui.adsabs.harvard.edu/abs/2018A&A...618A.151E},
 archiveprefix = {arXiv},
 author = {{Engler}, N. and {Schmid}, H.~M. and {Quanz}, S.~P. and {Avenhaus}, H. and {Bazzon}, A.},
 doi = {10.1051/0004-6361/201832674},
 eid = {A151},
 eprint = {1808.04373},
 journal = {\aap},
 month = {October},
 pages = {A151},
 primaryclass = {astro-ph.EP},
 title = {{Detection of scattered light from the hot dust in HD 172555}},
 volume = {618},
 year = {2018}
}

@article{Esposito_2020,
 adsurl = {https://ui.adsabs.harvard.edu/abs/2020AJ....160...24E},
 archiveprefix = {arXiv},
 author = {{Esposito}, Thomas M. and {Kalas}, Paul and {Fitzgerald}, Michael P. and {Millar-Blanchaer}, Maxwell A. and {Duch{e}ne}, Gaspard and {Patience}, Jennifer and {Hom}, Justin and {Perrin}, Marshall D. and {De Rosa}, Robert J. and {Chiang}, Eugene and {Czekala}, Ian and {Macintosh}, Bruce and {Graham}, James R. and {Ansdell}, Megan and {Arriaga}, Pauline and {Bruzzone}, Sebastian and {Bulger}, Joanna and {Chen}, Christine H. and {Cotten}, Tara and {Dong}, Ruobing and {Draper}, Zachary H. and {Follette}, Katherine B. and {Hung}, Li-Wei and {Lopez}, Ronald and {Matthews}, Brenda C. and {Mazoyer}, Johan and {Metchev}, Stan and {Rameau}, Julien and {Ren}, Bin and {Rice}, Malena and {Song}, Inseok and {Stahl}, Kevin and {Wang}, Jason and {Wolff}, Schuyler and {Zuckerman}, Ben and {Ammons}, S. Mark and {Bailey}, Vanessa P. and {Barman}, Travis and {Chilcote}, Jeffrey and {Doyon}, Rene and {Gerard}, Benjamin L. and {Goodsell}, Stephen J. and {Greenbaum}, Alexandra Z. and {Hibon}, Pascale and {Hinkley}, Sasha and {Ingraham}, Patrick and {Konopacky}, Quinn and {Maire}, J{\'e}r{o}me and {Marchis}, Franck and {Marley}, Mark S. and {Marois}, Christian and {Nielsen}, Eric L. and {Oppenheimer}, Rebecca and {Palmer}, David and {Poyneer}, Lisa and {Pueyo}, Laurent and {Rajan}, Abhijith and {Rantakyr{\"o}}, Fredrik T. and {Ruffio}, Jean-Baptiste and {Savransky}, Dmitry and {Schneider}, Adam C. and {Sivaramakrishnan}, Anand and {Soummer}, R{\'e}mi and {Thomas}, Sandrine and {Ward-Duong}, Kimberly},
 doi = {10.3847/1538-3881/ab9199},
 eid = {24},
 eprint = {2004.13722},
 journal = {\aj},
 month = {July},
 number = {1},
 pages = {24},
 primaryclass = {astro-ph.EP},
 title = {{Debris Disk Results from the Gemini Planet Imager Exoplanet Survey's Polarimetric Imaging Campaign}},
 volume = {160},
 year = {2020}
}

@article{Faramaz_2014,
 adsurl = {https://ui.adsabs.harvard.edu/abs/2014A&A...563A..72F},
 archiveprefix = {arXiv},
 author = {{Faramaz}, V. and {Beust}, H. and {Th{\'e}bault}, P. and {Augereau}, J. -C. and {Bonsor}, A. and {del Burgo}, C. and {Ertel}, S. and {Marshall}, J.~P. and {Milli}, J. and {Montesinos}, B. and {Mora}, A. and {Bryden}, G. and {Danchi}, W. and {Eiroa}, C. and {White}, G.~J. and {Wolf}, S.},
 doi = {10.1051/0004-6361/201322469},
 eid = {A72},
 eprint = {1312.5146},
 journal = {\aap},
 month = {March},
 pages = {A72},
 primaryclass = {astro-ph.EP},
 title = {{Can eccentric debris disks be long-lived?. A first numerical investigation and application to {\ensuremath{\zeta}}$^{2}$ Reticuli}},
 volume = {563},
 year = {2014}
}

@article{Feldt_2017,
 adsurl = {https://ui.adsabs.harvard.edu/abs/2017A&A...601A...7F},
 archiveprefix = {arXiv},
 author = {{Feldt}, M. and {Olofsson}, J. and {Boccaletti}, A. and {Maire}, A.~L. and {Milli}, J. and {Vigan}, A. and {Langlois}, M. and {Henning}, Th. and {Moor}, A. and {Bonnefoy}, M. and {Wahhaj}, Z. and {Desidera}, S. and {Gratton}, R. and {K{\'o}sp{\'a}l}, {\'A}. and {Abraham}, P. and {Menard}, F. and {Chauvin}, G. and {Lagrange}, A.~M. and {Mesa}, D. and {Salter}, G. and {Buenzli}, E. and {Lannier}, J. and {Perrot}, C. and {Peretti}, S. and {Sissa}, E.},
 doi = {10.1051/0004-6361/201629261},
 eid = {A7},
 eprint = {1612.07621},
 journal = {\aap},
 month = {May},
 pages = {A7},
 primaryclass = {astro-ph.SR},
 title = {{SPHERE/SHINE reveals concentric rings in the debris disk of HIP 73145}},
 volume = {601},
 year = {2017}
}

@inproceedings{Fischer_2014,
 adsurl = {https://ui.adsabs.harvard.edu/abs/2014prpl.conf..715F},
 archiveprefix = {arXiv},
 author = {{Fischer}, D.~A. and {Howard}, A.~W. and {Laughlin}, G.~P. and {Macintosh}, B. and {Mahadevan}, S. and {Sahlmann}, J. and {Yee}, J.~C.},
 booktitle = {Protostars and Planets VI},
 doi = {10.2458/azu_uapress_9780816531240-ch031},
 editor = {{Beuther}, Henrik and {Klessen}, Ralf S. and {Dullemond}, Cornelis P. and {Henning}, Thomas},
 eprint = {1505.06869},
 month = {January},
 pages = {715-737},
 primaryclass = {astro-ph.EP},
 title = {{Exoplanet Detection Techniques}},
 year = {2014}
}

@article{Follette_2023,
 adsurl = {https://ui.adsabs.harvard.edu/abs/2023PASP..135i3001F},
 archiveprefix = {arXiv},
 author = {{Follette}, Katherine B.},
 doi = {10.1088/1538-3873/aceb31},
 eid = {093001},
 eprint = {2308.01354},
 journal = {\pasp},
 month = {September},
 number = {1051},
 pages = {093001},
 primaryclass = {astro-ph.IM},
 title = {{An Introduction to High Contrast Differential Imaging of Exoplanets and Disks}},
 volume = {135},
 year = {2023}
}

@article{Franson_2023,
 adsurl = {https://ui.adsabs.harvard.edu/abs/2023ApJ...950L..19F},
 archiveprefix = {arXiv},
 author = {{Franson}, Kyle and {Bowler}, Brendan P. and {Zhou}, Yifan and {Pearce}, Tim D. and {Bardalez Gagliuffi}, Daniella C. and {Biddle}, Lauren I. and {Brandt}, Timothy D. and {Crepp}, Justin R. and {Dupuy}, Trent J. and {Faherty}, Jacqueline and {Jensen-Clem}, Rebecca and {Morgan}, Marvin and {Sanghi}, Aniket and {Theissen}, Christopher A. and {Tran}, Quang H. and {Wolf}, Trevor N.},
 doi = {10.3847/2041-8213/acd6f6},
 eid = {L19},
 eprint = {2302.05420},
 journal = {\apjl},
 month = {June},
 number = {2},
 pages = {L19},
 primaryclass = {astro-ph.EP},
 title = {{Astrometric Accelerations as Dynamical Beacons: A Giant Planet Imaged inside the Debris Disk of the Young Star AF Lep}},
 volume = {950},
 year = {2023}
}

@article{Friebe_2022,
 adsurl = {https://ui.adsabs.harvard.edu/abs/2022MNRAS.512.4441F},
 archiveprefix = {arXiv},
 author = {{Friebe}, Marc F. and {Pearce}, Tim D. and {L{\"o}hne}, Torsten},
 doi = {10.1093/mnras/stac664},
 eprint = {2203.03611},
 journal = {\mnras},
 month = {May},
 number = {3},
 pages = {4441-4454},
 primaryclass = {astro-ph.EP},
 title = {{Gap carving by a migrating planet embedded in a massive debris disc}},
 volume = {512},
 year = {2022}
}

@article{Gagne_2018,
 adsurl = {https://ui.adsabs.harvard.edu/abs/2018ApJ...860...43G},
 archiveprefix = {arXiv},
 author = {{Gagn{\'e}}, Jonathan and {Roy-Loubier}, Olivier and {Faherty}, Jacqueline K. and {Doyon}, Ren{\'e} and {Malo}, Lison},
 doi = {10.3847/1538-4357/aac2b8},
 eid = {43},
 eprint = {1804.03093},
 journal = {\apj},
 month = {June},
 number = {1},
 pages = {43},
 primaryclass = {astro-ph.SR},
 title = {{BANYAN. XII. New Members of Nearby Young Associations from GAIA-Tycho Data}},
 volume = {860},
 year = {2018}
}

@article{GaiaCollaboration_2023,
 adsurl = {https://ui.adsabs.harvard.edu/abs/2023A&A...674A...1G},
 archiveprefix = {arXiv},
 author = {{Gaia Collaboration} and {Vallenari}, A. and {Brown}, A.~G.~A. and {Prusti}, T. and {de Bruijne}, J.~H.~J. and {Arenou}, F. and {Babusiaux}, C. and {Biermann}, M. and {Creevey}, O.~L. and {Ducourant}, C. and {Evans}, D.~W. and {Eyer}, L. and {Guerra}, R. and {Hutton}, A. and {Jordi}, C. and {Klioner}, S.~A. and {Lammers}, U.~L. and {Lindegren}, L. and {Luri}, X. and {Mignard}, F. and {Panem}, C. and {Pourbaix}, D. and {Randich}, S. and {Sartoretti}, P. and {Soubiran}, C. and {Tanga}, P. and {Walton}, N.~A. and {Bailer-Jones}, C.~A.~L. and {Bastian}, U. and {Drimmel}, R. and {Jansen}, F. and {Katz}, D. and {Lattanzi}, M.~G. and {van Leeuwen}, F. and {Bakker}, J. and {Cacciari}, C. and {Casta{\~n}eda}, J. and {De Angeli}, F. and {Fabricius}, C. and {Fouesneau}, M. and {Fr{\'e}mat}, Y. and {Galluccio}, L. and {Guerrier}, A. and {Heiter}, U. and {Masana}, E. and {Messineo}, R. and {Mowlavi}, N. and {Nicolas}, C. and {Nienartowicz}, K. and {Pailler}, F. and {Panuzzo}, P. and {Riclet}, F. and {Roux}, W. and {Seabroke}, G.~M. and {Sordo}, R. and {Th{\'e}venin}, F. and {Gracia-Abril}, G. and {Portell}, J. and {Teyssier}, D. and {Altmann}, M. and {Andrae}, R. and {Audard}, M. and {Bellas-Velidis}, I. and {Benson}, K. and {Berthier}, J. and {Blomme}, R. and {Burgess}, P.~W. and {Busonero}, D. and {Busso}, G. and {C{\'a}novas}, H. and {Carry}, B. and {Cellino}, A. and {Cheek}, N. and {Clementini}, G. and {Damerdji}, Y. and {Davidson}, M. and {de Teodoro}, P. and {Nu{\~n}ez Campos}, M. and {Delchambre}, L. and {Dell'Oro}, A. and {Esquej}, P. and {Fern{\'a}ndez-Hern{\'a}ndez}, J. and {Fraile}, E. and {Garabato}, D. and {Garc{\'\i}a-Lario}, P. and {Gosset}, E. and {Haigron}, R. and {Halbwachs}, J.-L. and {Hambly}, N.~C. and {Harrison}, D.~L. and {Hern{\'a}ndez}, J. and {Hestroffer}, D. and {Hodgkin}, S.~T. and {Holl}, B. and {Jan{\ss}en}, K. and {Jevardat de Fombelle}, G. and {Jordan}, S. and {Krone-Martins}, A. and {Lanzafame}, A.~C. and {L{\"o}ffler}, W. and {Marchal}, O. and {Marrese}, P.~M. and {Moitinho}, A. and {Muinonen}, K. and {Osborne}, P. and {Pancino}, E. and {Pauwels}, T. and {Recio-Blanco}, A. and {Reyl{\'e}}, C. and {Riello}, M. and {Rimoldini}, L. and {Roegiers}, T. and {Rybizki}, J. and {Sarro}, L.~M. and {Siopis}, C. and {Smith}, M. and {Sozzetti}, A. and {Utrilla}, E. and {van Leeuwen}, M. and {Abbas}, U. and {{\'A}brah{\'a}m}, P. and {Abreu Aramburu}, A. and {Aerts}, C. and {Aguado}, J.~J. and {Ajaj}, M. and {Aldea-Montero}, F. and {Altavilla}, G. and {{\'A}lvarez}, M.~A. and {Alves}, J. and {Anders}, F. and {Anderson}, R.~I. and {Anglada Varela}, E. and {Antoja}, T. and {Baines}, D. and {Baker}, S.~G. and {Balaguer-N{\'u}{\~n}ez}, L. and {Balbinot}, E. and {Balog}, Z. and {Barache}, C. and {Barbato}, D. and {Barros}, M. and {Barstow}, M.~A. and {Bartolom{\'e}}, S. and {Bassilana}, J.-L. and {Bauchet}, N. and {Becciani}, U. and {Bellazzini}, M. and {Berihuete}, A. and {Bernet}, M. and {Bertone}, S. and {Bianchi}, L. and {Binnenfeld}, A. and {Blanco-Cuaresma}, S. and {Blazere}, A. and {Boch}, T. and {Bombrun}, A. and {Bossini}, D. and {Bouquillon}, S. and {Bragaglia}, A. and {Bramante}, L. and {Breedt}, E. and {Bressan}, A. and {Brouillet}, N. and {Brugaletta}, E. and {Bucciarelli}, B. and {Burlacu}, A. and {Butkevich}, A.~G. and {Buzzi}, R. and {Caffau}, E. and {Cancelliere}, R. and {Cantat-Gaudin}, T. and {Carballo}, R. and {Carlucci}, T. and {Carnerero}, M.~I. and {Carrasco}, J.~M. and {Casamiquela}, L. and {Castellani}, M. and {Castro-Ginard}, A. and {Chaoul}, L. and {Charlot}, P. and {Chemin}, L. and {Chiaramida}, V. and {Chiavassa}, A. and {Chornay}, N. and {Comoretto}, G. and {Contursi}, G. and {Cooper}, W.~J. and {Cornez}, T. and {Cowell}, S. and {Crifo}, F. and {Cropper}, M. and {Crosta}, M. and {Crowley}, C. and {Dafonte}, C. and {Dapergolas}, A. and {David}, M. and {David}, P. and {de Laverny}, P. and {De Luise}, F. and {De March}, R.},
 doi = {10.1051/0004-6361/202243940},
 eid = {A1},
 eprint = {2208.00211},
 journal = {\aap},
 month = {June},
 pages = {A1},
 primaryclass = {astro-ph.GA},
 shortauthor = {{Gaia DR3}},
 title = {{Gaia Data Release 3. Summary of the content and survey properties}},
 volume = {674},
 year = {2023}
}

@article{Gaspar_2013,
 adsurl = {https://ui.adsabs.harvard.edu/abs/2013ApJ...768...25G},
 archiveprefix = {arXiv},
 author = {{G{\'a}sp{\'a}r}, Andr{\'a}s and {Rieke}, George H. and {Balog}, Zolt{\'a}n},
 doi = {10.1088/0004-637X/768/1/25},
 eid = {25},
 eprint = {1211.1415},
 journal = {\apj},
 month = {May},
 number = {1},
 pages = {25},
 primaryclass = {astro-ph.SR},
 title = {{The Collisional Evolution of Debris Disks}},
 volume = {768},
 year = {2013}
}

@article{Gaspar_2023,
 adsurl = {https://ui.adsabs.harvard.edu/abs/2023NatAs...7..790G},
 archiveprefix = {arXiv},
 author = {{G{\'a}sp{\'a}r}, Andr{\'a}s and {Wolff}, Schuyler Grace and {Rieke}, George H. and {Leisenring}, Jarron M. and {Morrison}, Jane and {Su}, Kate Y.~L. and {Ward-Duong}, Kimberly and {Aguilar}, Jonathan and {Ygouf}, Marie and {Beichman}, Charles and {Llop-Sayson}, Jorge and {Bryden}, Geoffrey},
 doi = {10.1038/s41550-023-01962-6},
 eprint = {2305.03789},
 journal = {Nature Astronomy},
 month = {July},
 pages = {790-798},
 primaryclass = {astro-ph.EP},
 title = {{Spatially resolved imaging of the inner Fomalhaut disk using JWST/MIRI}},
 volume = {7},
 year = {2023}
}

@article{Geers_2007,
 adsurl = {https://ui.adsabs.harvard.edu/abs/2007A&A...476..279G},
 archiveprefix = {arXiv},
 author = {{Geers}, V.~C. and {van Dishoeck}, E.~F. and {Visser}, R. and {Pontoppidan}, K.~M. and {Augereau}, J.-C. and {Habart}, E. and {Lagrange}, A.~M.},
 doi = {10.1051/0004-6361:20078466},
 eprint = {0710.2825},
 journal = {\aap},
 month = {December},
 number = {1},
 pages = {279-289},
 primaryclass = {astro-ph},
 title = {{Spatially extended polycyclic aromatic hydrocarbons in circumstellar disks around T Tauri and Herbig Ae stars}},
 volume = {476},
 year = {2007}
}

@article{Girardi_2005,
 adsurl = {https://ui.adsabs.harvard.edu/abs/2005A&A...436..895G},
 archiveprefix = {arXiv},
 author = {{Girardi}, L. and {Groenewegen}, M.~A.~T. and {Hatziminaoglou}, E. and {da Costa}, L.},
 doi = {10.1051/0004-6361:20042352},
 eprint = {astro-ph/0504047},
 journal = {\aap},
 month = {June},
 number = {3},
 pages = {895-915},
 primaryclass = {astro-ph},
 title = {{Star counts in the Galaxy. Simulating from very deep to very shallow photometric surveys with the TRILEGAL code}},
 volume = {436},
 year = {2005}
}

@ARTICLE{Gladman_1993,
       author = {{Gladman}, Brett},
        title = "{Dynamics of Systems of Two Close Planets}",
      journal = {\icarus},
         year = 1993,
        month = nov,
       volume = {106},
       number = {1},
        pages = {247-263},
          doi = {10.1006/icar.1993.1169},
       adsurl = {https://ui.adsabs.harvard.edu/abs/1993Icar..106..247G}
}

@article{Golimowski_1993,
 adsurl = {https://ui.adsabs.harvard.edu/abs/1993AJ....105.1108G},
 author = {{Golimowski}, David A. and {Durrance}, Samuel T. and {Clampin}, Mark},
 doi = {10.1086/116498},
 journal = {\aj},
 month = {March},
 pages = {1108},
 title = {{Detection of an apparent Star 2.1 Arcsec from the Circumstellar Disk Candidate Epsilon Sagittarii}},
 volume = {105},
 year = {1993}
}

@article{Gray_2017,
 adsurl = {https://ui.adsabs.harvard.edu/abs/2017AJ....154...31G},
 author = {{Gray}, R.~O. and {Riggs}, Q.~S. and {Koen}, C. and {Murphy}, S.~J. and {Newsome}, I.~M. and {Corbally}, C.~J. and {Cheng}, K.-P. and {Neff}, J.~E.},
 doi = {10.3847/1538-3881/aa6d5e},
 eid = {31},
 journal = {\aj},
 month = {July},
 number = {1},
 pages = {31},
 title = {{The Discovery of {\ensuremath{\lambda}} Bootis Stars: The Southern Survey I}},
 volume = {154},
 year = {2017}
}

@article{Greaves_2014,
 adsurl = {https://ui.adsabs.harvard.edu/abs/2014MNRAS.438L..31G},
 archiveprefix = {arXiv},
 author = {{Greaves}, J.~S. and {Kennedy}, G.~M. and {Thureau}, N. and {Eiroa}, C. and {Marshall}, J.~P. and {Maldonado}, J. and {Matthews}, B.~C. and {Olofsson}, G. and {Barlow}, M.~J. and {Moro-Mart{\'\i}n}, A. and {Sibthorpe}, B. and {Absil}, O. and {Ardila}, D.~R. and {Booth}, M. and {Broekhoven-Fiene}, H. and {Brown}, D.~J.~A. and {Collier Cameron}, A. and {del Burgo}, C. and {Di Francesco}, J. and {Eisl{\"o}ffel}, J. and {Duch{e}ne}, G. and {Ertel}, S. and {Holland}, W.~S. and {Horner}, J. and {Kalas}, P. and {Kavelaars}, J.~J. and {Lestrade}, J.-F. and {Vican}, L. and {Wilner}, D.~J. and {Wolf}, S. and {Wyatt}, M.~C.},
 doi = {10.1093/mnrasl/slt153},
 eprint = {1310.6936},
 journal = {\mnras},
 month = {February},
 number = {1},
 pages = {L31-L35},
 primaryclass = {astro-ph.SR},
 title = {{Alignment in star-debris disc systems seen by Herschel}},
 volume = {438},
 year = {2014}
}

@article{Griffin_2010,
 adsurl = {https://ui.adsabs.harvard.edu/abs/2010A&A...518L...3G},
 archiveprefix = {arXiv},
 author = {{Griffin}, M.~J. and {Abergel}, A. and {Abreu}, A. and {Ade}, P.~A.~R. and {Andr{\'e}}, P. and {Augueres}, J.-L. and {Babbedge}, T. and {Bae}, Y. and {Baillie}, T. and {Baluteau}, J.-P. and {Barlow}, M.~J. and {Bendo}, G. and {Benielli}, D. and {Bock}, J.~J. and {Bonhomme}, P. and {Brisbin}, D. and {Brockley-Blatt}, C. and {Caldwell}, M. and {Cara}, C. and {Castro-Rodriguez}, N. and {Cerulli}, R. and {Chanial}, P. and {Chen}, S. and {Clark}, E. and {Clements}, D.~L. and {Clerc}, L. and {Coker}, J. and {Communal}, D. and {Conversi}, L. and {Cox}, P. and {Crumb}, D. and {Cunningham}, C. and {Daly}, F. and {Davis}, G.~R. and {de Antoni}, P. and {Delderfield}, J. and {Devin}, N. and {di Giorgio}, A. and {Didschuns}, I. and {Dohlen}, K. and {Donati}, M. and {Dowell}, A. and {Dowell}, C.~D. and {Duband}, L. and {Dumaye}, L. and {Emery}, R.~J. and {Ferlet}, M. and {Ferrand}, D. and {Fontignie}, J. and {Fox}, M. and {Franceschini}, A. and {Frerking}, M. and {Fulton}, T. and {Garcia}, J. and {Gastaud}, R. and {Gear}, W.~K. and {Glenn}, J. and {Goizel}, A. and {Griffin}, D.~K. and {Grundy}, T. and {Guest}, S. and {Guillemet}, L. and {Hargrave}, P.~C. and {Harwit}, M. and {Hastings}, P. and {Hatziminaoglou}, E. and {Herman}, M. and {Hinde}, B. and {Hristov}, V. and {Huang}, M. and {Imhof}, P. and {Isaak}, K.~J. and {Israelsson}, U. and {Ivison}, R.~J. and {Jennings}, D. and {Kiernan}, B. and {King}, K.~J. and {Lange}, A.~E. and {Latter}, W. and {Laurent}, G. and {Laurent}, P. and {Leeks}, S.~J. and {Lellouch}, E. and {Levenson}, L. and {Li}, B. and {Li}, J. and {Lilienthal}, J. and {Lim}, T. and {Liu}, S.~J. and {Lu}, N. and {Madden}, S. and {Mainetti}, G. and {Marliani}, P. and {McKay}, D. and {Mercier}, K. and {Molinari}, S. and {Morris}, H. and {Moseley}, H. and {Mulder}, J. and {Mur}, M. and {Naylor}, D.~A. and {Nguyen}, H. and {O'Halloran}, B. and {Oliver}, S. and {Olofsson}, G. and {Olofsson}, H.-G. and {Orfei}, R. and {Page}, M.~J. and {Pain}, I. and {Panuzzo}, P. and {Papageorgiou}, A. and {Parks}, G. and {Parr-Burman}, P. and {Pearce}, A. and {Pearson}, C. and {P{\'e}rez-Fournon}, I. and {Pinsard}, F. and {Pisano}, G. and {Podosek}, J. and {Pohlen}, M. and {Polehampton}, E.~T. and {Pouliquen}, D. and {Rigopoulou}, D. and {Rizzo}, D. and {Roseboom}, I.~G. and {Roussel}, H. and {Rowan-Robinson}, M. and {Rownd}, B. and {Saraceno}, P. and {Sauvage}, M. and {Savage}, R. and {Savini}, G. and {Sawyer}, E. and {Scharmberg}, C. and {Schmitt}, D. and {Schneider}, N. and {Schulz}, B. and {Schwartz}, A. and {Shafer}, R. and {Shupe}, D.~L. and {Sibthorpe}, B. and {Sidher}, S. and {Smith}, A. and {Smith}, A.~J. and {Smith}, D. and {Spencer}, L. and {Stobie}, B. and {Sudiwala}, R. and {Sukhatme}, K. and {Surace}, C. and {Stevens}, J.~A. and {Swinyard}, B.~M. and {Trichas}, M. and {Tourette}, T. and {Triou}, H. and {Tseng}, S. and {Tucker}, C. and {Turner}, A. and {Vaccari}, M. and {Valtchanov}, I. and {Vigroux}, L. and {Virique}, E. and {Voellmer}, G. and {Walker}, H. and {Ward}, R. and {Waskett}, T. and {Weilert}, M. and {Wesson}, R. and {White}, G.~J. and {Whitehouse}, N. and {Wilson}, C.~D. and {Winter}, B. and {Woodcraft}, A.~L. and {Wright}, G.~S. and {Xu}, C.~K. and {Zavagno}, A. and {Zemcov}, M. and {Zhang}, L. and {Zonca}, E.},
 doi = {10.1051/0004-6361/201014519},
 eid = {L3},
 eprint = {1005.5123},
 journal = {\aap},
 month = {July},
 pages = {L3},
 primaryclass = {astro-ph.IM},
 title = {{The Herschel-SPIRE instrument and its in-flight performance}},
 volume = {518},
 year = {2010}
}

@article{Guimaraes_2006,
 adsurl = {https://ui.adsabs.harvard.edu/abs/2006A&A...457..581G},
 author = {{Guimar{\~a}es}, M.~M. and {Alencar}, S.~H.~P. and {Corradi}, W.~J.~B. and {Vieira}, S.~L.~A.},
 doi = {10.1051/0004-6361:20065005},
 journal = {\aap},
 month = {October},
 number = {2},
 pages = {581-589},
 title = {{Stellar parameters and evidence of circumstellar activity for a sample of Herbig Ae/Be stars}},
 volume = {457},
 year = {2006}
}

@article{Hauschildt_2025,
 adsurl = {https://ui.adsabs.harvard.edu/abs/2025A&A...698A..47H},
 archiveprefix = {arXiv},
 author = {{Hauschildt}, P.~H. and {Barman}, T. and {Baron}, E. and {Aufdenberg}, J.~P. and {Schweitzer}, A.},
 doi = {10.1051/0004-6361/202554171},
 eid = {A47},
 eprint = {2504.17597},
 journal = {\aap},
 month = {June},
 pages = {A47},
 primaryclass = {astro-ph.SR},
 title = {{The NewEra model grid}},
 volume = {698},
 year = {2025}
}

@article{Herz2024,
 adsurl = {https://ui.adsabs.harvard.edu/abs/2024A&A...682A..92H},
 archiveprefix = {arXiv},
 author = {{Herz}, Philipp and {Samland}, Matthias and {Bailer-Jones}, Coryn A.~L.},
 doi = {10.1051/0004-6361/202348496},
 eid = {A92},
 eprint = {2312.04873},
 journal = {\aap},
 month = {February},
 pages = {A92},
 primaryclass = {astro-ph.EP},
 title = {{Distinguishing exoplanet companions from field stars in direct imaging using Gaia astrometry}},
 volume = {682},
 year = {2024}
}

@article{Holland_2017,
 adsurl = {https://ui.adsabs.harvard.edu/abs/2017MNRAS.470.3606H},
 archiveprefix = {arXiv},
 author = {{Holland}, Wayne S. and {Matthews}, Brenda C. and {Kennedy}, Grant M. and {Greaves}, Jane S. and {Wyatt}, Mark C. and {Booth}, Mark and {Bastien}, Pierre and {Bryden}, Geoff and {Butner}, Harold and {Chen}, Christine H. and {Chrysostomou}, Antonio and {Davies}, Claire L. and {Dent}, William R.~F. and {Di Francesco}, James and {Duch{e}ne}, Gaspard and {Gibb}, Andy G. and {Friberg}, Per and {Ivison}, Rob J. and {Jenness}, Tim and {Kavelaars}, JJ and {Lawler}, Samantha and {Lestrade}, Jean-Fran{\c{c}}ois and {Marshall}, Jonathan P. and {Moro-Martin}, Amaya and {Pani{\'c}}, Olja and {Phillips}, Neil and {Serjeant}, Stephen and {Schieven}, Gerald H. and {Sibthorpe}, Bruce and {Vican}, Laura and {Ward-Thompson}, Derek and {van der Werf}, Paul and {White}, Glenn J. and {Wilner}, David and {Zuckerman}, Ben},
 doi = {10.1093/mnras/stx1378},
 eprint = {1706.01218},
 journal = {\mnras},
 month = {September},
 number = {3},
 pages = {3606-3663},
 primaryclass = {astro-ph.SR},
 title = {{SONS: The JCMT legacy survey of debris discs in the submillimetre}},
 volume = {470},
 year = {2017}
}

@article{Hoogerwerf_2000,
 adsurl = {https://ui.adsabs.harvard.edu/abs/2000MNRAS.313...43H},
 author = {{Hoogerwerf}, Ronnie},
 doi = {10.1046/j.1365-8711.2000.03192.x},
 journal = {\mnras},
 month = {March},
 number = {1},
 pages = {43-65},
 title = {{OB association members in the ACT and TRC catalogues}},
 volume = {313},
 year = {2000}
}

@misc{Houk_1978,
 adsurl = {https://ui.adsabs.harvard.edu/abs/1978mcts.book.....H},
 author = {{Houk}, Nancy},
 title = {{Michigan catalogue of two-dimensional spectral types for the HD stars}},
 year = {1978}
}

@article{Hughes_2018,
 adsurl = {https://ui.adsabs.harvard.edu/abs/2018ARA&A..56..541H},
 archiveprefix = {arXiv},
 author = {{Hughes}, A. Meredith and {Duch{e}ne}, Gaspard and {Matthews}, Brenda C.},
 doi = {10.1146/annurev-astro-081817-052035},
 eprint = {1802.04313},
 journal = {\araa},
 month = {September},
 pages = {541-591},
 primaryclass = {astro-ph.EP},
 title = {{Debris Disks: Structure, Composition, and Variability}},
 volume = {56},
 year = {2018}
}

@article{ImazBlanco_2023,
 adsurl = {https://ui.adsabs.harvard.edu/abs/2023MNRAS.522.6150I},
 archiveprefix = {arXiv},
 author = {{Imaz Blanco}, Amaia and {Marino}, Sebastian and {Matr{\`a}}, Luca and {Booth}, Mark and {Carpenter}, John and {Faramaz}, Virginie and {Henning}, Thomas and {Hughes}, A. Meredith and {Kennedy}, Grant M. and {P{\'e}rez}, Sebasti{\'a}n and {Ricci}, Luca and {Wyatt}, Mark C.},
 doi = {10.1093/mnras/stad1221},
 eprint = {2304.12337},
 journal = {\mnras},
 month = {July},
 number = {4},
 pages = {6150-6169},
 primaryclass = {astro-ph.EP},
 title = {{Inner edges of planetesimal belts: collisionally eroded or truncated?}},
 volume = {522},
 year = {2023}
}

@article{Jancart_2005,
 adsurl = {https://ui.adsabs.harvard.edu/abs/2005A&A...442..365J},
 archiveprefix = {arXiv},
 author = {{Jancart}, S. and {Jorissen}, A. and {Babusiaux}, C. and {Pourbaix}, D.},
 doi = {10.1051/0004-6361:20053003},
 eprint = {astro-ph/0507695},
 journal = {\aap},
 month = {October},
 number = {1},
 pages = {365-380},
 primaryclass = {astro-ph},
 title = {{Astrometric orbits of S$_{B9}$ stars}},
 volume = {442},
 year = {2005}
}

@article{Janson_2013,
 adsurl = {https://ui.adsabs.harvard.edu/abs/2013ApJ...773...73J},
 archiveprefix = {arXiv},
 author = {{Janson}, Markus and {Brandt}, Timothy D. and {Moro-Mart{\'\i}n}, Amaya and {Usuda}, Tomonori and {Thalmann}, Christian and {Carson}, Joseph C. and {Goto}, Miwa and {Currie}, Thayne and {McElwain}, M.~W. and {Itoh}, Yoichi and {Fukagawa}, Misato and {Crepp}, Justin and {Kuzuhara}, Masayuki and {Hashimoto}, Jun and {Kudo}, Tomoyuki and {Kusakabe}, Nobuhiko and {Abe}, Lyu and {Brandner}, Wolfgang and {Egner}, Sebastian and {Feldt}, Markus and {Grady}, Carol A. and {Guyon}, Olivier and {Hayano}, Yutaka and {Hayashi}, Masahiro and {Hayashi}, Saeko and {Henning}, Thomas and {Hodapp}, Klaus W. and {Ishii}, Miki and {Iye}, Masanori and {Kandori}, Ryo and {Knapp}, Gillian R. and {Kwon}, Jungmi and {Matsuo}, Taro and {Miyama}, Shoken and {Morino}, Jun-Ichi and {Nishimura}, Tetsuro and {Pyo}, Tae-Soo and {Serabyn}, Eugene and {Suenaga}, Takuya and {Suto}, Hiroshi and {Suzuki}, Ryuji and {Takahashi}, Yasuhiro and {Takami}, Michihiro and {Takato}, Naruhisa and {Terada}, Hiroshi and {Tomono}, Daego and {Turner}, Edwin L. and {Watanabe}, Makoto and {Wisniewski}, John and {Yamada}, Toru and {Takami}, Hideki and {Tamura}, Motohide},
 doi = {10.1088/0004-637X/773/1/73},
 eid = {73},
 eprint = {1306.0581},
 journal = {\apj},
 month = {August},
 number = {1},
 pages = {73},
 primaryclass = {astro-ph.EP},
 title = {{The SEEDS Direct Imaging Survey for Planets and Scattered Dust Emission in Debris Disk Systems}},
 volume = {773},
 year = {2013}
}

@article{Jones_2023,
 adsurl = {https://ui.adsabs.harvard.edu/abs/2023ApJ...948..102J},
 archiveprefix = {arXiv},
 author = {{Jones}, Joshua W. and {Chiang}, Eugene and {Duch{\^e}ne}, Gaspard and {Kalas}, Paul and {Esposito}, Thomas M.},
 doi = {10.3847/1538-4357/acc466},
 eid = {102},
 eprint = {2303.10189},
 journal = {\apj},
 month = {May},
 number = {2},
 pages = {102},
 primaryclass = {astro-ph.EP},
 title = {{Giant Impacts and Debris Disk Morphology}},
 volume = {948},
 year = {2023}
}

@article{Kalas_2026,
 adsurl = {https://ui.adsabs.harvard.edu/abs/2026Sci...391..371K},
 archiveprefix = {arXiv},
 author = {{Kalas}, Paul and {Wang}, Jason J. and {Millar-Blanchaer}, Maxwell A. and {Ren}, Bin B. and {Wyatt}, Mark C. and {Kennedy}, Grant M. and {Sommer}, Maximilian and {Esposito}, Thomas M. and {De Rosa}, Robert J. and {Fitzgerald}, Michael},
 doi = {10.1126/science.adu6266},
 eprint = {2512.15861},
 journal = {Science},
 month = {January},
 number = {6783},
 pages = {371-373},
 primaryclass = {astro-ph.EP},
 title = {{A second planetesimal collision in the Fomalhaut system}},
 volume = {391},
 year = {2026}
}

@article{Kennedy_2015,
 adsurl = {https://ui.adsabs.harvard.edu/abs/2015MNRAS.447L..75K},
 archiveprefix = {arXiv},
 author = {{Kennedy}, G.~M.},
 doi = {10.1093/mnrasl/slu190},
 eprint = {1412.0674},
 journal = {\mnras},
 month = {February},
 pages = {L75-L79},
 primaryclass = {astro-ph.EP},
 title = {{Nature or nurture of coplanar Tatooines: the aligned circumbinary Kuiper Belt analogue around HD 131511.}},
 volume = {447},
 year = {2015}
}

@article{Keppler_2018,
 adsurl = {https://ui.adsabs.harvard.edu/abs/2018A&A...617A..44K},
 archiveprefix = {arXiv},
 author = {{Keppler}, M. and {Benisty}, M. and {M{\"u}ller}, A. and {Henning}, Th. and {van Boekel}, R. and {Cantalloube}, F. and {Ginski}, C. and {van Holstein}, R.~G. and {Maire}, A.-L. and {Pohl}, A. and {Samland}, M. and {Avenhaus}, H. and {Baudino}, J.-L. and {Boccaletti}, A. and {de Boer}, J. and {Bonnefoy}, M. and {Chauvin}, G. and {Desidera}, S. and {Langlois}, M. and {Lazzoni}, C. and {Marleau}, G.-D. and {Mordasini}, C. and {Pawellek}, N. and {Stolker}, T. and {Vigan}, A. and {Zurlo}, A. and {Birnstiel}, T. and {Brandner}, W. and {Feldt}, M. and {Flock}, M. and {Girard}, J. and {Gratton}, R. and {Hagelberg}, J. and {Isella}, A. and {Janson}, M. and {Juhasz}, A. and {Kemmer}, J. and {Kral}, Q. and {Lagrange}, A.-M. and {Launhardt}, R. and {Matter}, A. and {M{\'e}nard}, F. and {Milli}, J. and {Molli{\`e}re}, P. and {Olofsson}, J. and {P{\'e}rez}, L. and {Pinilla}, P. and {Pinte}, C. and {Quanz}, S.~P. and {Schmidt}, T. and {Udry}, S. and {Wahhaj}, Z. and {Williams}, J.~P. and {Buenzli}, E. and {Cudel}, M. and {Dominik}, C. and {Galicher}, R. and {Kasper}, M. and {Lannier}, J. and {Mesa}, D. and {Mouillet}, D. and {Peretti}, S. and {Perrot}, C. and {Salter}, G. and {Sissa}, E. and {Wildi}, F. and {Abe}, L. and {Antichi}, J. and {Augereau}, J.-C. and {Baruffolo}, A. and {Baudoz}, P. and {Bazzon}, A. and {Beuzit}, J.-L. and {Blanchard}, P. and {Brems}, S.~S. and {Buey}, T. and {De Caprio}, V. and {Carbillet}, M. and {Carle}, M. and {Cascone}, E. and {Cheetham}, A. and {Claudi}, R. and {Costille}, A. and {Delboulb{\'e}}, A. and {Dohlen}, K. and {Fantinel}, D. and {Feautrier}, P. and {Fusco}, T. and {Giro}, E. and {Gluck}, L. and {Gry}, C. and {Hubin}, N. and {Hugot}, E. and {Jaquet}, M. and {Le Mignant}, D. and {Llored}, M. and {Madec}, F. and {Magnard}, Y. and {Martinez}, P. and {Maurel}, D. and {Meyer}, M. and {M{\"o}ller-Nilsson}, O. and {Moulin}, T. and {Mugnier}, L. and {Orign{\'e}}, A. and {Pavlov}, A. and {Perret}, D. and {Petit}, C. and {Pragt}, J. and {Puget}, P. and {Rabou}, P. and {Ramos}, J. and {Rigal}, F. and {Rochat}, S. and {Roelfsema}, R. and {Rousset}, G. and {Roux}, A. and {Salasnich}, B. and {Sauvage}, J.-F. and {Sevin}, A. and {Soenke}, C. and {Stadler}, E. and {Suarez}, M. and {Turatto}, M. and {Weber}, L.},
 doi = {10.1051/0004-6361/201832957},
 eid = {A44},
 eprint = {1806.11568},
 journal = {\aap},
 month = {September},
 pages = {A44},
 primaryclass = {astro-ph.EP},
 title = {{Discovery of a planetary-mass companion within the gap of the transition disk around PDS 70}},
 volume = {617},
 year = {2018}
}

@ARTICLE{Kirsh_2009,
       author = {{Kirsh}, David R. and {Duncan}, Martin and {Brasser}, Ramon and {Levison}, Harold F.},
        title = "{Simulations of planet migration driven by planetesimal scattering}",
      journal = {\icarus},
         year = 2009,
        month = jan,
       volume = {199},
       number = {1},
        pages = {197-209},
          doi = {10.1016/j.icarus.2008.05.028},
       adsurl = {https://ui.adsabs.harvard.edu/abs/2009Icar..199..197K}
}

@article{Kluska_2018,
 adsurl = {https://ui.adsabs.harvard.edu/abs/2018ApJ...855...44K},
 archiveprefix = {arXiv},
 author = {{Kluska}, Jacques and {Kraus}, Stefan and {Davies}, Claire L. and {Harries}, Tim and {Willson}, Matthew and {Monnier}, John D. and {Aarnio}, Alicia and {Baron}, Fabien and {Millan-Gabet}, Rafael and {Ten Brummelaar}, Theo and {Che}, Xiao and {Hinkley}, Sasha and {Preibisch}, Thomas and {Sturmann}, Judit and {Sturmann}, Laszlo and {Touhami}, Yamina},
 doi = {10.3847/1538-4357/aaacd3},
 eid = {44},
 eprint = {1803.02419},
 journal = {\apj},
 month = {March},
 number = {1},
 pages = {44},
 primaryclass = {astro-ph.SR},
 title = {{A Multi-instrument and Multi-wavelength High Angular Resolution Study of MWC 614: Quantum Heated Particles Inside the Disk Cavity}},
 volume = {855},
 year = {2018}
}

@article{Koerner_2010,
 adsurl = {https://ui.adsabs.harvard.edu/abs/2010ApJ...710L..26K},
 author = {{Koerner}, D.~W. and {Kim}, S. and {Trilling}, D.~E. and {Larson}, H. and {Cotera}, A. and {Stapelfeldt}, K.~R. and {Wahhaj}, Z. and {Fajardo-Acosta}, S. and {Padgett}, D. and {Backman}, D.},
 doi = {10.1088/2041-8205/710/1/L26},
 journal = {\apjl},
 month = {February},
 number = {1},
 pages = {L26-L29},
 title = {{New Debris Disk Candidates Around 49 Nearby Stars}},
 volume = {710},
 year = {2010}
}

@article{Krist_2010,
 adsurl = {https://ui.adsabs.harvard.edu/abs/2010AJ....140.1051K},
 archiveprefix = {arXiv},
 author = {{Krist}, John E. and {Stapelfeldt}, Karl R. and {Bryden}, Geoffrey and {Rieke}, George H. and {Su}, K.~Y.~L. and {Chen}, Christine H. and {Beichman}, Charles A. and {Hines}, Dean C. and {Rebull}, Luisa M. and {Tanner}, Angelle and {Trilling}, David E. and {Clampin}, Mark and {G{\'a}sp{\'a}r}, Andr{\'a}s},
 doi = {10.1088/0004-6256/140/4/1051},
 eprint = {1008.2793},
 journal = {\aj},
 month = {October},
 number = {4},
 pages = {1051-1061},
 primaryclass = {astro-ph.SR},
 title = {{HST and Spitzer Observations of the HD 207129 Debris Ring}},
 volume = {140},
 year = {2010}
}

@article{Krivov_2010,
 adsurl = {https://ui.adsabs.harvard.edu/abs/2010RAA....10..383K},
 archiveprefix = {arXiv},
 author = {{Krivov}, Alexander V.},
 doi = {10.1088/1674-4527/10/5/001},
 eprint = {1003.5229},
 journal = {Research in Astronomy and Astrophysics},
 month = {May},
 number = {5},
 pages = {383-414},
 primaryclass = {astro-ph.EP},
 title = {{Debris disks: seeing dust, thinking of planetesimals and planets}},
 volume = {10},
 year = {2010}
}

@article{Krivov_2021,
 adsurl = {https://ui.adsabs.harvard.edu/abs/2021MNRAS.500..718K},
 archiveprefix = {arXiv},
 author = {{Krivov}, Alexander V. and {Wyatt}, Mark C.},
 doi = {10.1093/mnras/staa2385},
 eprint = {2008.07406},
 journal = {\mnras},
 month = {January},
 number = {1},
 pages = {718-735},
 primaryclass = {astro-ph.EP},
 title = {{Solution to the debris disc mass problem: planetesimals are born small?}},
 volume = {500},
 year = {2021}
}

@article{Lafreniere_2009,
 adsurl = {https://ui.adsabs.harvard.edu/abs/2009ApJ...694L.148L},
 archiveprefix = {arXiv},
 author = {{Lafreni{\`e}re}, David and {Marois}, Christian and {Doyon}, Ren{\'e} and {Barman}, Travis},
 doi = {10.1088/0004-637X/694/2/L148},
 eprint = {0902.3247},
 journal = {\apjl},
 month = {April},
 number = {2},
 pages = {L148-L152},
 primaryclass = {astro-ph.EP},
 title = {{HST/NICMOS Detection of HR 8799 b in 1998}},
 volume = {694},
 year = {2009}
}

@article{Lagrange_2010,
 adsurl = {https://ui.adsabs.harvard.edu/abs/2010Sci...329...57L},
 archiveprefix = {arXiv},
 author = {{Lagrange}, A.-M. and {Bonnefoy}, M. and {Chauvin}, G. and {Apai}, D. and {Ehrenreich}, D. and {Boccaletti}, A. and {Gratadour}, D. and {Rouan}, D. and {Mouillet}, D. and {Lacour}, S. and {Kasper}, M.},
 doi = {10.1126/science.1187187},
 eprint = {1006.3314},
 journal = {Science},
 month = {July},
 number = {5987},
 pages = {57},
 primaryclass = {astro-ph.EP},
 title = {{A Giant Planet Imaged in the Disk of the Young Star {\ensuremath{\beta}} Pictoris}},
 volume = {329},
 year = {2010}
}

@ARTICLE{Brandt_2021_HGCA,
       author = {{Brandt}, Timothy D.},
        title = "{The Hipparcos-Gaia Catalog of Accelerations: Gaia EDR3 Edition}",
      journal = {\apjs},
         year = 2021,
        month = jun,
       volume = {254},
       number = {2},
          eid = {42},
        pages = {42},
          doi = {10.3847/1538-4365/abf93c},
archivePrefix = {arXiv},
       eprint = {2105.11662},
 primaryClass = {astro-ph.GA},
       adsurl = {https://ui.adsabs.harvard.edu/abs/2021ApJS..254...42B}
}

@ARTICLE{Chambers_1996,
       author = {{Chambers}, J.~E. and {Wetherill}, G.~W. and {Boss}, A.~P.},
        title = "{The Stability of Multi-Planet Systems}",
      journal = {\icarus},
         year = 1996,
        month = feb,
       volume = {119},
       number = {2},
        pages = {261-268},
          doi = {10.1006/icar.1996.0019},
       adsurl = {https://ui.adsabs.harvard.edu/abs/1996Icar..119..261C}
}

@article{Lagrange_2025,
 adsurl = {https://ui.adsabs.harvard.edu/abs/2025Natur.642..905L},
 archiveprefix = {arXiv},
 author = {{Lagrange}, A.-M. and {Wilkinson}, C. and {M{a}lin}, M. and {Boccaletti}, A. and {Perrot}, C. and {Matr{\`a}}, L. and {Combes}, F. and {Beust}, H. and {Rouan}, D. and {Chomez}, A. and {Milli}, J. and {Charnay}, B. and {Mazevet}, S. and {Flasseur}, O. and {Olofsson}, J. and {Bayo}, A. and {Kral}, Q. and {Carter}, A. and {Crotts}, K.~A. and {Delorme}, P. and {Chauvin}, G. and {Thebault}, P. and {Rubini}, P. and {Kiefer}, F. and {Radcliffe}, A. and {Mazoyer}, J. and {Bodrito}, T. and {Stasevic}, S. and {Langlois}, M.},
 doi = {10.1038/s41586-025-09150-4},
 eprint = {2502.15081},
 journal = {\nat},
 month = {June},
 number = {8069},
 pages = {905-908},
 primaryclass = {astro-ph.EP},
 title = {{Evidence for a sub-Jovian planet in the young TWA 7 disk}},
 volume = {642},
 year = {2025}
}

@article{Launhardt_2020,
 adsurl = {https://ui.adsabs.harvard.edu/abs/2020A&A...635A.162L},
 archiveprefix = {arXiv},
 author = {{Launhardt}, R. and {Henning}, Th. and {Quirrenbach}, A. and {S{\'e}gransan}, D. and {Avenhaus}, H. and {van Boekel}, R. and {Brems}, S.~S. and {Cheetham}, A.~C. and {Cugno}, G. and {Girard}, J. and {Godoy}, N. and {Kennedy}, G.~M. and {Maire}, A.-L. and {Metchev}, S. and {M{\"u}ller}, A. and {Musso Barcucci}, A. and {Olofsson}, J. and {Pepe}, F. and {Quanz}, S.~P. and {Queloz}, D. and {Reffert}, S. and {Rickman}, E.~L. and {Ruh}, H.~L. and {Samland}, M.},
 doi = {10.1051/0004-6361/201937000},
 eid = {A162},
 eprint = {2002.01807},
 journal = {\aap},
 month = {March},
 pages = {A162},
 primaryclass = {astro-ph.EP},
 title = {{ISPY-NACO Imaging Survey for Planets around Young stars. Survey description and results from the first 2.5 years of observations}},
 volume = {635},
 year = {2020}
}

@article{Lazzoni_2018,
 adsurl = {https://ui.adsabs.harvard.edu/abs/2018A&A...611A..43L},
 archiveprefix = {arXiv},
 author = {{Lazzoni}, C. and {Desidera}, S. and {Marzari}, F. and {Boccaletti}, A. and {Langlois}, M. and {Mesa}, D. and {Gratton}, R. and {Kral}, Q. and {Pawellek}, N. and {Olofsson}, J. and {Bonnefoy}, M. and {Chauvin}, G. and {Lagrange}, A.~M. and {Vigan}, A. and {Sissa}, E. and {Antichi}, J. and {Avenhaus}, H. and {Baruffolo}, A. and {Baudino}, J.~L. and {Bazzon}, A. and {Beuzit}, J.~L. and {Biller}, B. and {Bonavita}, M. and {Brandner}, W. and {Bruno}, P. and {Buenzli}, E. and {Cantalloube}, F. and {Cascone}, E. and {Cheetham}, A. and {Claudi}, R.~U. and {Cudel}, M. and {Daemgen}, S. and {De Caprio}, V. and {Delorme}, P. and {Fantinel}, D. and {Farisato}, G. and {Feldt}, M. and {Galicher}, R. and {Ginski}, C. and {Girard}, J. and {Giro}, E. and {Janson}, M. and {Hagelberg}, J. and {Henning}, T. and {Incorvaia}, S. and {Kasper}, M. and {Kopytova}, T. and {LeCoroller}, H. and {Lessio}, L. and {Ligi}, R. and {Maire}, A.~L. and {M{\'e}nard}, F. and {Meyer}, M. and {Milli}, J. and {Mouillet}, D. and {Peretti}, S. and {Perrot}, C. and {Rouan}, D. and {Samland}, M. and {Salasnich}, B. and {Salter}, G. and {Schmidt}, T. and {Scuderi}, S. and {Sezestre}, E. and {Turatto}, M. and {Udry}, S. and {Wildi}, F. and {Zurlo}, A.},
 doi = {10.1051/0004-6361/201731426},
 eid = {A43},
 eprint = {1710.03019},
 journal = {\aap},
 month = {March},
 pages = {A43},
 primaryclass = {astro-ph.EP},
 title = {{Dynamical models to explain observations with SPHERE in planetary systems with double debris belts}},
 volume = {611},
 year = {2018}
}

@article{Lebreton_2012,
 adsurl = {https://ui.adsabs.harvard.edu/abs/2012A&A...539A..17L},
 archiveprefix = {arXiv},
 author = {{Lebreton}, J. and {Augereau}, J.-C. and {Thi}, W.-F. and {Roberge}, A. and {Donaldson}, J. and {Schneider}, G. and {Maddison}, S.~T. and {M{\'e}nard}, F. and {Riviere-Marichalar}, P. and {Mathews}, G.~S. and {Kamp}, I. and {Pinte}, C. and {Dent}, W.~R.~F. and {Barrado}, D. and {Duch{\^e}ne}, G. and {Gonzalez}, J.-F. and {Grady}, C.~A. and {Meeus}, G. and {Pantin}, E. and {Williams}, J.~P. and {Woitke}, P.},
 doi = {10.1051/0004-6361/201117714},
 eid = {A17},
 eprint = {1112.3398},
 journal = {\aap},
 month = {March},
 pages = {A17},
 primaryclass = {astro-ph.EP},
 title = {{An icy Kuiper belt around the young solar-type star HD 181327}},
 volume = {539},
 year = {2012}
}

@article{Lee_2016,
 adsurl = {https://ui.adsabs.harvard.edu/abs/2016ApJ...827..125L},
 archiveprefix = {arXiv},
 author = {{Lee}, Eve J. and {Chiang}, Eugene},
 doi = {10.3847/0004-637X/827/2/125},
 eid = {125},
 eprint = {1605.06118},
 journal = {\apj},
 month = {August},
 number = {2},
 pages = {125},
 primaryclass = {astro-ph.EP},
 title = {{A Primer on Unifying Debris Disk Morphologies}},
 volume = {827},
 year = {2016}
}

@inproceedings{Lenzen_2003,
 adsurl = {https://ui.adsabs.harvard.edu/abs/2003SPIE.4841..944L},
 author = {{Lenzen}, Rainer and {Hartung}, Markus and {Brandner}, Wolfgang and {Finger}, Gert and {Hubin}, Norbert N. and {Lacombe}, Francois and {Lagrange}, Anne-Marie and {Lehnert}, Matthew D. and {Moorwood}, Alan F.~M. and {Mouillet}, David},
 booktitle = {Instrument Design and Performance for Optical/Infrared Ground-based Telescopes},
 doi = {10.1117/12.460044},
 editor = {{Iye}, Masanori and {Moorwood}, Alan F.~M.},
 month = {March},
 pages = {944-952},
 series = {Society of Photo-Optical Instrumentation Engineers (SPIE) Conference Series},
 title = {{NAOS-CONICA first on sky results in a variety of observing modes}},
 volume = {4841},
 year = {2003}
}

@article{Lombart_2020,
 adsurl = {https://ui.adsabs.harvard.edu/abs/2020A&A...639A..54L},
 archiveprefix = {arXiv},
 author = {{Lombart}, M. and {Chauvin}, G. and {Rojo}, P. and {Lagadec}, E. and {Delorme}, P. and {Beust}, H. and {Bonnefoy}, M. and {Galicher}, R. and {Gratton}, R. and {Mesa}, D. and {Bonavita}, M. and {Allard}, F. and {Bayo}, A. and {Boccaletti}, A. and {Desidera}, S. and {Girard}, J. and {Jenkins}, J.~S. and {Klahr}, H. and {Laibe}, G. and {Lagrange}, A.-M. and {Lazzoni}, C. and {Marleau}, G.-D. and {Minniti}, D. and {Mordasini}, C.},
 doi = {10.1051/0004-6361/202037961},
 eid = {A54},
 eprint = {2005.08850},
 journal = {\aap},
 month = {July},
 pages = {A54},
 primaryclass = {astro-ph.EP},
 title = {{VLT/SPHERE survey for exoplanets around young early-type stars, including systems with multi-belt architectures}},
 volume = {639},
 year = {2020}
}

@article{Macintosh_2014,
 adsurl = {https://ui.adsabs.harvard.edu/abs/2014PNAS..11112661M},
 archiveprefix = {arXiv},
 author = {{Macintosh}, Bruce and {Graham}, James R. and {Ingraham}, Patrick and {Konopacky}, Quinn and {Marois}, Christian and {Perrin}, Marshall and {Poyneer}, Lisa and {Bauman}, Brian and {Barman}, Travis and {Burrows}, Adam S. and {Cardwell}, Andrew and {Chilcote}, Jeffrey and {De Rosa}, Robert J. and {Dillon}, Daren and {Doyon}, Rene and {Dunn}, Jennifer and {Erikson}, Darren and {Fitzgerald}, Michael P. and {Gavel}, Donald and {Goodsell}, Stephen and {Hartung}, Markus and {Hibon}, Pascale and {Kalas}, Paul and {Larkin}, James and {Maire}, Jerome and {Marchis}, Franck and {Marley}, Mark S. and {McBride}, James and {Millar-Blanchaer}, Max and {Morzinski}, Katie and {Norton}, Andrew and {Oppenheimer}, B.~R. and {Palmer}, David and {Patience}, Jennifer and {Pueyo}, Laurent and {Rantakyro}, Fredrik and {Sadakuni}, Naru and {Saddlemyer}, Leslie and {Savransky}, Dmitry and {Serio}, Andrew and {Soummer}, Remi and {Sivaramakrishnan}, Anand and {Song}, Inseok and {Thomas}, Sandrine and {Wallace}, J. Kent and {Wiktorowicz}, Sloane and {Wolff}, Schuyler},
 doi = {10.1073/pnas.1304215111},
 eprint = {1403.7520},
 journal = {PNAS},
 month = {September},
 number = {35},
 pages = {12661-12666},
 primaryclass = {astro-ph.EP},
 title = {{First light of the Gemini Planet Imager}},
 volume = {111},
 year = {2014}
}

@inproceedings{Maire_2010,
 adsurl = {https://ui.adsabs.harvard.edu/abs/2010SPIE.7735E..31M},
 author = {{Maire}, J{\'e}r{\^o}me and {Perrin}, Marshall D. and {Doyon}, Ren{\'e} and {Artigau}, Etienne and {Dunn}, Jennifer and {Gavel}, Donald T. and {Graham}, James R. and {Lafreni{\`e}re}, David and {Larkin}, James E. and {Lavigne}, Jean-Fran{\c{c}}ois and {Macintosh}, Bruce A. and {Marois}, Christian and {Oppenheimer}, Ben and {Palmer}, David W. and {Poyneer}, Lisa A. and {Thibault}, Simon and {V{\'e}ran}, Jean-Pierre},
 booktitle = {Ground-based and Airborne Instrumentation for Astronomy III},
 doi = {10.1117/12.858028},
 editor = {{McLean}, Ian S. and {Ramsay}, Suzanne K. and {Takami}, Hideki},
 eid = {773531},
 month = {July},
 pages = {773531},
 series = {Society of Photo-Optical Instrumentation Engineers (SPIE) Conference Series},
 title = {{Data reduction pipeline for the Gemini Planet Imager}},
 volume = {7735},
 year = {2010}
}

@inproceedings{Maire_2012,
 adsurl = {https://ui.adsabs.harvard.edu/abs/2012SPIE.8451E..3GM},
 author = {{Maire}, J{\'e}r{\^o}me and {Perrin}, Marshall D. and {Doyon}, Ren{\'e} and {Chilcote}, Jeffrey and {Larkin}, James E. and {Weiss}, Jason L. and {Marois}, Christian and {Konopacky}, Quinn M. and {Millar-Blanchaer}, Maxwell and {Graham}, James R. and {Dunn}, Jennifer and {Galicher}, Raphael and {Marchis}, Franck and {Wiktorowicz}, Sloane J. and {Labrie}, Kathleen and {Thomas}, Sandrine J. and {Goodsell}, Stephen J. and {Rantakyro}, Fredrik T. and {Palmer}, David W. and {Macintosh}, Bruce A.},
 booktitle = {Software and Cyberinfrastructure for Astronomy II},
 doi = {10.1117/12.926202},
 editor = {{Radziwill}, Nicole M. and {Chiozzi}, Gianluca},
 eid = {84513G},
 month = {September},
 pages = {84513G},
 series = {Society of Photo-Optical Instrumentation Engineers (SPIE) Conference Series},
 title = {{Test results for the Gemini Planet Imager data reduction pipeline}},
 volume = {8451},
 year = {2012}
}

@article{Malin2024,
 adsurl = {https://ui.adsabs.harvard.edu/abs/2024A&A...690A.316M},
 archiveprefix = {arXiv},
 author = {{M{\^a}lin}, Mathilde and {Boccaletti}, Anthony and {Perrot}, Cl{\'e}ment and {Baudoz}, Pierre and {Rouan}, Daniel and {Lagage}, Pierre-Olivier and {Waters}, Rens and {G{\"u}del}, Manuel and {Henning}, Thomas and {Vandenbussche}, Bart and {Absil}, Olivier and {Barrado}, David and {Cossou}, Christophe and {Decin}, Leen and {Glauser}, Adrian M. and {Pye}, John and {Olofsson}, Goran and {Glasse}, Alistair and {Lahuis}, Fred and {Patapis}, Polychronis and {Royer}, Pierre and {Scheithauer}, Silvia and {Whiteford}, Niall and {Serabyn}, Eugene and {Choquet}, Elodie and {Colina}, Luis and {Ostlin}, G{\"o}ran and {Ray}, Tom P. and {Wright}, Gillian},
 doi = {10.1051/0004-6361/202450470},
 eid = {A316},
 eprint = {2408.16843},
 journal = {\aap},
 month = {October},
 pages = {A316},
 primaryclass = {astro-ph.EP},
 title = {{Unveiling the HD 95086 system at mid-infrared wavelengths with JWST/MIRI}},
 volume = {690},
 year = {2024}
}

@article{Mamajek_2008,
 adsurl = {https://ui.adsabs.harvard.edu/abs/2008ApJ...687.1264M},
 archiveprefix = {arXiv},
 author = {{Mamajek}, Eric E. and {Hillenbrand}, Lynne A.},
 doi = {10.1086/591785},
 eprint = {0807.1686},
 journal = {\apj},
 month = {November},
 number = {2},
 pages = {1264-1293},
 primaryclass = {astro-ph},
 title = {{Improved Age Estimation for Solar-Type Dwarfs Using Activity-Rotation Diagnostics}},
 volume = {687},
 year = {2008}
}

@article{Mannings_1998,
 adsurl = {https://ui.adsabs.harvard.edu/abs/1998ApJ...497..330M},
 author = {{Mannings}, Vincent and {Barlow}, Michael J.},
 doi = {10.1086/305432},
 journal = {\apj},
 month = {April},
 number = {1},
 pages = {330-341},
 title = {{Candidate Main-Sequence Stars with Debris Disks: A New Sample of Vega-like Sources}},
 volume = {497},
 year = {1998}
}

@article{Marino_2016,
 adsurl = {https://ui.adsabs.harvard.edu/abs/2016MNRAS.460.2933M},
 archiveprefix = {arXiv},
 author = {{Marino}, S. and {Matr{\`a}}, L. and {Stark}, C. and {Wyatt}, M.~C. and {Casassus}, S. and {Kennedy}, G. and {Rodriguez}, D. and {Zuckerman}, B. and {Perez}, S. and {Dent}, W.~R.~F. and {Kuchner}, M. and {Hughes}, A.~M. and {Schneider}, G. and {Steele}, A. and {Roberge}, A. and {Donaldson}, J. and {Nesvold}, E.},
 doi = {10.1093/mnras/stw1216},
 eprint = {1605.05331},
 journal = {\mnras},
 month = {August},
 number = {3},
 pages = {2933-2944},
 primaryclass = {astro-ph.EP},
 title = {{Exocometary gas in the HD 181327 debris ring}},
 volume = {460},
 year = {2016}
}

@article{Marino_2026_121617,
 adsurl = {https://ui.adsabs.harvard.edu/abs/2026A&A...705A.202M},
 archiveprefix = {arXiv},
 author = {{Marino}, S. and {Gupta}, V. and {Weber}, P. and {Pearce}, T.~D. and {Brennan}, A. and {P{\'e}rez}, S. and {Mac Manamon}, S. and {Matr{\`a}}, L. and {Milli}, J. and {Booth}, M. and {del Burgo}, C. and {Cataldi}, G. and {Chiang}, E. and {Han}, Y. and {Henning}, Th. and {Hughes}, A.~M. and {Jankovic}, M.~R. and {K{\'o}sp{\'a}l}, {\'A}. and {Lovell}, J.~B. and {Luppe}, P. and {Mansell}, E. and {MacGregor}, M.~A. and {Mo{\'o}r}, A. and {Olofsson}, J. and {Sefilian}, A.~A. and {Wilner}, D.~J. and {Wyatt}, M.~C. and {Zawadzki}, B.},
 doi = {10.1051/0004-6361/202556493},
 eid = {A202},
 eprint = {2601.11711},
 journal = {\aap},
 month = {January},
 pages = {A202},
 primaryclass = {astro-ph.EP},
 title = {{The ALMA survey to Resolve exoKuiper belt Substructures (ARKS): VIII. A dust arc and non-Keplerian gas kinematics in HD 121617}},
 volume = {705},
 year = {2026}
}

@article{Marino_2026_ARKSOverview,
 adsurl = {https://ui.adsabs.harvard.edu/abs/2026A&A...705A.195M},
 archiveprefix = {arXiv},
 author = {{Marino}, S. and {Matr{\`a}}, L. and {Hughes}, A.~M. and {Ehrhardt}, J. and {Kennedy}, G.~M. and {del Burgo}, C. and {Brennan}, A. and {Han}, Y. and {Jankovic}, M.~R. and {Lovell}, J.~B. and {Mac Manamon}, S. and {Milli}, J. and {Weber}, P. and {Zawadzki}, B. and {Bendahan-West}, R. and {Fehr}, A. and {Mansell}, E. and {Olofsson}, J. and {Pearce}, T.~D. and {Bayo}, A. and {Matthews}, B.~C. and {L{\"o}hne}, T. and {Wyatt}, M.~C. and {{\'A}brah{\'a}m}, P. and {Bonduelle}, M. and {Booth}, M. and {Cataldi}, G. and {Carpenter}, J.~M. and {Chiang}, E. and {Ertel}, S. and {Hales}, A.~S. and {Henning}, Th. and {K{\'o}sp{\'a}l}, {\'A}. and {Krivov}, A.~V. and {Luppe}, P. and {MacGregor}, M.~A. and {Marshall}, J.~P. and {Mo{\'o}r}, A. and {P{\'e}rez}, S. and {Sefilian}, A.~A. and {Sepulveda}, A.~G. and {Wilner}, D.~J.},
 doi = {10.1051/0004-6361/202556489},
 eid = {A195},
 eprint = {2601.11708},
 journal = {\aap},
 month = {January},
 pages = {A195},
 primaryclass = {astro-ph.EP},
 sortyear = {2026-1},
 title = {{The ALMA survey to Resolve exoKuiper belt Substructures (ARKS): I. Motivation, sample, data reduction, and results overview}},
 volume = {705},
 year = {2026}
}

@article{Marois_2006,
 adsurl = {https://ui.adsabs.harvard.edu/abs/2006ApJ...641..556M},
 archiveprefix = {arXiv},
 author = {{Marois}, Christian and {Lafreni{\`e}re}, David and {Doyon}, Ren{\'e} and {Macintosh}, Bruce and {Nadeau}, Daniel},
 doi = {10.1086/500401},
 eprint = {astro-ph/0512335},
 journal = {\apj},
 month = {April},
 number = {1},
 pages = {556-564},
 primaryclass = {astro-ph},
 title = {{Angular Differential Imaging: A Powerful High-Contrast Imaging Technique}},
 volume = {641},
 year = {2006}
}

@article{Marois_2008,
 adsurl = {https://ui.adsabs.harvard.edu/abs/2008Sci...322.1348M},
 archiveprefix = {arXiv},
 author = {{Marois}, Christian and {Macintosh}, Bruce and {Barman}, Travis and {Zuckerman}, B. and {Song}, Inseok and {Patience}, Jennifer and {Lafreni{\`e}re}, David and {Doyon}, Ren{\'e}},
 doi = {10.1126/science.1166585},
 eprint = {0811.2606},
 journal = {Science},
 month = {November},
 number = {5906},
 pages = {1348},
 primaryclass = {astro-ph},
 title = {{Direct Imaging of Multiple Planets Orbiting the Star HR 8799}},
 volume = {322},
 year = {2008}
}

@article{Marois_2010,
 adsurl = {https://ui.adsabs.harvard.edu/abs/2010Natur.468.1080M},
 archiveprefix = {arXiv},
 author = {{Marois}, Christian and {Zuckerman}, B. and {Konopacky}, Quinn M. and {Macintosh}, Bruce and {Barman}, Travis},
 doi = {10.1038/nature09684},
 eprint = {1011.4918},
 journal = {\nat},
 month = {December},
 number = {7327},
 pages = {1080-1083},
 primaryclass = {astro-ph.EP},
 title = {{Images of a fourth planet orbiting HR 8799}},
 volume = {468},
 year = {2010}
}

@article{Marshall_2011,
 adsurl = {https://ui.adsabs.harvard.edu/abs/2011A&A...529A.117M},
 archiveprefix = {arXiv},
 author = {{Marshall}, J.~P. and {L{\"o}hne}, T. and {Montesinos}, B. and {Krivov}, A.~V. and {Eiroa}, C. and {Absil}, O. and {Bryden}, G. and {Maldonado}, J. and {Mora}, A. and {Sanz-Forcada}, J. and {Ardila}, D. and {Augereau}, J.-Ch. and {Bayo}, A. and {Del Burgo}, C. and {Danchi}, W. and {Ertel}, S. and {Fedele}, D. and {Fridlund}, M. and {Lebreton}, J. and {Gonz{\'a}lez-Garc{\'\i}a}, B.~M. and {Liseau}, R. and {Meeus}, G. and {M{\"u}ller}, S. and {Pilbratt}, G.~L. and {Roberge}, A. and {Stapelfeldt}, K. and {Th{\'e}bault}, P. and {White}, G.~J. and {Wolf}, S.},
 doi = {10.1051/0004-6361/201116673},
 eid = {A117},
 eprint = {1104.3560},
 journal = {\aap},
 month = {May},
 pages = {A117},
 primaryclass = {astro-ph.SR},
 title = {{A Herschel resolved far-infrared dust ring around HD 207129}},
 volume = {529},
 year = {2011}
}

@article{Marshall_2021,
 adsurl = {https://ui.adsabs.harvard.edu/abs/2021MNRAS.501.6168M},
 author = {{Marshall}, J.~P. and {Wang}, L. and {Kennedy}, G.~M. and {Zeegers}, S.~T. and {Scicluna}, P.},
 doi = {10.1093/mnras/staa3917},
 journal = {\mnras},
 month = {March},
 number = {4},
 pages = {6168-6180},
 title = {{A search for trends in spatially resolved debris discs at far-infrared wavelengths}},
 volume = {501},
 year = {2021}
}

@article{Marshall_2026,
 adsurl = {https://ui.adsabs.harvard.edu/abs/2026MNRAS.545f1990M},
 archiveprefix = {arXiv},
 author = {{Marshall}, Jonathan P. and {Mu{\~n}oz-Guti{\'e}rrez}, Marco A. and {Sefilian}, Antranik A. and {Peimbert}, A.},
 doi = {10.1093/mnras/staf1990},
 eid = {staf1990},
 eprint = {2509.22822},
 journal = {\mnras},
 month = {January},
 number = {1},
 pages = {staf1990},
 primaryclass = {astro-ph.EP},
 title = {{Testing the impact of planet-stirring, self-stirring, and mixed-stirring on debris disc architecture: a case study of HD 16743}},
 volume = {545},
 year = {2026}
}

@article{Matthews_2017,
 adsurl = {https://ui.adsabs.harvard.edu/abs/2017ApJ...843L..12M},
 archiveprefix = {arXiv},
 author = {{Matthews}, Elisabeth and {Hinkley}, Sasha and {Vigan}, Arthur and {Kennedy}, Grant and {Rizzuto}, Aaron and {Stapelfeldt}, Karl and {Mawet}, Dimitri and {Booth}, Mark and {Chen}, Christine and {Jang-Condell}, Hannah},
 doi = {10.3847/2041-8213/aa7943},
 eid = {L12},
 eprint = {1706.04624},
 journal = {\apjl},
 month = {July},
 number = {1},
 pages = {L12},
 primaryclass = {astro-ph.EP},
 title = {{The First Scattered-light Image of the Debris Disk around the Sco-Cen Target HD 129590}},
 volume = {843},
 year = {2017}
}

@article{Matthews_2018,
 adsurl = {https://ui.adsabs.harvard.edu/abs/2018MNRAS.480.2757M},
 archiveprefix = {arXiv},
 author = {{Matthews}, Elisabeth and {Hinkley}, Sasha and {Vigan}, Arthur and {Kennedy}, Grant and {Sutlieff}, Ben and {Wickenden}, Dawn and {Treves}, Sam and {David}, Trevor and {Meshkat}, Tiffany and {Mawet}, Dimitri and {Morales}, Farisa and {Shannon}, Andrew and {Stapelfeldt}, Karl},
 doi = {10.1093/mnras/sty1778},
 eprint = {1807.01326},
 journal = {\mnras},
 month = {October},
 number = {2},
 pages = {2757-2783},
 primaryclass = {astro-ph.EP},
 title = {{Constraining the presence of giant planets in two-belt debris disc systems with VLT/SPHERE direct imaging and dynamical arguments}},
 volume = {480},
 year = {2018}
}

@article{Mawet_2005,
 adsurl = {https://ui.adsabs.harvard.edu/abs/2005ApJ...633.1191M},
 author = {{Mawet}, D. and {Riaud}, P. and {Absil}, O. and {Surdej}, J.},
 doi = {10.1086/462409},
 journal = {\apj},
 month = {November},
 number = {2},
 pages = {1191-1200},
 title = {{Annular Groove Phase Mask Coronagraph}},
 volume = {633},
 year = {2005}
}

@article{Mawet_2014,
 adsurl = {https://ui.adsabs.harvard.edu/abs/2014ApJ...792...97M},
 archiveprefix = {arXiv},
 author = {{Mawet}, D. and {Milli}, J. and {Wahhaj}, Z. and {Pelat}, D. and {Absil}, O. and {Delacroix}, C. and {Boccaletti}, A. and {Kasper}, M. and {Kenworthy}, M. and {Marois}, C. and {Mennesson}, B. and {Pueyo}, L.},
 doi = {10.1088/0004-637X/792/2/97},
 eid = {97},
 eprint = {1407.2247},
 journal = {\apj},
 month = {September},
 number = {2},
 pages = {97},
 primaryclass = {astro-ph.IM},
 title = {{Fundamental Limitations of High Contrast Imaging Set by Small Sample Statistics}},
 volume = {792},
 year = {2014}
}

@article{Mawet_2019,
 adsurl = {https://ui.adsabs.harvard.edu/abs/2019AJ....157...33M},
 archiveprefix = {arXiv},
 author = {{Mawet}, Dimitri and {Hirsch}, Lea and {Lee}, Eve J. and {Ruffio}, Jean-Baptiste and {Bottom}, Michael and {Fulton}, Benjamin J. and {Absil}, Olivier and {Beichman}, Charles and {Bowler}, Brendan and {Bryan}, Marta and {Choquet}, Elodie and {Ciardi}, David and {Christiaens}, Valentin and {Defr{\`e}re}, Denis and {Gomez Gonzalez}, Carlos Alberto and {Howard}, Andrew W. and {Huby}, Elsa and {Isaacson}, Howard and {Jensen-Clem}, Rebecca and {Kosiarek}, Molly and {Marcy}, Geoff and {Meshkat}, Tiffany and {Petigura}, Erik and {Reggiani}, Maddalena and {Ruane}, Garreth and {Serabyn}, Eugene and {Sinukoff}, Evan and {Wang}, Ji and {Weiss}, Lauren and {Ygouf}, Marie},
 doi = {10.3847/1538-3881/aaef8a},
 eid = {33},
 eprint = {1810.03794},
 journal = {\aj},
 month = {January},
 number = {1},
 pages = {33},
 primaryclass = {astro-ph.EP},
 title = {{Deep Exploration of {\ensuremath{\in}} Eridani with Keck Ms-band Vortex Coronagraphy and Radial Velocities: Mass and Orbital Parameters of the Giant Exoplanet}},
 volume = {157},
 year = {2019}
}

@article{Mesa_2023,
 adsurl = {https://ui.adsabs.harvard.edu/abs/2023A&A...672A..93M},
 archiveprefix = {arXiv},
 author = {{Mesa}, D. and {Gratton}, R. and {Kervella}, P. and {Bonavita}, M. and {Desidera}, S. and {D'Orazi}, V. and {Marino}, S. and {Zurlo}, A. and {Rigliaco}, E.},
 doi = {10.1051/0004-6361/202345865},
 eid = {A93},
 eprint = {2302.06213},
 journal = {\aap},
 month = {April},
 pages = {A93},
 primaryclass = {astro-ph.EP},
 title = {{AF Lep b: The lowest-mass planet detected by coupling astrometric and direct imaging data}},
 volume = {672},
 year = {2023}
}

@article{Meshkat_2017,
 adsurl = {https://ui.adsabs.harvard.edu/abs/2017AJ....154..245M},
 archiveprefix = {arXiv},
 author = {{Meshkat}, Tiffany and {Mawet}, Dimitri and {Bryan}, Marta L. and {Hinkley}, Sasha and {Bowler}, Brendan P. and {Stapelfeldt}, Karl R. and {Batygin}, Konstantin and {Padgett}, Deborah and {Morales}, Farisa Y. and {Serabyn}, Eugene and {Christiaens}, Valentin and {Brandt}, Timothy D. and {Wahhaj}, Zahed},
 doi = {10.3847/1538-3881/aa8e9a},
 eid = {245},
 eprint = {1710.04185},
 journal = {\aj},
 month = {December},
 number = {6},
 pages = {245},
 primaryclass = {astro-ph.EP},
 title = {{A Direct Imaging Survey of Spitzer-detected Debris Disks: Occurrence of Giant Planets in Dusty Systems}},
 volume = {154},
 year = {2017}
}

@article{Milli_2024,
 adsurl = {https://ui.adsabs.harvard.edu/abs/2024A&A...683A..22M},
 author = {{Milli}, J. and {Choquet}, E. and {Tazaki}, R. and {M{\'e}nard}, F. and {Augereau}, J.-C. and {Olofsson}, J. and {Th{\'e}bault}, P. and {Poch}, O. and {Levasseur-Regourd}, A.-C. and {Lasue}, J. and {Renard}, J.~B. and {Hadamcik}, E. and {Baruteau}, C. and {Schmid}, H.~M. and {Engler}, N. and {van Holstein}, R.~G. and {Zubko}, E. and {Lagrange}, A.~M. and {Marino}, S. and {Pinte}, C. and {Dominik}, C. and {Boccaletti}, A. and {Langlois}, M. and {Zurlo}, A. and {Desgrange}, C. and {Gluck}, L. and {Mouillet}, D. and {Costille}, A. and {Sauvage}, J.~F.},
 doi = {10.1051/0004-6361/202347933},
 eid = {A22},
 journal = {\aap},
 month = {March},
 pages = {A22},
 title = {{The polarisation properties of the HD 181327 debris ring. Evidence for sub-micron particles from scattered light observations}},
 volume = {683},
 year = {2024}
}

@article{Milli_2026,
 adsurl = {https://ui.adsabs.harvard.edu/abs/2026A&A...705A.199M},
 archiveprefix = {arXiv},
 author = {{Milli}, J. and {Olofsson}, J. and {Bonduelle}, M. and {Bendahan-West}, R. and {Marshall}, J.~P. and {Choquet}, E. and {Sefilian}, A.~A. and {Han}, Y. and {Zawadzki}, B. and {Mac Manamon}, S. and {Mansell}, E. and {del Burgo}, C. and {Carpenter}, J.~M. and {Hughes}, A.~M. and {Booth}, M. and {Chiang}, E. and {Ertel}, S. and {Esposito}, Th. M. and {Henning}, Th. and {Hom}, J. and {Jankovic}, M.~R. and {Krivov}, A.~V. and {Lovell}, J.~B. and {Luppe}, P. and {MacGregor}, M.~A. and {Marino}, S. and {Matthews}, B.~C. and {Matr{\`a}}, L. and {Mo{\'o}r}, A. and {Pawellek}, N. and {Pearce}, T.~D. and {P{\'e}rez}, S. and {Squicciarini}, V. and {Weber}, P. and {Wilner}, D.~J. and {Wyatt}, M.~C.},
 doi = {10.1051/0004-6361/202556523},
 eid = {A199},
 eprint = {2601.12586},
 journal = {\aap},
 month = {January},
 pages = {A199},
 primaryclass = {astro-ph.EP},
 title = {{The ALMA survey to Resolve exoKuiper belt Substructures (ARKS): V. Comparison between scattered light and thermal emission}},
 volume = {705},
 year = {2026}
}

@article{Moor_2017,
 adsurl = {https://ui.adsabs.harvard.edu/abs/2017ApJ...849..123M},
 archiveprefix = {arXiv},
 author = {{Mo{\'o}r}, Attila and {Cur{\'e}}, Michel and {K{\'o}sp{\'a}l}, {\'A}gnes and {{\'A}brah{\'a}m}, P{\'e}ter and {Csengeri}, Timea and {Eiroa}, Carlos and {Gunawan}, Diah and {Henning}, Thomas and {Hughes}, A. Meredith and {Juh{\'a}sz}, Attila and {Pawellek}, Nicole and {Wyatt}, Mark},
 doi = {10.3847/1538-4357/aa8e4e},
 eid = {123},
 eprint = {1709.08414},
 journal = {\apj},
 month = {November},
 number = {2},
 pages = {123},
 primaryclass = {astro-ph.SR},
 title = {{Molecular Gas in Debris Disks around Young A-type Stars}},
 volume = {849},
 year = {2017}
}

@misc{Moshir_1989,
 adsurl = {https://ui.adsabs.harvard.edu/abs/1989ifss.book.....M},
 author = {{Moshir}, M.},
 title = {{IRAS Faint Source Survey, Explanatory supplement version 1 and tape}},
 year = {1989}
}

@article{Mouillet_1997,
 adsurl = {https://ui.adsabs.harvard.edu/abs/1997MNRAS.292..896M},
 archiveprefix = {arXiv},
 author = {{Mouillet}, D. and {Larwood}, J.~D. and {Papaloizou}, J.~C.~B. and {Lagrange}, A.~M.},
 doi = {10.1093/mnras/292.4.896},
 eprint = {astro-ph/9705100},
 journal = {\mnras},
 month = {December},
 number = {4},
 pages = {896-904},
 primaryclass = {astro-ph},
 title = {{A planet on an inclined orbit as an explanation of the warp in the beta Pictoris disc}},
 volume = {292},
 year = {1997}
}

@article{MussoBarcucci_2019,
 adsurl = {https://ui.adsabs.harvard.edu/abs/2019A&A...627A..77M},
 archiveprefix = {arXiv},
 author = {{Musso Barcucci}, Arianna and {Launhardt}, Ralf and {Kennedy}, Grant M. and {Avenhaus}, Henning and {Brems}, Stefan S. and {van Boekel}, Roy and {Cantalloube}, F. and {Cheetham}, Anthony and {Cugno}, Gabriele and {Girard}, Julien and {Godoy}, Nicol{\'a}s and {Henning}, Thomas K. and {Metchev}, Stanimir and {M{\"u}ller}, Andr{\'e} and {Olofsson}, Johan and {Pepe}, Francesco and {Quanz}, Sascha P. and {Quirrenbach}, Andreas and {Reffert}, Sabine and {Rickman}, Emily L. and {Samland}, Matthias and {Segransan}, Damien},
 doi = {10.1051/0004-6361/201935146},
 eid = {A77},
 eprint = {1906.01391},
 journal = {\aap},
 month = {July},
 pages = {A77},
 primaryclass = {astro-ph.EP},
 title = {{ISPY - NaCo Imaging Survey for Planets around Young stars. Discovery of an M dwarf in the gap between HD 193571 and its debris ring}},
 volume = {627},
 year = {2019}
}

@article{MussoBarcucci_2021,
 adsurl = {https://ui.adsabs.harvard.edu/abs/2021A&A...645A..88M},
 author = {{Musso Barcucci}, Arianna and {Launhardt}, Ralf and {M{\"u}ller}, Andr{\'e} and {Kennedy}, Grant M. and {van Boekel}, Roy and {Henning}, Thomas and {Ruh}, Henrik L. and {Marino}, Sebastian and {Pearce}, Tim D. and {Brems}, Stefan S. and {Ertel}, Steve and {Spalding}, Eckhart A.},
 doi = {10.1051/0004-6361/202039541},
 eid = {A88},
 journal = {\aap},
 month = {January},
 pages = {A88},
 title = {{LIStEN: L' band Imaging Survey for Exoplanets in the North}},
 volume = {645},
 year = {2021}
}

@article{Mustill_2009,
 adsurl = {https://ui.adsabs.harvard.edu/abs/2009MNRAS.399.1403M},
 archiveprefix = {arXiv},
 author = {{Mustill}, Alexander J. and {Wyatt}, Mark C.},
 doi = {10.1111/j.1365-2966.2009.15360.x},
 eprint = {0907.1389},
 journal = {\mnras},
 month = {November},
 number = {3},
 pages = {1403-1414},
 primaryclass = {astro-ph.EP},
 title = {{Debris disc stirring by secular perturbations from giant planets}},
 volume = {399},
 year = {2009}
}

@article{nelder1965simplex,
 author = {Nelder, John A and Mead, Roger},
 doi = {10.1093/comjnl/7.4.308},
 journal = {The Computer Journal},
 number = {4},
 pages = {308--313},
 publisher = {Oxford University Press},
 title = {A simplex method for function minimization},
 volume = {7},
 year = {1965}
}

@article{Nidever_2002,
 adsurl = {https://ui.adsabs.harvard.edu/abs/2002ApJS..141..503N},
 archiveprefix = {arXiv},
 author = {{Nidever}, David L. and {Marcy}, Geoffrey W. and {Butler}, R. Paul and {Fischer}, Debra A. and {Vogt}, Steven S.},
 doi = {10.1086/340570},
 eprint = {astro-ph/0112477},
 journal = {\apjs},
 month = {August},
 number = {2},
 pages = {503-522},
 primaryclass = {astro-ph},
 title = {{Radial Velocities for 889 Late-Type Stars}},
 volume = {141},
 year = {2002}
}

@article{Nielsen_2017,
 adsurl = {https://ui.adsabs.harvard.edu/abs/2017AJ....154..218N},
 archiveprefix = {arXiv},
 author = {{Nielsen}, Eric L. and {Rosa}, Robert J. De and {Rameau}, Julien and {Wang}, Jason J. and {Esposito}, Thomas M. and {Millar-Blanchaer}, Maxwell A. and {Marois}, Christian and {Vigan}, Arthur and {Ammons}, S. Mark and {Artigau}, Etienne and {Bailey}, Vanessa P. and {Blunt}, Sarah and {Bulger}, Joanna and {Chilcote}, Jeffrey and {Cotten}, Tara and {Doyon}, Ren{\'e} and {Duch{\^e}ne}, Gaspard and {Fabrycky}, Daniel and {Fitzgerald}, Michael P. and {Follette}, Katherine B. and {Gerard}, Benjamin L. and {Goodsell}, Stephen J. and {Graham}, James R. and {Greenbaum}, Alexandra Z. and {Hibon}, Pascale and {Hinkley}, Sasha and {Hung}, Li-Wei and {Ingraham}, Patrick and {Jensen-Clem}, Rebecca and {Kalas}, Paul and {Konopacky}, Quinn and {Larkin}, James E. and {Macintosh}, Bruce and {Maire}, J{\'e}r{\^o}me and {Marchis}, Franck and {Metchev}, Stanimir and {Morzinski}, Katie M. and {Murray-Clay}, Ruth A. and {Oppenheimer}, Rebecca and {Palmer}, David and {Patience}, Jennifer and {Perrin}, Marshall and {Poyneer}, Lisa and {Pueyo}, Laurent and {Rafikov}, Roman R. and {Rajan}, Abhijith and {Rantakyr{\"o}}, Fredrik T. and {Ruffio}, Jean-Baptiste and {Savransky}, Dmitry and {Schneider}, Adam C. and {Sivaramakrishnan}, Anand and {Song}, Inseok and {Soummer}, Remi and {Thomas}, Sandrine and {Wallace}, J. Kent and {Ward-Duong}, Kimberly and {Wiktorowicz}, Sloane and {Wolff}, Schuyler},
 doi = {10.3847/1538-3881/aa8a69},
 eid = {218},
 eprint = {1705.06851},
 journal = {\aj},
 month = {December},
 number = {6},
 pages = {218},
 primaryclass = {astro-ph.EP},
 title = {{Evidence That the Directly Imaged Planet HD 131399 Ab Is a Background Star}},
 volume = {154},
 year = {2017}
}

@article{Nissen_2020,
 adsurl = {https://ui.adsabs.harvard.edu/abs/2020A&A...640A..81N},
 archiveprefix = {arXiv},
 author = {{Nissen}, P.~E. and {Christensen-Dalsgaard}, J. and {Mosumgaard}, J.~R. and {Silva Aguirre}, V. and {Spitoni}, E. and {Verma}, K.},
 doi = {10.1051/0004-6361/202038300},
 eid = {A81},
 eprint = {2006.06013},
 journal = {\aap},
 month = {August},
 pages = {A81},
 primaryclass = {astro-ph.SR},
 title = {{High-precision abundances of elements in solar-type stars. Evidence of two distinct sequences in abundance-age relations}},
 volume = {640},
 year = {2020}
}

@ARTICLE{Olofsson_2026,
author = {{Olofsson}, J. and {Jankovic}, M.~R. and {Marino}, S. and {Krivov}, A.~V. and {Bonduelle}, M. and {Cataldi}, G. and {Han}, Y. and {Hughes}, A.~M. and {L{\"o}hne}, T. and {Mac Manamon}, S. and {Mansell}, E. and {Matr{\`a}}, L. and {Milli}, J. and {Sefilian}, A.~A. and {Th{\'e}bault}, P. and {Zawadzki}, B. and {Booth}, M. and {del Burgo}, C. and {Carpenter}, J.~M. and {Henning}, Th. and {Lovell}, J.~B. and {Pearce}, T. and {P{\'e}rez}, S. and {Wilner}, D.~J. and {Wyatt}, M.~C.},
title = "{The ALMA survey to Resolve exoKuiper belt Substructures (ARKS): XI. Gas-dust interactions and radial offsets between micron and millimetre-sized grains}",
journal = {\aap},
 year = 2026,
month = jun,
volume = {710},
  eid = {A265},
pages = {A265},
  doi = {10.1051/0004-6361/202659039},
archivePrefix = {arXiv},
eprint = {2605.03009},
}

@article{Pascual_2016,
 adsurl = {https://ui.adsabs.harvard.edu/abs/2016A&A...586A...6P},
 archiveprefix = {arXiv},
 author = {{Pascual}, N. and {Montesinos}, B. and {Meeus}, G. and {Marshall}, J.~P. and {Mendigut{\'\i}a}, I. and {Sandell}, G.},
 doi = {10.1051/0004-6361/201526605},
 eid = {A6},
 eprint = {1511.08356},
 journal = {\aap},
 month = {February},
 pages = {A6},
 primaryclass = {astro-ph.SR},
 title = {{The far-infrared behaviour of Herbig Ae/Be discs: Herschel PACS photometry}},
 volume = {586},
 year = {2016}
}

@article{Paunzen_2022,
 adsurl = {https://ui.adsabs.harvard.edu/abs/2022A&A...661A..89P},
 archiveprefix = {arXiv},
 author = {{Paunzen}, E.},
 doi = {10.1051/0004-6361/202142355},
 eid = {A89},
 eprint = {2111.04810},
 journal = {\aap},
 month = {May},
 pages = {A89},
 primaryclass = {astro-ph.SR},
 title = {{Catalogue of stars measured in the Geneva seven-colour photometric system}},
 volume = {661},
 year = {2022}
}

@article{Pawellek_2021,
 adsurl = {https://ui.adsabs.harvard.edu/abs/2021MNRAS.502.5390P},
 archiveprefix = {arXiv},
 author = {{Pawellek}, Nicole and {Wyatt}, Mark and {Matr{\`a}}, Luca and {Kennedy}, Grant and {Yelverton}, Ben},
 doi = {10.1093/mnras/stab269},
 eprint = {2101.12049},
 journal = {\mnras},
 month = {April},
 number = {4},
 pages = {5390-5416},
 primaryclass = {astro-ph.EP},
 title = {{A {\ensuremath{\sim}}75 per cent occurrence rate of debris discs around F stars in the {\ensuremath{\beta}} Pic moving group}},
 volume = {502},
 year = {2021}
}

@article{Pearce_2014,
 adsurl = {https://ui.adsabs.harvard.edu/abs/2014MNRAS.443.2541P},
 archiveprefix = {arXiv},
 author = {{Pearce}, Tim D. and {Wyatt}, Mark C.},
 doi = {10.1093/mnras/stu1302},
 eprint = {1406.7294},
 journal = {\mnras},
 month = {September},
 number = {3},
 pages = {2541-2560},
 primaryclass = {astro-ph.EP},
 title = {{Dynamical evolution of an eccentric planet and a less massive debris disc}},
 volume = {443},
 year = {2014}
}

@article{Pearce_2015,
 adsurl = {https://ui.adsabs.harvard.edu/abs/2015MNRAS.453.3329P},
 archiveprefix = {arXiv},
 author = {{Pearce}, Tim D. and {Wyatt}, Mark C.},
 doi = {10.1093/mnras/stv1847},
 eprint = {1507.04367},
 journal = {\mnras},
 month = {November},
 number = {3},
 pages = {3329-3340},
 primaryclass = {astro-ph.EP},
 title = {{Double-ringed debris discs could be the work of eccentric planets: explaining the strange morphology of HD 107146}},
 volume = {453},
 year = {2015}
}

@article{Pearce_2022,
 adsurl = {https://ui.adsabs.harvard.edu/abs/2022A&A...659A.135P},
 archiveprefix = {arXiv},
 author = {{Pearce}, Tim D. and {Launhardt}, Ralf and {Ostermann}, Robert and {Kennedy}, Grant M. and {Gennaro}, Mario and {Booth}, Mark and {Krivov}, Alexander V. and {Cugno}, Gabriele and {Henning}, Thomas K. and {Quirrenbach}, Andreas and {Barcucci}, Arianna Musso and {Matthews}, Elisabeth C. and {Ruh}, Henrik L. and {Stone}, Jordan M.},
 doi = {10.1051/0004-6361/202142720},
 eid = {A135},
 eprint = {2201.08369},
 journal = {\aap},
 month = {March},
 pages = {A135},
 primaryclass = {astro-ph.EP},
 title = {{Planet populations inferred from debris discs. Insights from 178 debris systems in the ISPY, LEECH, and LIStEN planet-hunting surveys}},
 volume = {659},
 year = {2022}
}

@article{Pearce_2024,
 adsurl = {https://ui.adsabs.harvard.edu/abs/2024MNRAS.527.3876P},
 archiveprefix = {arXiv},
 author = {{Pearce}, Tim D. and {Krivov}, Alexander V. and {Sefilian}, Antranik A. and {Jankovic}, Marija R. and {L{\"o}hne}, Torsten and {Morgner}, Tobias and {Wyatt}, Mark C. and {Booth}, Mark and {Marino}, Sebastian},
 doi = {10.1093/mnras/stad3462},
 eprint = {2311.04265},
 journal = {\mnras},
 month = {January},
 number = {2},
 pages = {3876-3899},
 primaryclass = {astro-ph.EP},
 title = {{The effect of sculpting planets on the steepness of debris-disc inner edges}},
 volume = {527},
 year = {2024}
}

@article{Pearce_2025,
 adsurl = {https://ui.adsabs.harvard.edu/abs/2025MNRAS.544.1447P},
 archiveprefix = {arXiv},
 author = {{Pearce}, Tim D. and {L{\"o}hne}, Torsten and {Krivov}, Alexander V.},
 doi = {10.1093/mnras/staf1735},
 eprint = {2510.07187},
 journal = {\mnras},
 month = {December},
 number = {2},
 pages = {1447-1462},
 primaryclass = {astro-ph.EP},
 title = {{Fomalhaut's debris disc is not dominated by primordial Plutos}},
 volume = {544},
 year = {2025}
}

@article{Pecaut_2013,
 adsurl = {https://ui.adsabs.harvard.edu/abs/2013ApJS..208....9P},
 archiveprefix = {arXiv},
 author = {{Pecaut}, Mark J. and {Mamajek}, Eric E.},
 doi = {10.1088/0067-0049/208/1/9},
 eid = {9},
 eprint = {1307.2657},
 journal = {\apjs},
 month = {September},
 number = {1},
 pages = {9},
 primaryclass = {astro-ph.SR},
 title = {{Intrinsic Colors, Temperatures, and Bolometric Corrections of Pre-main-sequence Stars}},
 volume = {208},
 year = {2013}
}

@article{Pecaut_2016,
 adsurl = {https://ui.adsabs.harvard.edu/abs/2016MNRAS.461..794P},
 archiveprefix = {arXiv},
 author = {{Pecaut}, Mark J. and {Mamajek}, Eric E.},
 doi = {10.1093/mnras/stw1300},
 eprint = {1605.08789},
 journal = {\mnras},
 month = {September},
 number = {1},
 pages = {794-815},
 primaryclass = {astro-ph.SR},
 title = {{The star formation history and accretion-disc fraction among the K-type members of the Scorpius-Centaurus OB association}},
 volume = {461},
 year = {2016}
}

@inproceedings{Perrin_2014,
 adsurl = {https://ui.adsabs.harvard.edu/abs/2014SPIE.9147E..3JP},
 archiveprefix = {arXiv},
 author = {{Perrin}, Marshall D. and {Maire}, J{\'e}r{\^o}me and {Ingraham}, Patrick and {Savransky}, Dmitry and {Millar-Blanchaer}, Max and {Wolff}, Schuyler G. and {Ruffio}, Jean-Baptiste and {Wang}, Jason J. and {Draper}, Zachary H. and {Sadakuni}, Naru and {Marois}, Christian and {Rajan}, Abhijith and {Fitzgerald}, Michael P. and {Macintosh}, Bruce and {Graham}, James R. and {Doyon}, Ren{\'e} and {Larkin}, James E. and {Chilcote}, Jeffrey K. and {Goodsell}, Stephen J. and {Palmer}, David W. and {Labrie}, Kathleen and {Beaulieu}, Mathilde and {De Rosa}, Robert J. and {Greenbaum}, Alexandra Z. and {Hartung}, Markus and {Hibon}, Pascale and {Konopacky}, Quinn and {Lafreniere}, David and {Lavigne}, Jean-Francois and {Marchis}, Franck and {Patience}, Jenny and {Pueyo}, Laurent and {Rantakyr{\"o}}, Fredrik T. and {Soummer}, R{\'e}mi and {Sivaramakrishnan}, Anand and {Thomas}, Sandrine and {Ward-Duong}, Kimberly and {Wiktorowicz}, Sloane},
 booktitle = {Ground-based and Airborne Instrumentation for Astronomy V},
 doi = {10.1117/12.2055246},
 editor = {{Ramsay}, Suzanne K. and {McLean}, Ian S. and {Takami}, Hideki},
 eid = {91473J},
 eprint = {1407.2301},
 month = {July},
 pages = {91473J},
 primaryclass = {astro-ph.IM},
 series = {Society of Photo-Optical Instrumentation Engineers (SPIE) Conference Series},
 title = {{Gemini Planet Imager observational calibrations I: Overview of the GPI data reduction pipeline}},
 volume = {9147},
 year = {2014}
}

@article{Perrot_2023,
 adsurl = {https://ui.adsabs.harvard.edu/abs/2023A&A...673A..39P},
 archiveprefix = {arXiv},
 author = {{Perrot}, Cl{\'e}ment and {Olofsson}, Johan and {Kral}, Quentin and {Th{\'e}bault}, Philippe and {Montesinos}, Mat{\'\i}as and {Kennedy}, Grant and {Bayo}, Amelia and {Iglesias}, Daniela and {van Holstein}, Rob and {Pinte}, Christophe},
 doi = {10.1051/0004-6361/202244694},
 eid = {A39},
 eprint = {2302.07057},
 journal = {\aap},
 month = {May},
 pages = {A39},
 primaryclass = {astro-ph.EP},
 title = {{Morphology of the gas-rich debris disk around HD 121617 with SPHERE observations in polarized light}},
 volume = {673},
 year = {2023}
}

@article{Perryman_1997,
 adsurl = {https://ui.adsabs.harvard.edu/abs/1997A&A...323L..49P},
 author = {{Perryman}, M.~A.~C. and {Lindegren}, L. and {Kovalevsky}, J. and {Hoeg}, E. and {Bastian}, U. and {Bernacca}, P.~L. and {Cr{\'e}z{\'e}}, M. and {Donati}, F. and {Grenon}, M. and {Grewing}, M. and {van Leeuwen}, F. and {van der Marel}, H. and {Mignard}, F. and {Murray}, C.~A. and {Le Poole}, R.~S. and {Schrijver}, H. and {Turon}, C. and {Arenou}, F. and {Froeschl{\'e}}, M. and {Petersen}, C.~S.},
 journal = {\aap},
 month = {July},
 pages = {L49-L52},
 title = {{The HIPPARCOS Catalogue}},
 volume = {323},
 year = {1997}
}

@article{Phillips_2020,
 adsurl = {https://ui.adsabs.harvard.edu/abs/2020A&A...637A..38P},
 archiveprefix = {arXiv},
 author = {{Phillips}, M.~W. and {Tremblin}, P. and {Baraffe}, I. and {Chabrier}, G. and {Allard}, N.~F. and {Spiegelman}, F. and {Goyal}, J.~M. and {Drummond}, B. and {H{\'e}brard}, E.},
 doi = {10.1051/0004-6361/201937381},
 eid = {A38},
 eprint = {2003.13717},
 journal = {\aap},
 month = {May},
 pages = {A38},
 primaryclass = {astro-ph.SR},
 title = {{A new set of atmosphere and evolution models for cool T-Y brown dwarfs and giant exoplanets}},
 volume = {637},
 year = {2020}
}

@article{Poblete_2023,
 adsurl = {https://ui.adsabs.harvard.edu/abs/2023MNRAS.526.2017P},
 archiveprefix = {arXiv},
 author = {{Poblete}, Pedro P. and {L{\"o}hne}, Torsten and {Pearce}, Tim D. and {Sefilian}, Antranik A.},
 doi = {10.1093/mnras/stad2827},
 eprint = {2309.06994},
 journal = {\mnras},
 month = {December},
 number = {2},
 pages = {2017-2031},
 primaryclass = {astro-ph.EP},
 title = {{Self-gravity of debris discs can strongly change the outcomes of interactions with inclined planets}},
 volume = {526},
 year = {2023}
}

@article{Poblete_2025,
 adsurl = {https://ui.adsabs.harvard.edu/abs/2025A&A...700A.148P},
 archiveprefix = {arXiv},
 author = {{Poblete}, Pedro P. and {Pearce}, Tim D. and {Charalambous}, Carolina},
 doi = {10.1051/0004-6361/202554802},
 eid = {A148},
 eprint = {2507.01589},
 journal = {\aap},
 month = {August},
 pages = {A148},
 primaryclass = {astro-ph.EP},
 title = {{The HR 8799 debris disk: Shaped by planetary migration and a possible fifth outermost planet}},
 volume = {700},
 year = {2025}
}

@article{Poglitsch_2010,
 adsurl = {https://ui.adsabs.harvard.edu/abs/2010A&A...518L...2P},
 archiveprefix = {arXiv},
 author = {{Poglitsch}, A. and {Waelkens}, C. and {Geis}, N. and {Feuchtgruber}, H. and {Vandenbussche}, B. and {Rodriguez}, L. and {Krause}, O. and {Renotte}, E. and {van Hoof}, C. and {Saraceno}, P. and {Cepa}, J. and {Kerschbaum}, F. and {Agn{\`e}se}, P. and {Ali}, B. and {Altieri}, B. and {Andreani}, P. and {Augueres}, J.-L. and {Balog}, Z. and {Barl}, L. and {Bauer}, O.~H. and {Belbachir}, N. and {Benedettini}, M. and {Billot}, N. and {Boulade}, O. and {Bischof}, H. and {Blommaert}, J. and {Callut}, E. and {Cara}, C. and {Cerulli}, R. and {Cesarsky}, D. and {Contursi}, A. and {Creten}, Y. and {De Meester}, W. and {Doublier}, V. and {Doumayrou}, E. and {Duband}, L. and {Exter}, K. and {Genzel}, R. and {Gillis}, J.-M. and {Gr{\"o}zinger}, U. and {Henning}, T. and {Herreros}, J. and {Huygen}, R. and {Inguscio}, M. and {Jakob}, G. and {Jamar}, C. and {Jean}, C. and {de Jong}, J. and {Katterloher}, R. and {Kiss}, C. and {Klaas}, U. and {Lemke}, D. and {Lutz}, D. and {Madden}, S. and {Marquet}, B. and {Martignac}, J. and {Mazy}, A. and {Merken}, P. and {Montfort}, F. and {Morbidelli}, L. and {M{\"u}ller}, T. and {Nielbock}, M. and {Okumura}, K. and {Orfei}, R. and {Ottensamer}, R. and {Pezzuto}, S. and {Popesso}, P. and {Putzeys}, J. and {Regibo}, S. and {Reveret}, V. and {Royer}, P. and {Sauvage}, M. and {Schreiber}, J. and {Stegmaier}, J. and {Schmitt}, D. and {Schubert}, J. and {Sturm}, E. and {Thiel}, M. and {Tofani}, G. and {Vavrek}, R. and {Wetzstein}, M. and {Wieprecht}, E. and {Wiezorrek}, E.},
 doi = {10.1051/0004-6361/201014535},
 eid = {L2},
 eprint = {1005.1487},
 journal = {\aap},
 month = {July},
 pages = {L2},
 primaryclass = {astro-ph.IM},
 title = {{The Photodetector Array Camera and Spectrometer (PACS) on the Herschel Space Observatory}},
 volume = {518},
 year = {2010}
}

@article{Racine_1999,
 adsurl = {https://ui.adsabs.harvard.edu/abs/1999PASP..111..587R},
 author = {{Racine}, Ren{\'e} and {Walker}, Gordon A.~H. and {Nadeau}, Daniel and {Doyon}, Ren{\'e} and {Marois}, Christian},
 doi = {10.1086/316367},
 journal = {\pasp},
 month = {May},
 number = {759},
 pages = {587-594},
 title = {{Speckle Noise and the Detection of Faint Companions}},
 volume = {111},
 year = {1999}
}

@article{Ramirez_2012,
 adsurl = {https://ui.adsabs.harvard.edu/abs/2012ApJ...756...46R},
 archiveprefix = {arXiv},
 author = {{Ram{\'\i}rez}, I. and {Fish}, J.~R. and {Lambert}, D.~L. and {Allende Prieto}, C.},
 doi = {10.1088/0004-637X/756/1/46},
 eid = {46},
 eprint = {1207.0499},
 journal = {\apj},
 month = {September},
 number = {1},
 pages = {46},
 primaryclass = {astro-ph.SR},
 title = {{Lithium Abundances in nearby FGK Dwarf and Subgiant Stars: Internal Destruction, Galactic Chemical Evolution, and Exoplanets}},
 volume = {756},
 year = {2012}
}

@article{Ren_2019,
 adsurl = {https://ui.adsabs.harvard.edu/abs/2019ApJ...882...64R},
 archiveprefix = {arXiv},
 author = {{Ren}, Bin and {Choquet}, {\'E}lodie and {Perrin}, Marshall D. and {Duch{\^e}ne}, Gaspard and {Debes}, John H. and {Pueyo}, Laurent and {Rice}, Malena and {Chen}, Christine and {Schneider}, Glenn and {Esposito}, Thomas M. and {Poteet}, Charles A. and {Wang}, Jason J. and {Ammons}, S. Mark and {Ansdell}, Megan and {Arriaga}, Pauline and {Bailey}, Vanessa P. and {Barman}, Travis and {Sebasti{\'a}n Bruzzone}, Juan and {Bulger}, Joanna and {Chilcote}, Jeffrey and {Cotten}, Tara and {De Rosa}, Robert J. and {Doyon}, Rene and {Fitzgerald}, Michael P. and {Follette}, Katherine B. and {Goodsell}, Stephen J. and {Gerard}, Benjamin L. and {Graham}, James R. and {Greenbaum}, Alexandra Z. and {Hagan}, J. Brendan and {Hibon}, Pascale and {Hines}, Dean C. and {Hung}, Li-Wei and {Ingraham}, Patrick and {Kalas}, Paul and {Konopacky}, Quinn and {Larkin}, James E. and {Macintosh}, Bruce and {Maire}, J{\'e}r{\^o}me and {Marchis}, Franck and {Marois}, Christian and {Mazoyer}, Johan and {M{\'e}nard}, Fran{\c{c}}ois and {Metchev}, Stanimir and {Millar-Blanchaer}, Maxwell A. and {Mittal}, Tushar and {Moerchen}, Margaret and {Nielsen}, Eric L. and {N'Diaye}, Mamadou and {Oppenheimer}, Rebecca and {Palmer}, David and {Patience}, Jennifer and {Pinte}, Christophe and {Poyneer}, Lisa and {Rajan}, Abhijith and {Rameau}, Julien and {Rantakyr{\"o}}, Fredrik T. and {Ruffio}, Jean-Baptiste and {Ryan}, Dominic and {Savransky}, Dmitry and {Schneider}, Adam C. and {Sivaramakrishnan}, Anand and {Song}, Inseok and {Soummer}, R{\'e}mi and {Stark}, Christopher and {Thomas}, Sandrine and {Vigan}, Arthur and {Wallace}, J. Kent and {Ward-Duong}, Kimberly and {Wiktorowicz}, Sloane and {Wolff}, Schuyler and {Ygouf}, Marie and {Norman}, Colin},
 doi = {10.3847/1538-4357/ab3403},
 eid = {64},
 eprint = {1908.00006},
 journal = {\apj},
 month = {September},
 number = {1},
 pages = {64},
 primaryclass = {astro-ph.EP},
 title = {{An Exo-Kuiper Belt with an Extended Halo around HD 191089 in Scattered Light}},
 volume = {882},
 year = {2019}
}

@article{Rich_2022,
 adsurl = {https://ui.adsabs.harvard.edu/abs/2022AJ....164..109R},
 archiveprefix = {arXiv},
 author = {{Rich}, Evan A. and {Monnier}, John D. and {Aarnio}, Alicia and {Laws}, Anna S.~E. and {Setterholm}, Benjamin R. and {Wilner}, David J. and {Calvet}, Nuria and {Harries}, Tim and {Miller}, Chris and {Davies}, Claire L. and {Adams}, Fred C. and {Andrews}, Sean M. and {Bae}, Jaehan and {Espaillat}, Catherine and {Greenbaum}, Alexandra Z. and {Hinkley}, Sasha and {Kraus}, Stefan and {Hartmann}, Lee and {Isella}, Andrea and {McClure}, Melissa and {Oppenheimer}, Rebecca and {P{\'e}rez}, Laura M. and {Zhu}, Zhaohuan},
 doi = {10.3847/1538-3881/ac7be4},
 eid = {109},
 eprint = {2206.05815},
 journal = {\aj},
 month = {September},
 number = {3},
 pages = {109},
 primaryclass = {astro-ph.EP},
 title = {{Gemini-LIGHTS: Herbig Ae/Be and Massive T Tauri Protoplanetary Disks Imaged with Gemini Planet Imager}},
 volume = {164},
 year = {2022}
}

@article{Rivera_2010,
 adsurl = {https://ui.adsabs.harvard.edu/abs/2010ApJ...708.1492R},
 archiveprefix = {arXiv},
 author = {{Rivera}, Eugenio J. and {Butler}, R. Paul and {Vogt}, Steven S. and {Laughlin}, Gregory and {Henry}, Gregory W. and {Meschiari}, Stefano},
 doi = {10.1088/0004-637X/708/2/1492},
 eprint = {0912.2566},
 journal = {\apj},
 month = {January},
 number = {2},
 pages = {1492-1499},
 primaryclass = {astro-ph.EP},
 title = {{A Super-Earth Orbiting the Nearby Sun-like Star HD 1461}},
 volume = {708},
 year = {2010}
}

@misc{Rodrigo_2012,
 adsurl = {https://ui.adsabs.harvard.edu/abs/2012ivoa.rept.1015R},
 author = {{Rodrigo}, Carlos and {Solano}, Enrique and {Bayo}, Amelia},
 doi = {10.5479/ADS/bib/2012ivoa.rept.1015R},
 howpublished = {IVOA Working Draft 15 October 2012},
 title = {{SVO Filter Profile Service Version 1.0}},
 year = {2012}
}

@inproceedings{Rodrigo_2020,
 adsurl = {https://ui.adsabs.harvard.edu/abs/2020sea..confE.182R},
 author = {{Rodrigo}, C. and {Solano}, E.},
 booktitle = {XIV.0 Scientific Meeting (virtual) of the Spanish Astronomical Society},
 eid = {182},
 month = {July},
 pages = {182},
 title = {{The SVO Filter Profile Service}},
 year = {2020}
}

@article{Rodrigo_2024,
 adsurl = {https://ui.adsabs.harvard.edu/abs/2024A&A...689A..93R},
 archiveprefix = {arXiv},
 author = {{Rodrigo}, Carlos and {Cruz}, Patricia and {Aguilar}, John F. and {Aller}, Alba and {Solano}, Enrique and {G{\'a}lvez-Ortiz}, Maria Cruz and {Jim{\'e}nez-Esteban}, Francisco and {Mas-Buitrago}, Pedro and {Bayo}, Amelia and {Cort{\'e}s-Contreras}, Miriam and {Murillo-Ojeda}, Raquel and {Bonoli}, Silvia and {Cenarro}, Javier and {Dupke}, Renato and {L{\'o}pez-Sanjuan}, Carlos and {Mar{\'\i}n-Franch}, Antonio and {de Oliveira}, Claudia Mendes and {Moles}, Mariano and {Taylor}, Keith and {Varela}, Jes{\'u}s and {Rami{\'o}}, H{\'e}ctor V{\'a}zquez},
 doi = {10.1051/0004-6361/202449998},
 eid = {A93},
 eprint = {2406.03310},
 journal = {\aap},
 month = {September},
 pages = {A93},
 primaryclass = {astro-ph.SR},
 title = {{Photometric segregation of dwarf and giant FGK stars using the SVO Filter Profile Service and photometric tools}},
 volume = {689},
 year = {2024}
}

@inproceedings{Rousset_2003,
 adsurl = {https://ui.adsabs.harvard.edu/abs/2003SPIE.4839..140R},
 author = {{Rousset}, Gerard and {Lacombe}, Francois and {Puget}, Pascal and {Hubin}, Norbert N. and {Gendron}, Eric and {Fusco}, Thierry and {Arsenault}, Robin and {Charton}, Julien and {Feautrier}, Philippe and {Gigan}, Pierre and {Kern}, Pierre Y. and {Lagrange}, Anne-Marie and {Madec}, Pierre-Yves and {Mouillet}, David and {Rabaud}, Didier and {Rabou}, Patrick and {Stadler}, Eric and {Zins}, Gerard},
 booktitle = {Adaptive Optical System Technologies II},
 doi = {10.1117/12.459332},
 editor = {{Wizinowich}, Peter L. and {Bonaccini}, Domenico},
 month = {February},
 pages = {140-149},
 series = {Society of Photo-Optical Instrumentation Engineers (SPIE) Conference Series},
 title = {{NAOS, the first AO system of the VLT: on-sky performance}},
 volume = {4839},
 year = {2003}
}

@article{Samland2025arXiv,
 adsurl = {https://ui.adsabs.harvard.edu/abs/2025arXiv250908044S},
 archiveprefix = {arXiv},
 author = {{Samland}, Matthias},
 doi = {10.48550/arXiv.2509.08044},
 eid = {arXiv:2509.08044},
 eprint = {2509.08044},
 journal = {arXiv e-prints},
 month = {September},
 pages = {arXiv:2509.08044},
 primaryclass = {astro-ph.IM},
 title = {{spherical: A Comprehensive Database and Automated Pipeline for VLT/SPHERE High-Contrast Imaging}},
 year = {2025}
}

@article{Samland_2021,
 adsurl = {https://ui.adsabs.harvard.edu/abs/2021A&A...646A..24S},
 archiveprefix = {arXiv},
 author = {{Samland}, M. and {Bouwman}, J. and {Hogg}, D.~W. and {Brandner}, W. and {Henning}, T. and {Janson}, M.},
 doi = {10.1051/0004-6361/201937308},
 eid = {A24},
 eprint = {2011.12311},
 journal = {\aap},
 month = {February},
 pages = {A24},
 primaryclass = {astro-ph.EP},
 title = {{TRAP: a temporal systematics model for improved direct detection of exoplanets at small angular separations}},
 volume = {646},
 year = {2021}
}

@article{Samland_2022,
 adsurl = {https://ui.adsabs.harvard.edu/abs/2022A&A...668A..84S},
 archiveprefix = {arXiv},
 author = {{Samland}, M. and {Brandt}, T.~D. and {Milli}, J. and {Delorme}, P. and {Vigan}, A.},
 doi = {10.1051/0004-6361/202244587},
 eid = {A84},
 eprint = {2210.06390},
 journal = {\aap},
 month = {December},
 pages = {A84},
 primaryclass = {astro-ph.EP},
 title = {{Spectral cube extraction for the VLT/SPHERE IFS. Open-source pipeline with full forward modeling and improved sensitivity}},
 volume = {668},
 year = {2022}
}

@article{Sanghi_2026,
 adsurl = {https://ui.adsabs.harvard.edu/abs/2026AJ....171..225S},
 archiveprefix = {arXiv},
 author = {{Sanghi}, Aniket and {Mang}, James and {Llop-Sayson}, Jorge and {Mamajek}, Eric E. and {Thompson}, William and {Sur}, Ankan and {Beichman}, Charles and {Bryden}, Geoffrey and {G{\'a}sp{\'a}r}, Andr{\'a}s and {Leisenring}, Jarron and et al.},
 doi = {10.3847/1538-3881/ae4909},
 eid = {225},
 eprint = {2602.23423},
 journal = {\aj},
 month = {April},
 number = {4},
 pages = {225},
 primaryclass = {astro-ph.EP},
 title = {{Worlds Next Door. III. Indirect Evidence for Enhanced Atmospheric Metallicity and/or the Presence of Water Clouds in the Nearest Jupiter-analog ϵ Eri b}},
 volume = {171},
 year = {2026}
}

@article{Schneider_1999,
 adsurl = {https://ui.adsabs.harvard.edu/abs/1999ApJ...513L.127S},
 archiveprefix = {arXiv},
 author = {{Schneider}, Glenn and {Smith}, Bradford A. and {Becklin}, E.~E. and {Koerner}, David W. and {Meier}, Roland and {Hines}, Dean C. and {Lowrance}, Patrick J. and {Terrile}, Richard J. and {Thompson}, Rodger I. and {Rieke}, Marcia},
 doi = {10.1086/311921},
 eprint = {astro-ph/9901218},
 journal = {\apjl},
 month = {March},
 number = {2},
 pages = {L127-L130},
 primaryclass = {astro-ph},
 title = {{NICMOS Imaging of the HR 4796A Circumstellar Disk}},
 volume = {513},
 year = {1999}
}

@article{Schneider_2006,
 adsurl = {https://ui.adsabs.harvard.edu/abs/2006ApJ...650..414S},
 archiveprefix = {arXiv},
 author = {{Schneider}, Glenn and {Silverstone}, Murray D. and {Hines}, Dean C. and {Augereau}, Jean-Charles and {Pinte}, Christophe and {M{\'e}nard}, Fran{\c{c}}ois and {Krist}, John and {Clampin}, Mark and {Grady}, Carol and {Golimowski}, David and {Ardila}, David and {Henning}, Thomas and {Wolf}, Sebastian and {Rodmann}, Jens},
 doi = {10.1086/506507},
 eprint = {astro-ph/0606213},
 journal = {\apj},
 month = {October},
 number = {1},
 pages = {414-431},
 primaryclass = {astro-ph},
 title = {{Discovery of an 86 AU Radius Debris Ring around HD 181327}},
 volume = {650},
 year = {2006}
}

@article{Schneider_2014,
 adsurl = {https://ui.adsabs.harvard.edu/abs/2014AJ....148...59S},
 archiveprefix = {arXiv},
 author = {{Schneider}, Glenn and {Grady}, Carol A. and {Hines}, Dean C. and {Stark}, Christopher C. and {Debes}, John H. and {Carson}, Joe and {Kuchner}, Marc J. and {Perrin}, Marshall D. and {Weinberger}, Alycia J. and {Wisniewski}, John P. and {Silverstone}, Murray D. and {Jang-Condell}, Hannah and {Henning}, Thomas and {Woodgate}, Bruce E. and {Serabyn}, Eugene and {Moro-Martin}, Amaya and {Tamura}, Motohide and {Hinz}, Phillip M. and {Rodigas}, Timothy J.},
 doi = {10.1088/0004-6256/148/4/59},
 eid = {59},
 eprint = {1406.7303},
 journal = {\aj},
 month = {October},
 number = {4},
 pages = {59},
 primaryclass = {astro-ph.IM},
 title = {{Probing for Exoplanets Hiding in Dusty Debris Disks: Disk Imaging, Characterization, and Exploration with HST/STIS Multi-roll Coronagraphy}},
 volume = {148},
 year = {2014}
}

@article{Sefilian_2021,
 adsurl = {https://ui.adsabs.harvard.edu/abs/2021ApJ...910...13S},
 archiveprefix = {arXiv},
 author = {{Sefilian}, Antranik A. and {Rafikov}, Roman R. and {Wyatt}, Mark C.},
 doi = {10.3847/1538-4357/abda46},
 eid = {13},
 eprint = {2010.15617},
 journal = {\apj},
 month = {March},
 number = {1},
 pages = {13},
 primaryclass = {astro-ph.EP},
 title = {{Formation of Gaps in Self-gravitating Debris Disks by Secular Resonance in a Single-planet System. I. A Simplified Model}},
 volume = {910},
 year = {2021}
}

@article{Sefilian_2025,
 adsurl = {https://ui.adsabs.harvard.edu/abs/2025MNRAS.543.3123S},
 archiveprefix = {arXiv},
 author = {{Sefilian}, Antranik A. and {Kratter}, Kaitlin M. and {Wyatt}, Mark C. and {Petrovich}, Cristobal and {Th{\'e}bault}, Philippe and {Malhotra}, Renu and {Faramaz-Gorka}, Virginie},
 doi = {10.1093/mnras/staf1555},
 eprint = {2505.09578},
 journal = {\mnras},
 month = {November},
 number = {4},
 pages = {3123-3151},
 primaryclass = {astro-ph.EP},
 title = {{The vertical structure of debris discs and the role of disc gravity: a primer using a simplified model}},
 volume = {543},
 year = {2025}
}

@article{Shannon_2016,
 adsurl = {https://ui.adsabs.harvard.edu/abs/2016MNRAS.462L.116S},
 archiveprefix = {arXiv},
 author = {{Shannon}, Andrew and {Bonsor}, Amy and {Kral}, Quentin and {Matthews}, Elisabeth},
 doi = {10.1093/mnrasl/slw143},
 eprint = {1607.04282},
 journal = {\mnras},
 month = {October},
 number = {1},
 pages = {L116-L120},
 primaryclass = {astro-ph.EP},
 title = {{The unseen planets of double belt debris disc systems}},
 volume = {462},
 year = {2016}
}

@article{Sierchio_2014,
 adsurl = {https://ui.adsabs.harvard.edu/abs/2014ApJ...785...33S},
 archiveprefix = {arXiv},
 author = {{Sierchio}, J.~M. and {Rieke}, G.~H. and {Su}, K.~Y.~L. and {G{\'a}sp{\'a}r}, Andras},
 doi = {10.1088/0004-637X/785/1/33},
 eid = {33},
 eprint = {1402.6308},
 journal = {\apj},
 month = {April},
 number = {1},
 pages = {33},
 primaryclass = {astro-ph.EP},
 title = {{The Decay of Debris Disks around Solar-type Stars}},
 volume = {785},
 year = {2014}
}

@inproceedings{Skemer_2014,
 adsurl = {https://ui.adsabs.harvard.edu/abs/2014ebi..conf..4.4S},
 author = {{Skemer}, A.},
 booktitle = {Search for Life Beyond the Solar System. Exoplanets, Biosignatures \& Instruments},
 editor = {{Apai}, Daniel and {Gabor}, Pavel},
 eid = {4.4},
 month = {March},
 pages = {4.4},
 title = {{LEECH: LBTI Exozodi Exoplanet Common Hunt}},
 year = {2014}
}

@article{Slettebak_1975,
 adsurl = {https://ui.adsabs.harvard.edu/abs/1975ApJ...197..137S},
 author = {{Slettebak}, A.},
 doi = {10.1086/153493},
 journal = {\apj},
 month = {April},
 pages = {137-138},
 title = {{Some interesting bright southern stars of early type.}},
 volume = {197},
 year = {1975}
}

@article{student_1908,
 author = {Student},
 doi = {10.1093/biomet/6.1.1},
 eprint = {https://academic.oup.com/biomet/article-pdf/6/1/1/49729532/biomet_6_1_1.pdf},
 issn = {0006-3444},
 journal = {Biometrika},
 month = {03},
 number = {1},
 pages = {1-25},
 title = {The Probable Error of a Mean},
 url = {https://doi.org/10.1093/biomet/6.1.1},
 volume = {6},
 year = {1908}
}
\bibliographystyle{aa_url}
\begin{appendix}
\onecolumn
\nolinenumbers
\section{PSF-subtracted residual images}
We show examples of the PCA and \fours residual images in \autoref{fig: all residual images} for 3 of our 13 stars. For PCA, we show the full FoV captured by \irdis in the top row, while for \fours, due to computational resources, we are only able to reduce the central $\approx 1 \times 1$~\si{\arcsec} FoV, shown in the bottom row. The images for all stars, including FITS files, are available on Zenodo.
\begin{figure*}[ht]
    \centering
    \begin{subfigure}{0.33\textwidth}
        \centering
        \includegraphics[width=\linewidth]{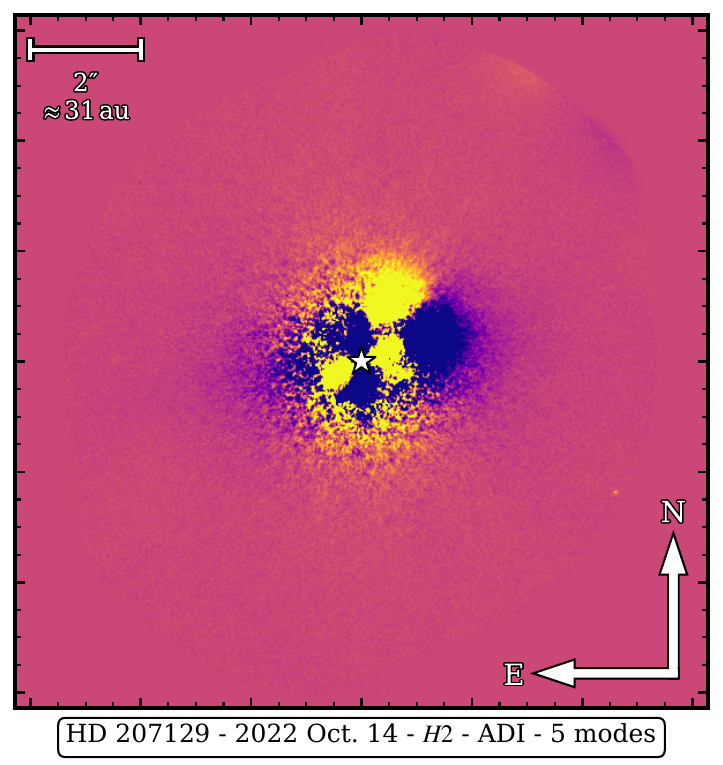}
    \end{subfigure}
    \hfill
    \begin{subfigure}{0.33\textwidth}
        \centering
        \includegraphics[width=\linewidth]{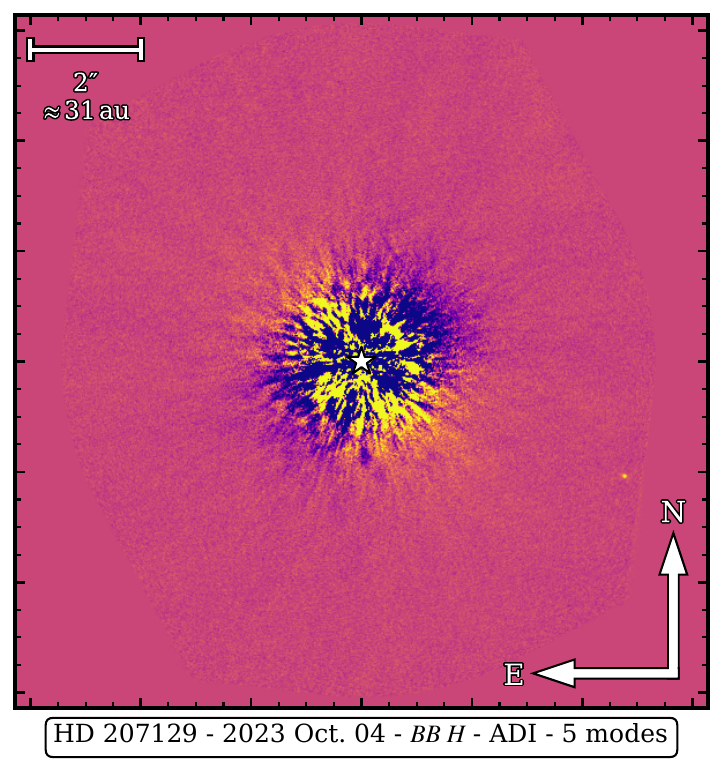}
    \end{subfigure}
    \hfill
    \begin{subfigure}{0.33\textwidth}
        \centering
        \includegraphics[width=\linewidth]{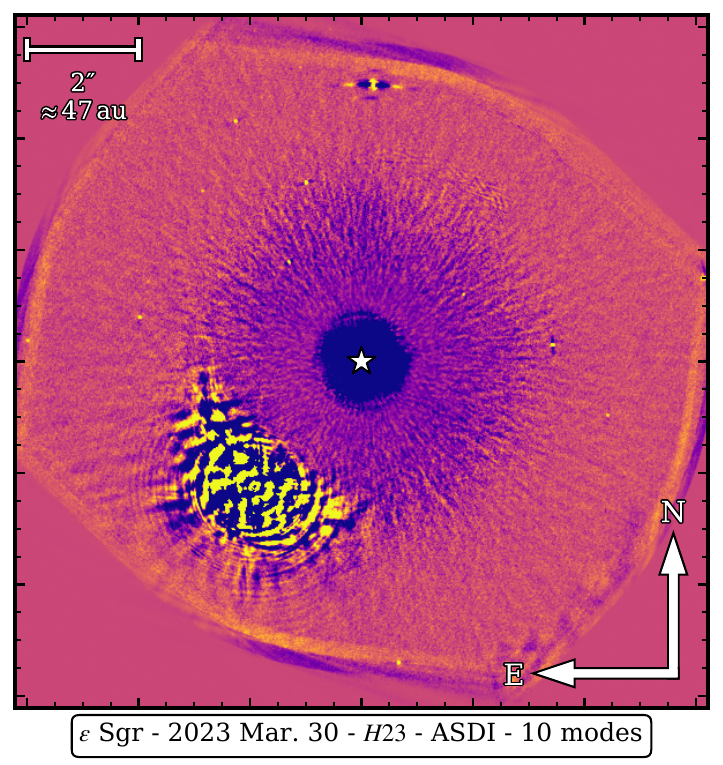}
    \end{subfigure}
    \begin{subfigure}{0.33\textwidth}
        \centering
        \includegraphics[width=\linewidth]{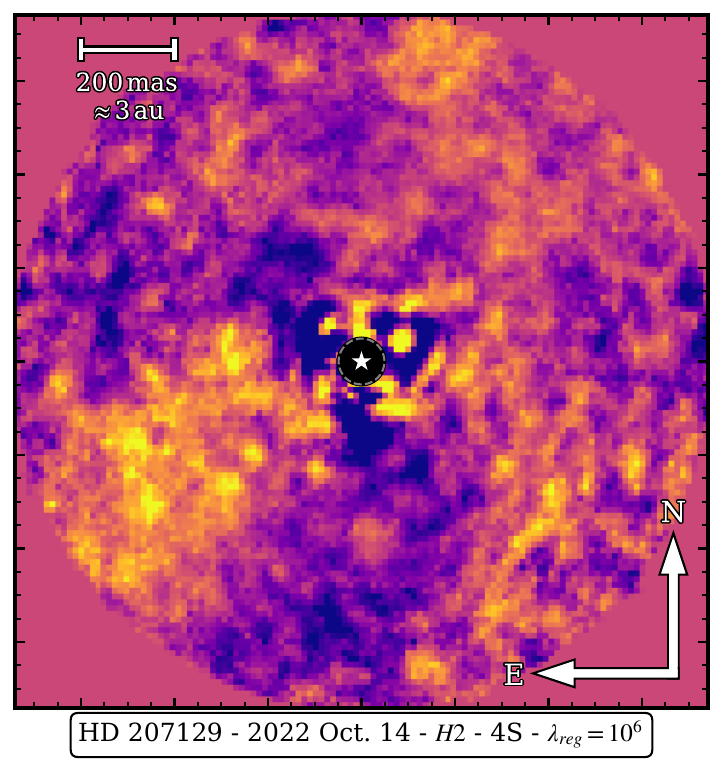}
    \end{subfigure}
    \hfill
    \begin{subfigure}{0.33\textwidth}
        \centering
        \includegraphics[width=\linewidth]{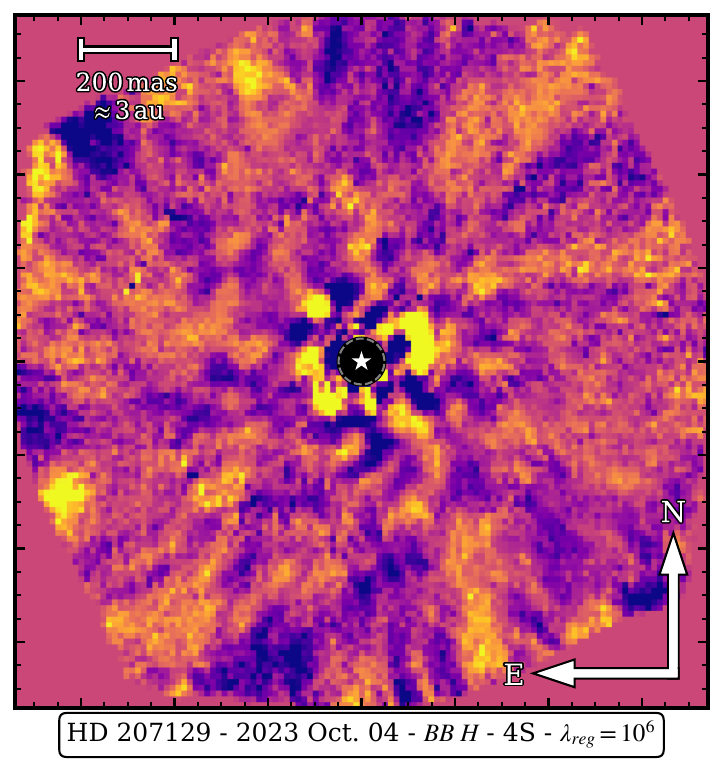}
    \end{subfigure}
    \hfill
    \begin{subfigure}{0.33\textwidth}
        \centering
        \includegraphics[width=\linewidth]{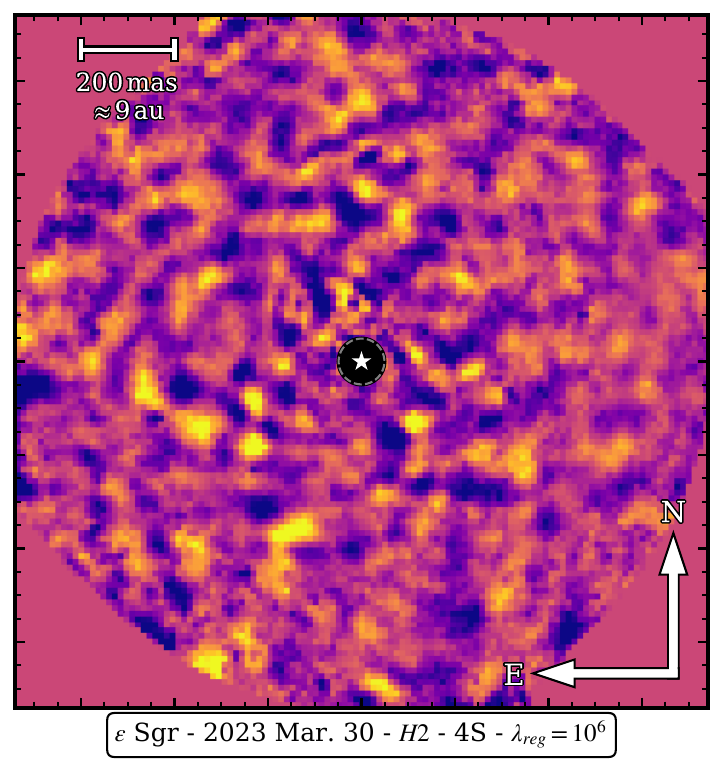}
    \end{subfigure}
    \caption{\textit{Top}: Individual ADI residual images when using PCA with 5 modes. \textit{Bottom}: \fours residuals with a value of $\lambda_\text{reg} = 10^6$. With the exception of the top-right panel showing an SDI+ADI reduction of $\epsilon$~Sgr with 10 PCA modes. The SDI reduction here allows for an improved subtraction of scattered light originating from the secondary star around \SI{3}{\arcsec} southeast of the primary star; we additionally removed residuals from over-subtracted light in the center. The bar at the top of each panel indicates the image size and the corresponding projected separation from the host star, which is centered. All images are aligned such that north is up and east to the left. For \fours we only compute the central $\approx 1 \times 1$~\si{\arcsec} FoV and indicate the central 1 FWHM, which is blocked by the coronagraph with a black mask.} 
    \label{fig: all residual images}
\end{figure*}
\FloatBarrier
\section[Speckle nature of the point-like object near HD 1461]{Speckle nature of the point-like object near HD\,1461}
To test the hypothesis that the point-like feature is created by a temporary speckle, the observation cubes can be split into parts and analyzed individually. This way, if, say, the signal comes from a speckle only in the first half of the sequence, the resulting images would show a signal in only the first-half residual and not in the second half residual. 

To test this robustly for our 2022 dataset, we applied the method to both the original cube and a cleaned cube with an injected signal at the same radial distance but with a \SI{180}{\degree} offset. The results for both datasets are shown in \autoref{fig: HD 1461 split_cube} for the real cube (top row) and the fake injected cube (bottom row), respectively. The first image displays the residuals for the full cube. The middle and right panels show the residuals for only the first and second halves, respectively. Comparing the three panels of the original cube, one can see that the feature has a slightly increased brightness in the first half and then vanishes in the second half. The same is observable for the fake planet, although the increase and decrease in brightness are not as strong as for the real cube.  While this is generally consistent with speckle behavior, the results from the fake injected cube raise questions about the test's reliability in this case, as the injected signal should not be affected. The observed phenomenon can be explained by examining the parallactic angle rotation for the HD\,1461 dataset. In \autoref{fig: HD 1461 rotation}, we show the rotation in pixels at a separation of 21.4 pixels against frame number. The second half of the cube experiences only 25\% of the total rotation, which, expressed in FWHM, is only 2 for the $H2$ image, resulting in severe self-subtraction and potentially explaining the observed trends. However, due to slight differences observed between the real and fake injected datasets, this does not fully rule out the possibility of speckle noise. We also tested splitting the cubes differently, but unfortunately, this led to the same results, and so subdividing the cube cannot help differentiate planets and speckles. With the non-detection in the new 2025 epoch, we therefore conclude that the signal is actually a result of non-Gaussian residual noise. 
\begin{figure*}[ht]
    \centering
    \makebox[\linewidth][c]{
    \includegraphics[width=0.92\linewidth]{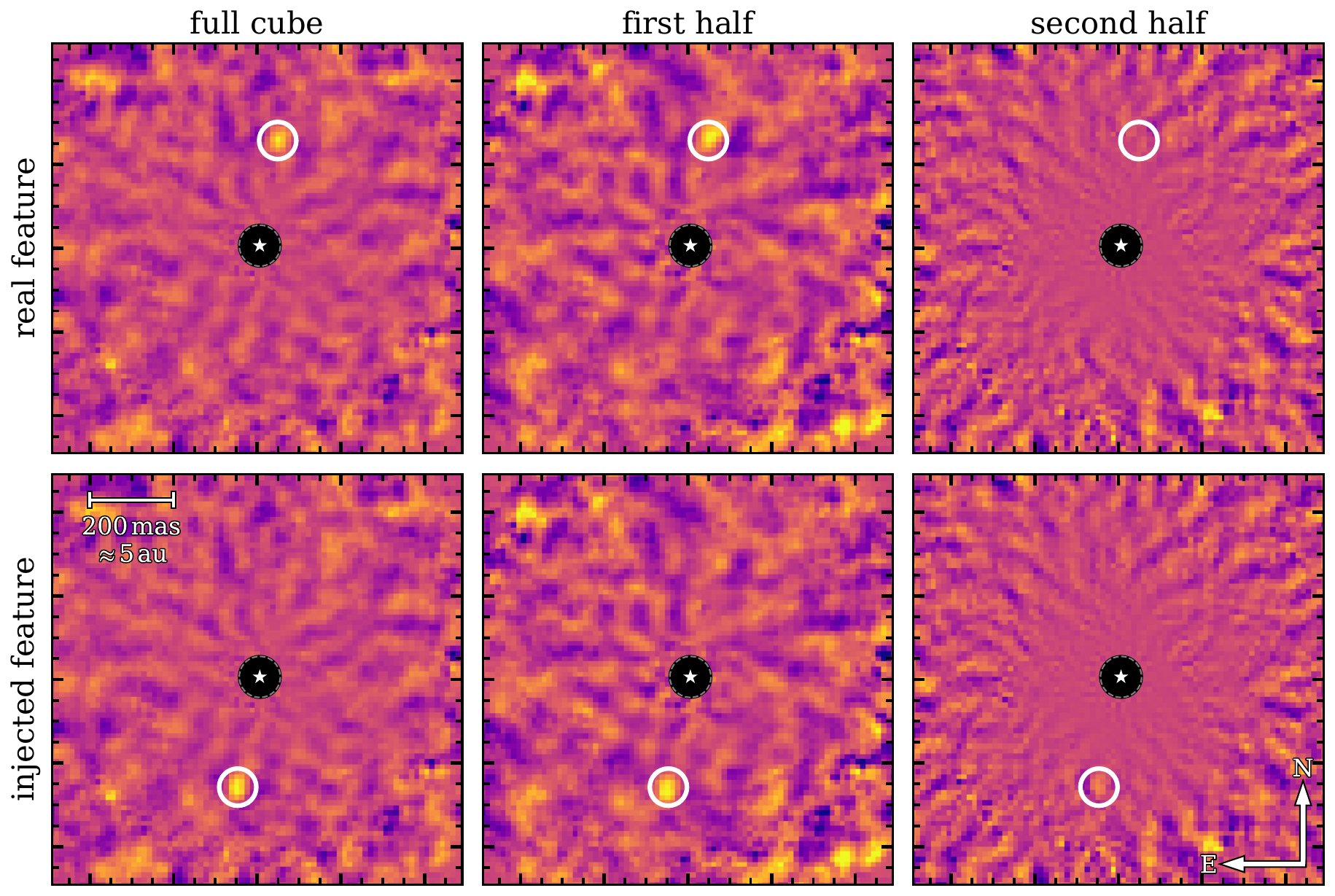}}
    \vspace{-0.5cm}
    \caption{\fours residuals for splitting the science cube. \textit{Left to right}: Residual for a full cube, the first 90 images in the cube, and the last  90 images in the cube. \textit{Top row}: Original point-like feature. \textit{Bottom row}: Signal subtracted out. Instead, a fake signal with a \SI{180}{\degree} offset to the original one is injected into the cube. For both features, the \snr in the residual appears slightly stronger in the first half than in the full cube, but vanishes nearly entirely in the second half.}
    \label{fig: HD 1461 split_cube}
    \vspace{-0.5cm}
\end{figure*}

\begin{figure*}[h!]
  \begin{minipage}[c]{0.62\textwidth}
    \includegraphics[width=\linewidth]{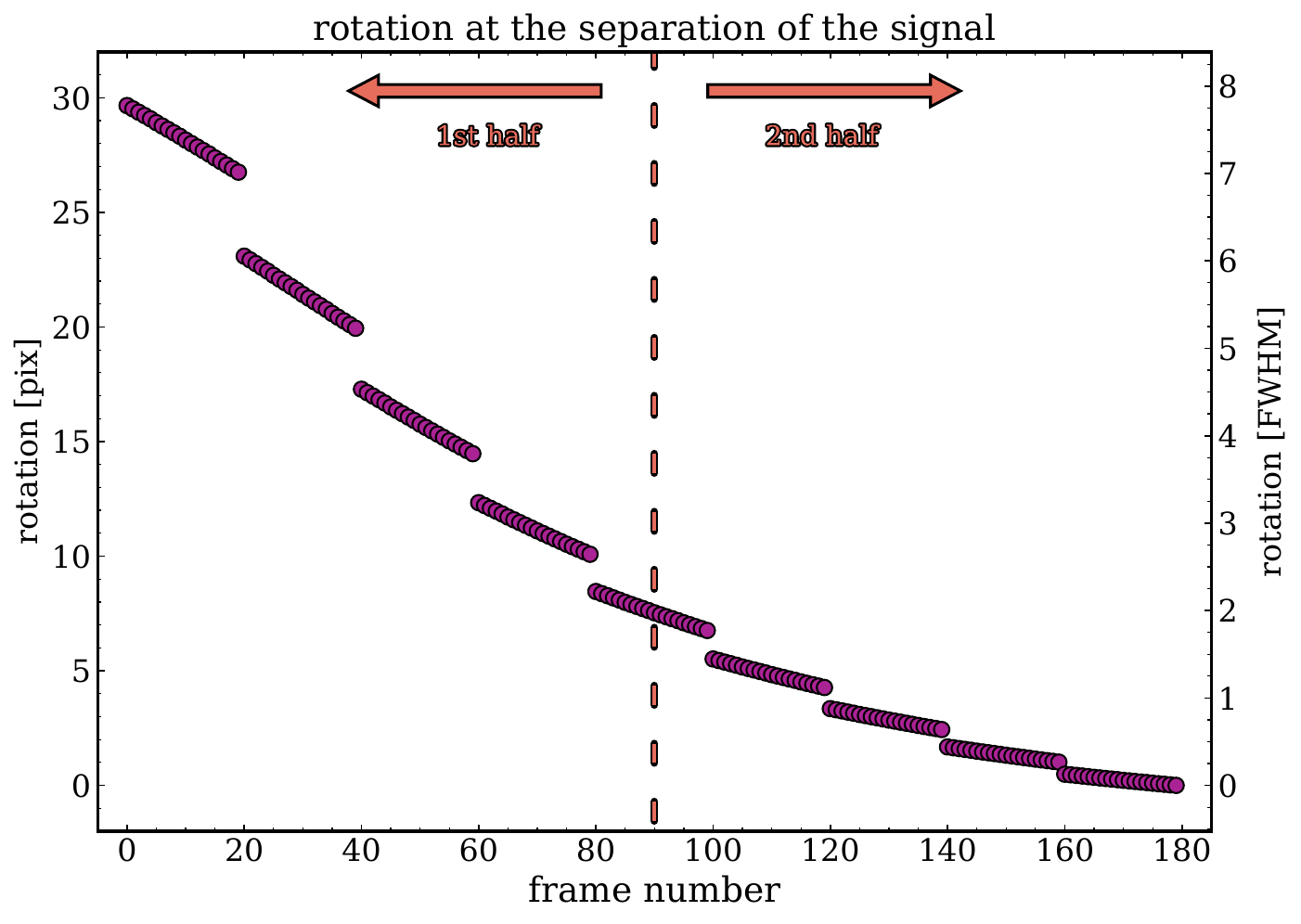}
  \end{minipage}\hfill
  \begin{minipage}[c]{0.35\textwidth}
    \caption{Parallactic angle rotation for the HD\,1461 dataset. The angle is converted to pixel separation at a radius of the detected source (21.4 pixels from the frame center) to show how much the tentative companion candidate would move from frame to frame. The y-axis on the right shows the rotation in units of FWHM (3.8 pixels) for the $H2$ filter frame. The largest separations are achieved at the beginning of the dataset. By frame 90, which represents the temporal mean shown by the orange line, 75\% of the rotation is already achieved. Over the last 90 frames, only a rotation of \SI{20}{\degree} is performed, resulting in a shift of only $\approx$\,8 pixels, which corresponds to $\approx$\,2 FWHM. The result of this is that the second half suffers from severe self-subtraction, observed in the right panel of \protect\autoref{fig: HD 1461 split_cube}.}
    \label{fig: HD 1461 rotation}
  \end{minipage}
  \vspace{-0.2cm}
\end{figure*}
\FloatBarrier

\section[HD 59659 orbital analyses]{HD\,59659 orbital analyses}\label{sec: HD 59659 orbital analyses}
To assess the possibility that our detected signal is real, we fit possible orbits, including RVs from RVSPY, our \sphere astrometric data point, and HGCA. We fix the priors on the primary mass to the literature value and on the secondary mass to the mass inferred from photometry using ATMO planetary evolution models \citep{Phillips_2020}. We show the full corner plot in \autoref{fig: HD 59659 rotation} a). The corresponding orbits are shown in \autoref{fig: HD 59659 rotation} b) with the fits to the RV data in panel c). The fits generally converge but yield a highly eccentric solution, with our detection near apastron. The RV plot in panel c) also shows that the eccentricity appears to be forced by the offset of the first batch of data points relative to the later ones. To test this hypothesis, we performed a second orbit fit, excluding this first batch. We find that, in this case, the MCMC walkers do not converge to a distinct set of orbits and instead yield many possible configurations. We attribute this to insufficient information, as in this case, the RVs are consistent with a flat line and thus do not provide strong evidence of a planetary signature. Hence, as HD\,59659 also shows no strong HGCA, a single astrometric point does not significantly constrain the orbit. 

In summary, this does not allow for a definite conclusion; our detected signal, together with all RV data points, results in a stable, possible configuration of orbits. However, multiple questions arise, including the high eccentricity and inclination, as well as the similarity between the point-like and false-positive cases in HD\,1461, casting doubt on the reliability of the detected signal. Follow-up observations reaching even deeper limits, rather close to the star, are required to fully determine the true nature of this source.
\begin{figure*}[ht!]
    \centering
    \makebox[\linewidth][c]{
    \includegraphics[width=\linewidth]{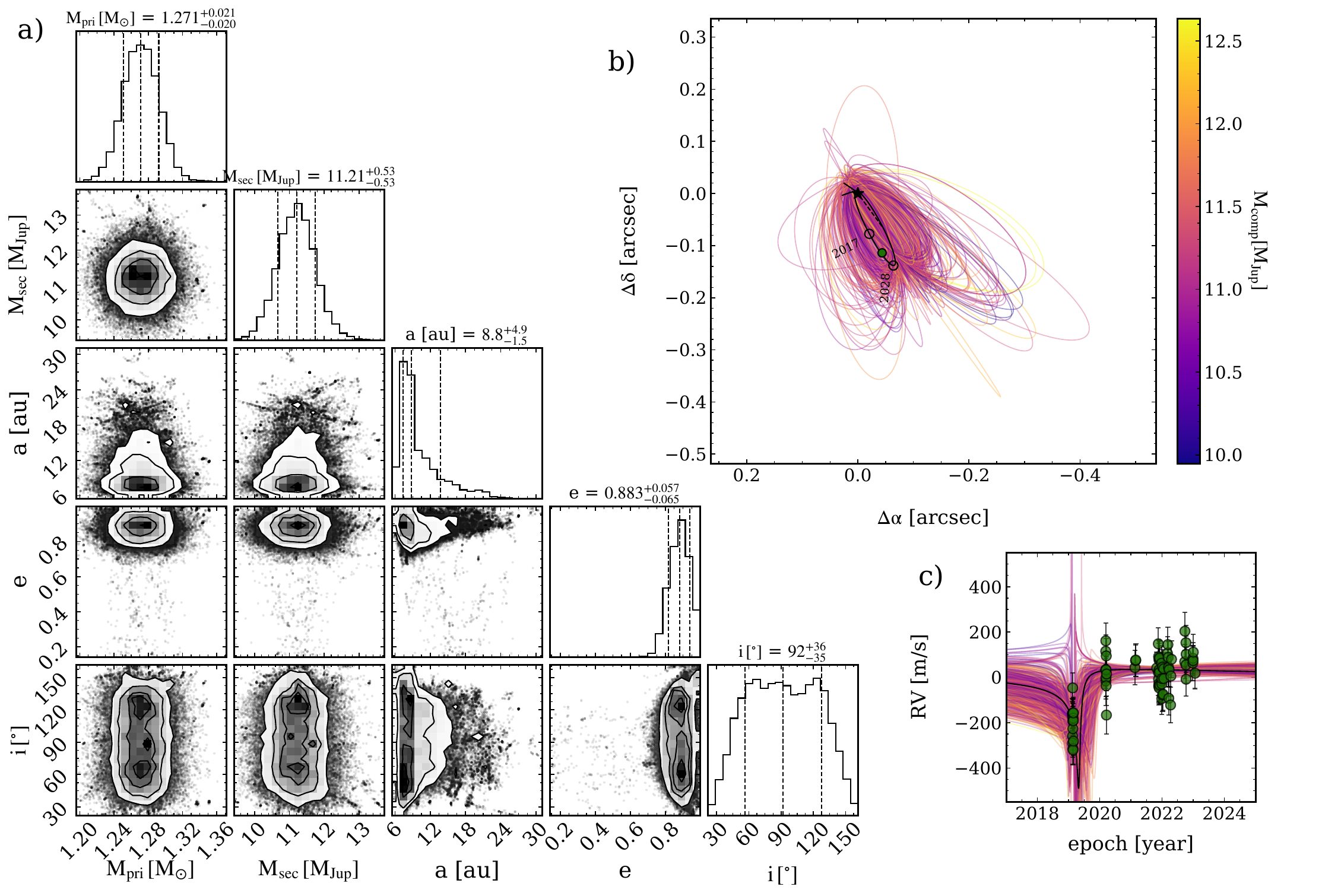}}
    \vspace{-0.5cm}
    \caption{Orbit fit, including RVs and the imaging data for the \snr$=6.0$ signal detected in the HD\,59659 residuals. \textit{Panel (a)}: Full corner plot of the posterior. \textit{Panels (b)} and\textit{ (c)}: Orbits fitted to the astrometric value as well as the RVs. The color bar shown next to panel (b) also applies to the radial velocities in panel (c).}
    \label{fig: HD 59659 rotation}
    \vspace{-0.5cm}
\end{figure*}
\vspace{-0.5cm}
\FloatBarrier
\section{Common proper motion tests}\label{sec: Common Proper Motion (CPM) Tests}
\begin{figure*}[!htbp]
    \centering
    \includegraphics[width=0.95\linewidth]{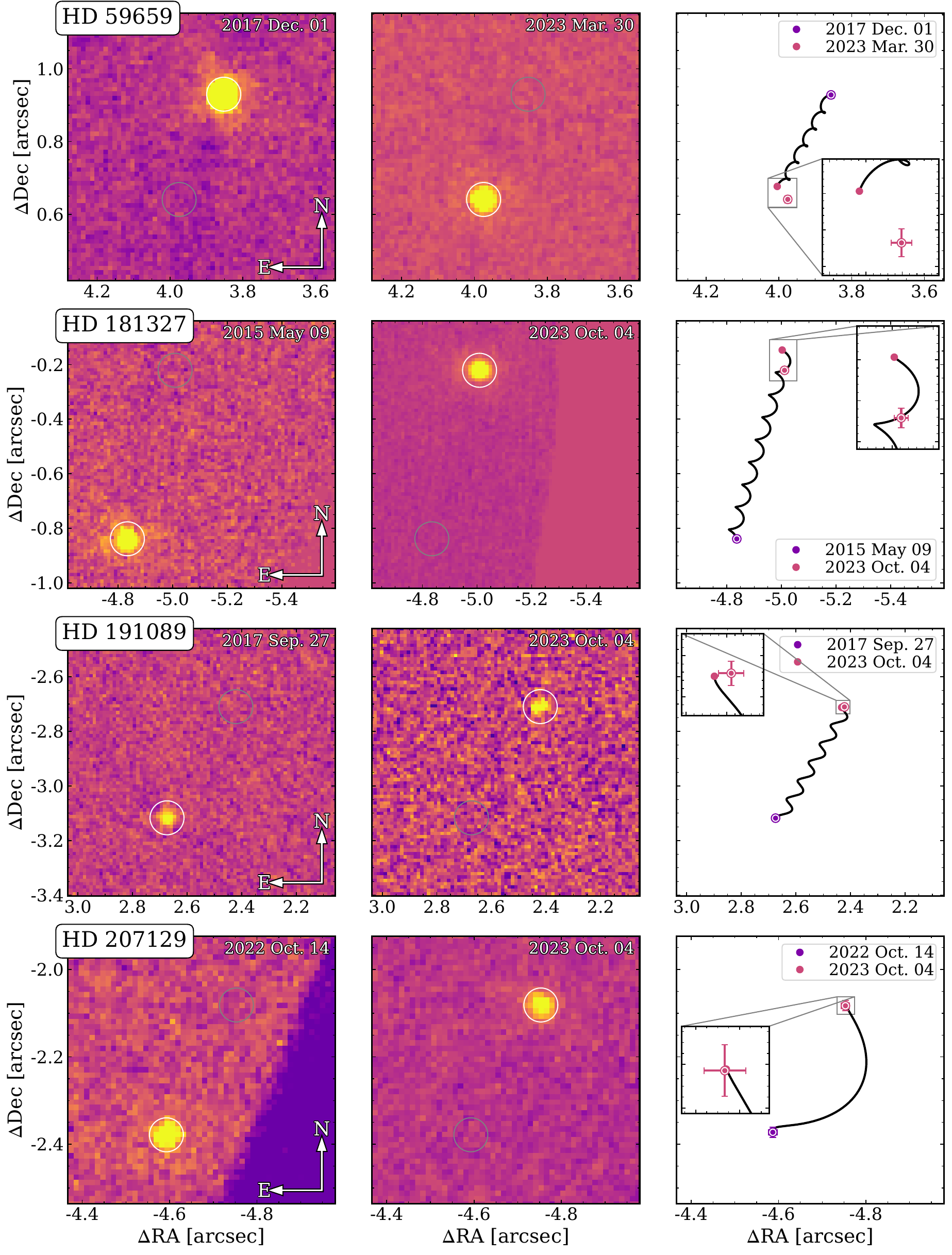}
    \vspace{-0.1cm}
    \caption{\textit{Left} and \textit{center}: Residual images around the locations of the detected source. The measured location in each image is indicated by the white circle, while the measured position from the other one is marked by the gray circle. \textit{Right}: Extracted astrometry from both epochs marked with different colors. The relative motion of a stationary background object starting from the first epoch is indicated by the black line. The colored marker on the line shows the predicted location at the second epoch.}
    \label{fig: CPM}
\end{figure*}
To validate the background origins of multiple sources visible in our data, we reanalyzed archival \sphere data using our pipeline REACHER to ensure consistency and avoid bias. We present CPM tests for targets with a single background source and show cutout images of the source detections in two epochs, along with an astrometry plot. There, we indicate the two measured positions, as well as the predicted background motion due to Earth's barycentric motion and the proper motion of the host star.
\FloatBarrier
\section{PCA contrast limits}
In \autoref{fig: individual contrast curves}, we show examples of the individual contrast curves achievable for each of our datasets. We compute these using the \texttt{applefy} package \citep{Bonse_2023}, which allows for individual injection of fake companions into the datasets, PSF subtraction post-processing, and \snr calculation in the residual images. We start by cleaning our raw cubes by injecting negative signals at locations where we detect point-like objects, producing signal-free raw cubes. In these, we then inject positive fake signals with a contrast of 10 mag at separations in steps of 1 FWHM, starting at 1 FWHM from the center out to a projected separation of \SI{500}{\AU}; hence, this differs for each cube. For each separation, we inject the companions at three different position angles. These cubes are then processed with PCA as implemented in VIP \citep{GomezGonzalez_2017, Christiaens_2023} using the same settings as explained in Sect. \ref{sec: IRDIS PCA}. To compute the \snr we use aperture sum + search function in \texttt{applefy} while for the noise we use only aperture sum. We compute the final contrast limits assuming Gaussian noise and using a t-test as defined by \citet{Mawet_2014} with a 5$\sigma$ confidence level. The full set of contrast curves for all datasets is accessible online on Zenodo.
\begin{figure*}[ht]
    \centering
    \begin{subfigure}{0.33\textwidth}
        \centering
        \includegraphics[width=\linewidth]{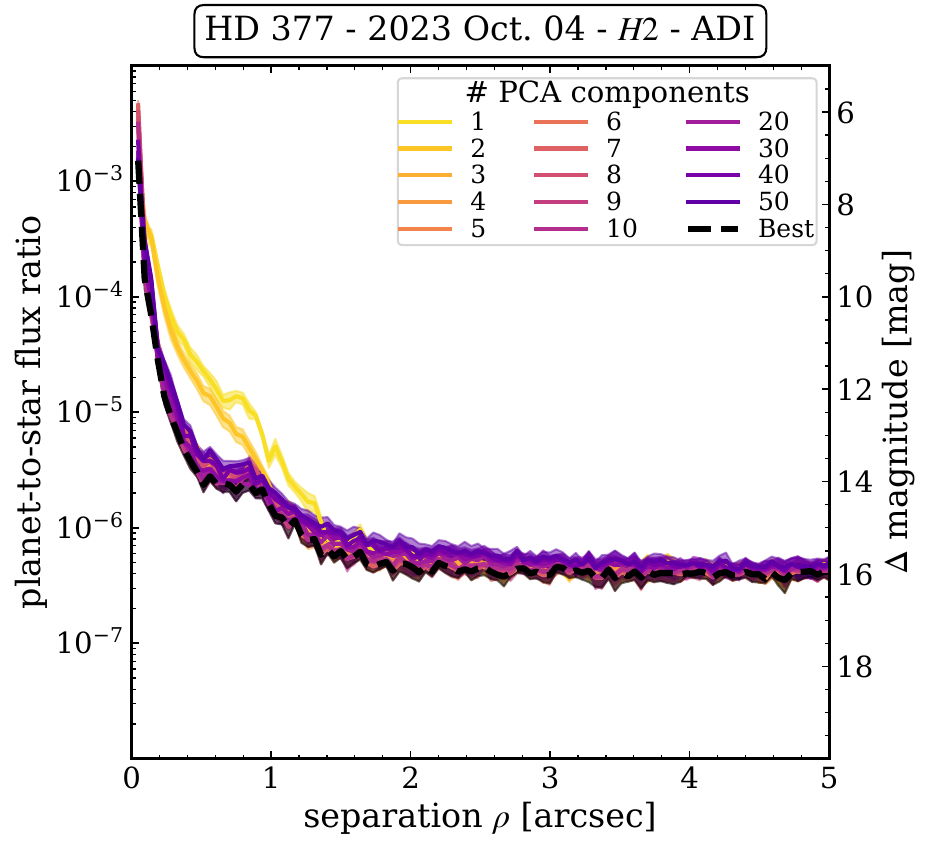}
    \end{subfigure}
    \hfill
    \begin{subfigure}{0.33\textwidth}
        \centering
        \includegraphics[width=\linewidth]{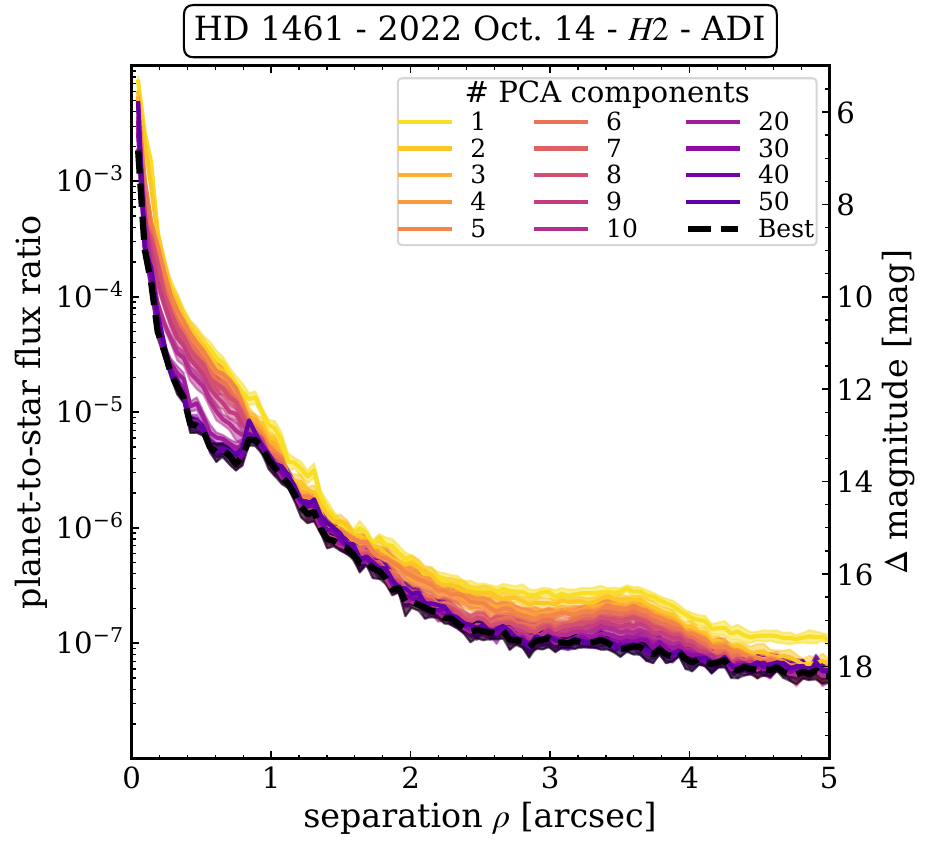}
    \end{subfigure}
    \hfill
    \begin{subfigure}{0.33\textwidth}
        \centering
        \includegraphics[width=\linewidth]{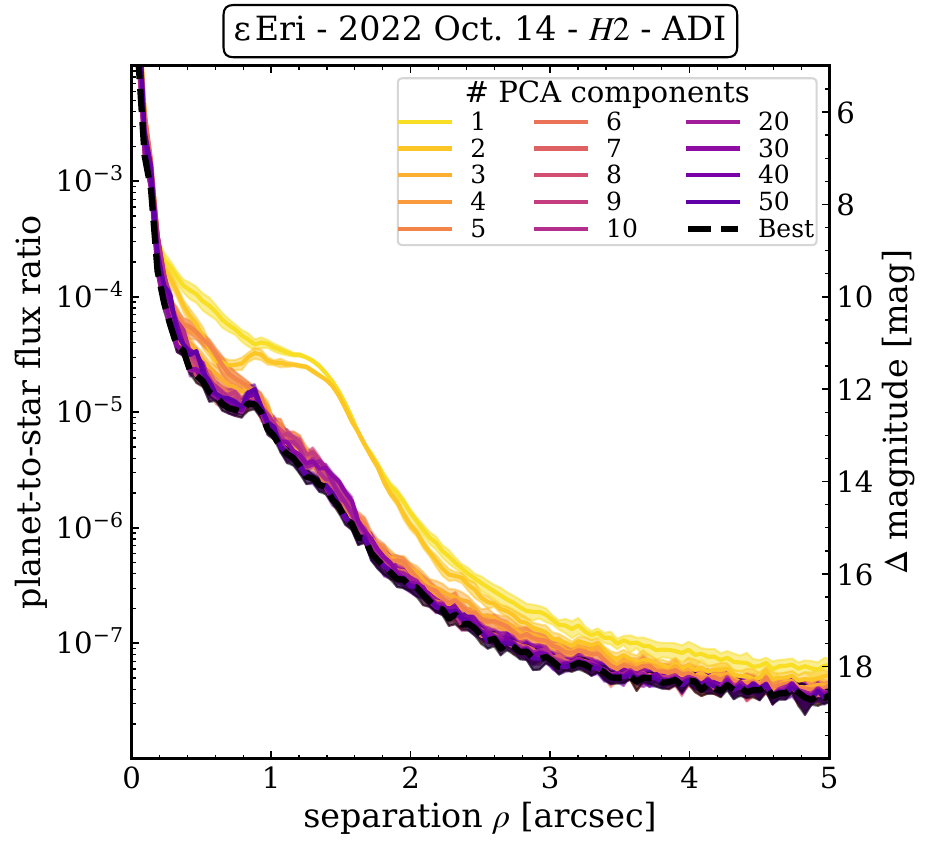}
    \end{subfigure}
    \caption{Individual contrast curves for PCA reductions with different numbers of PCA modes (colored). The black line corresponds to the overall best achievable contrast. Within the AO control radius, all datasets show consistency following the same curve. In the outer, background-limited regime, a higher contrast is achievable only for brighter stars. All curves are available in the online version.} 
    \label{fig: individual contrast curves}
\end{figure*}
\vspace{-0.5cm}
\FloatBarrier
\section[Atmospheric characterization of HD 101412 B and C]{Atmospheric characterization of HD\,101412 B and C}\label{sec: Atmospheric Characterization of HD 101412 B and C}
The atmospheric fitting was performed using the \texttt{species} package \citep{Stolker_2020}, which determines best-fit grid models via Bayesian inference with nested sampling. This approach samples a multimodal distribution and estimates the marginalized likelihood, enabling model comparison through the Bayes factor. The spectroscopic and photometric data from \ifs and \irdis $K1$ and $K2$ bands, as well as \naco\ $L'$ photometry, were included in the fit. A weighting factor was applied to both the spectra and photometry to account for the width of the wavelength bins and the FWHM of the respective filter profiles. The spectral resolution of the model spectra was set to match the instrument resolution of \ifs ($R = 40$). The model spectra were fit against four parameters: the effective temperature ($T_\text{eff}$), surface gravity ($\log g$), radius ($R$), and parallax ($\varpi$). 

Prior to the atmospheric fitting, we compared the available photometry against AMES-Cond \citep{Allard_2012} evolutionary model tracks to obtain approximate estimates of the masses, the effective temperature, surface gravity, and radius of each companion, which were subsequently used to inform our choice of atmospheric model family and to set the priors for the fit. We find that for these young ages, the companions tend to be inflated compared to main-sequence stars. We consequently adopt this inflation when informing our radius priors.

The parallax of the HD\,101412 system was taken from \textit{Gaia} DR3 \citep{GaiaCollaboration_2023} and set as a Gaussian prior of $\varpi = 2.43 \pm 0.01$\ \si{\mas} for both companion fits. Based on the findings from our comparison to evolutionary tracks, for HD\,101412 B we use a Gaussian prior of $R = 6.0 \pm 2.0\,$\si{\Rjup}, while uniform priors were set for $T_\text{eff}$ in the range $2600$--$5000$\,\si{\K} and for $\log g$ in the range $2.5$--$5.5$, covering the full surface gravity range provided by the model grid. For HD\,101412 C, the same uniform priors were applied, with a Gaussian radius prior of $R = 4.0 \pm 1.0\,$\si{\Rjup}. The resulting corner plots for HD\,101412 B and C are shown in \autoref{fig: corner plots} in the top-right and lower-left corners, respectively. Without the priors on the radii, we observe that the models tend toward smaller, non-inflated radii, which then results in higher temperatures to cover for the total observed luminosity.

Given the large uncertainties on the $\log g$, we are unable to determine precise masses for both companions using the fitted spectra. Hence, we adopted the values determined from the comparison of the photometry to AMES-Cond \citep{Allard_2012} evolutionary tracks. We compute individual values for each photometric filter ($K1$, $K2$, and $L'$) and also apply synthetic filters for $Y$, $J$, and $H$ bands on the IFS spectrum and also consider these as individual values. We compute the mean and error as given in \autoref{tab: HD101412 properties}.
\begin{figure}[!ht]
    \centering
    \includegraphics[width=\linewidth]{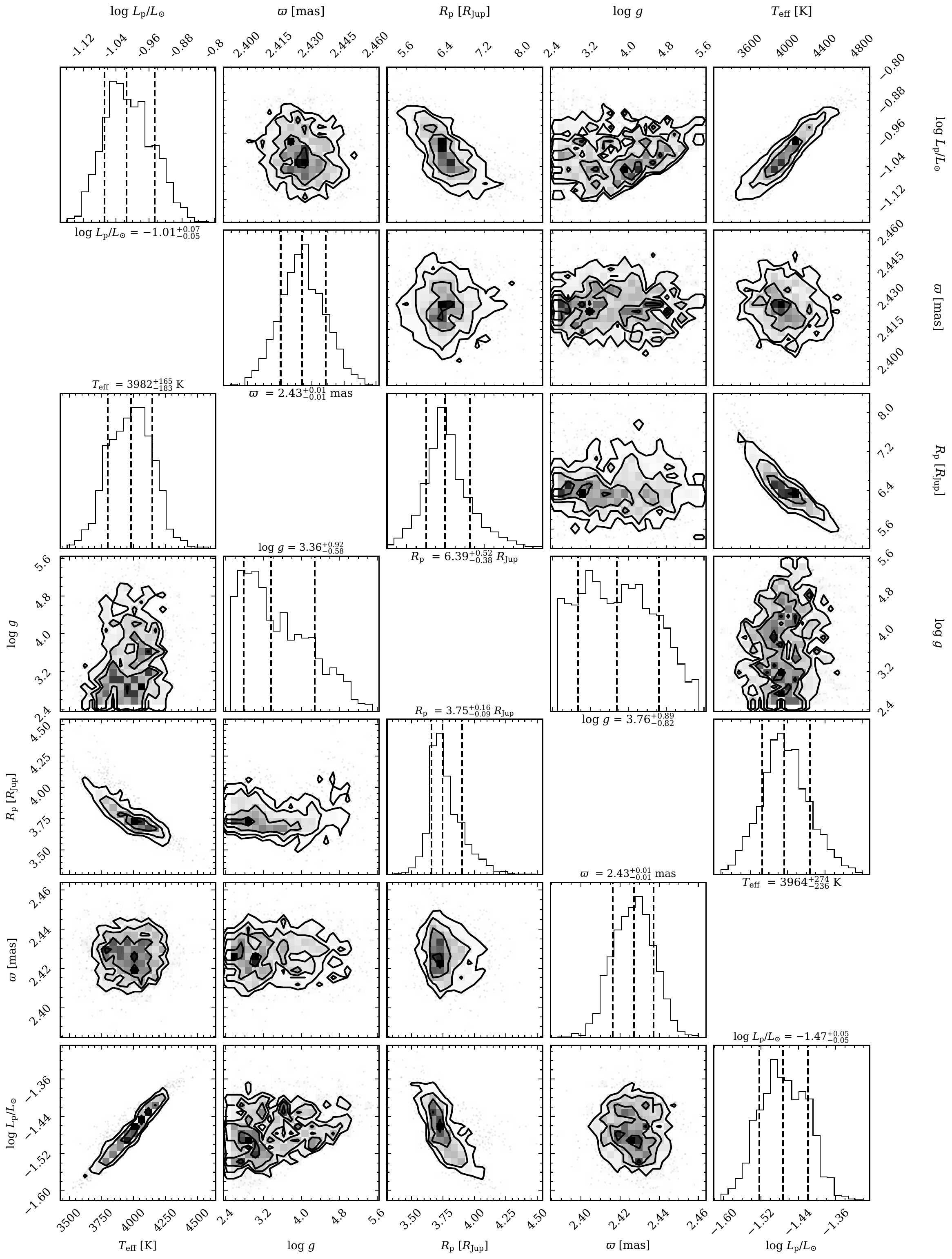}
    \vspace{-0.5cm}
    \caption{Corner plots of the \texttt{species} posteriors. Companion B is the top right, C the lower-left corner. \label{fig: corner plots}}
\end{figure}
\newpage
\FloatBarrier
\section{Observations}
\autoref{table: observations} lists all observed targets with their corresponding observation nights and filter settings, reference target for RDI if applicable, integration time, field rotation angle, and average seeing.
\begin{table}[H]
\caption{Observed targets in the \ispysphere sample.}
\label{table: observations}  
\resizebox{\textwidth}{!}{
\begin{tabular}{l c c l l l c c c c c}
\hline  
\hline  
Target & \makecell{RA \\ {[hh:mm:ss]}} & \makecell{Dec. \\ {[dd:mm:ss]}} & \makecell{Observation \\ night} & Filter & \makecell{Reference \\ target} & \makecell{Int. \\ time \\ {[h]}} & \makecell{Exp. \\ time \\ {[s]}} & \makecell{Rot. \\ angle \\ {[deg]}} & \makecell{Seeing \\ {[arcsec]}} \\
\hline
HD\,377 & 00:08:25.74 & +06:37:00.48 & 2023 October 04 & $H2$/$H3$ & HD\,342 & 0.9 &  8 & 37 & 0.50\\
HD\,1461 & 00:18:41.86 & -08:03:10.80 & 2022 October 14 & $H2$/$H3$ & HD\,2195 & 1.6 & 32 & 80 & 0.64 \\
\makecell[c]{\vdots} & \makecell[c]{\vdots}  & \makecell[c]{\vdots}  & 2025 November 24 & \makecell[c]{\vdots}  & \makecell[c]{\vdots} & 0.8 & 8  & 49 & 0.58 \\
\makecell[c]{\vdots} & \makecell[c]{\vdots}  & \makecell[c]{\vdots}  & 2025 November 25 & \makecell[c]{\vdots}  & \makecell[c]{\vdots} & 0.6 & 8  & 41 & 0.53 \\
$\epsilon$~Eri & 03:32:55.84 & -09:27:29.73 & 2022 October 14 & $H2$/$H3$ & ... & 1.1 &  1 & 74 & 0.55 \\
HD\,59659 & 07:28:30.06 & -49:08:58.89 & 2023 March 30 & $H2$/$H3$ & HD\,58826 & 0.6 &  8 & 36 & 0.80 \\
HD\,101412 & 11:39:44.45 & -60:10:27.71 & 2023 March 30 & $K1$/$K2$ & TYC\,8972-2028-1 & 0.6 &  8 & 19 & 1.06 \\
HD\,105912 & 12:11:21.79 & -03:46:43.93 & 2023 March 30 & $H2$/$H3$ & HD\,106976 & 0.4 &  8 & 110 & 0.77 \\
HD\,121617 & 13:57:41.12 & -47:00:34.25 & 2023 March 30 & $H2$/$H3$ & HD\,120275A & 1.1 &  8 & 39 & 0.86 \\
HD\,131511 & 14:53:23.76 & +19:09:10.07 & 2023 March 30 & $H2$/$H3$ & HD\,131302 & 1.1 &  8 & 35 & 0.85 \\
$\epsilon$~Sgr & 18:24:10.31 & -34:23:04.61 & 2023 March 30 & $H2$/$H3$ & ... & 1.8 &  4 & 39 & 0.84 \\
HD\,179218 & 19:11:11.25 & +15:47:15.63 & 2023 October 04 & $H2$/$H3$ & ... & 0.6 &  8 & 15 & 0.49 \\
HD\,181327 & 19:22:58.94 & -54:32:16.97 & 2023 October 04 & $BB~H$  & HD\,180134 & 1.4 &  8 & 61 & 0.63 \\
HD\,191089 & 20:09:05.21 & -26:13:26.52 & 2023 October 04 & $H2$/$H3$ & HD\,191052 & 1.1 &  8 & 15 & 0.46 \\
HD\,207129 & 21:48:15.75 & -47:18:13.01 & 2022 October 14 & $H2$/$H3$ & HD\,207400 & 1.4 & 32 & 79 & 0.79 \\
\makecell[c]{\vdots} & \makecell[c]{\vdots}  & \makecell[c]{\vdots}  & 2023 October 04 & $BB~H$  &  HD\,206690 & 1.0 & 8  & 36 & 0.92 \\
\hline
\end{tabular}}
\tablefoot{Names and coordinates for the observed target stars in the \ispysphere sample as well as the dates the observations were carried out, the \irdis filters that were used, names of reference stars if observed, total integration time in hours, single-frame exposure time in seconds, parallactic angle rotation that was carried out, and the seeing FWHM (averaged over the observations) in units of arcsec.}    
\end{table}
\end{appendix}
\end{document}